\documentclass[11pt,a4paper]{article}
\pdfoutput=1 
             
\usepackage{graphicx}
             
\usepackage[utf8x]{inputenc}	
\usepackage[english]{babel}
\usepackage{paralist}
\usepackage{amsmath}
\numberwithin{equation}{section}
\usepackage{amssymb}
\usepackage{amsthm}
\usepackage{MnSymbol}
\usepackage{mathtools}		
\usepackage{dsfont}		
\usepackage{stmaryrd}		
\usepackage{cite}		
\usepackage[svgnames]{xcolor}
\usepackage{enumerate}
\usepackage{setspace}
\usepackage{booktabs}
\usepackage{multirow}
\usepackage{tikz}
\usetikzlibrary{shapes}
\usetikzlibrary{positioning}
\usetikzlibrary{chains}
\usetikzlibrary{arrows,fit,decorations.pathreplacing, shapes.misc}
\usetikzlibrary{decorations.markings}
\tikzstyle{every picture}+=[remember picture]
\tikzstyle{na} = [baseline=-.5ex]
\usepackage[font=small,labelfont=bf]{caption}
\usepackage[font=small,labelfont=bf]{subcaption}
\usepackage[pdftex,breaklinks,colorlinks=true,urlcolor=RoyalBlue,linkcolor=blue,
citecolor=blue]{hyperref}

\def\Tr{{\rm Tr}}
\newcommand{\PE}{\mathrm{PE}}
\newcommand{\diff}{\mathrm{d}}
\newcommand{\im}{\mathrm{i}}

\newcommand{\Ncal}{\mathcal{N}}
\newcommand{\Scal}{\mathcal{S}}
\newcommand{\Lcal}{\mathcal{L}}
\newcommand{\Ocal}{\mathcal{O}}
\newcommand{\Wcal}{\mathcal{W}}

\newcommand{\C}{\mathbb{C}}
\newcommand{\R}{\mathbb{R}}
\newcommand{\T}{\mathbb{T}}
\newcommand{\Z}{\mathbb{Z}}

\newcommand{\surm}{\mathrm{SU}}
\newcommand{\urm}{\mathrm{U}}

\newcommand{\sorm}{\mathrm{SO}}

\newcommand{\str}{\mathrm{str}}
\newcommand{\Deff}{\mathrm{def}}
\newcommand{\sh}{\mathrm{sh}}

\catcode`@=11
\allowdisplaybreaks

\begin{document}
\begin{titlepage}
\setcounter{page}{0}

\begin{center}

{\Large\bf 
Elliptic Quantum Curves of Class $\Scal_k$
}

\vspace{15mm}

{\large Jin Chen${}^{1}$},\ 
{\large Babak Haghighat${}^{1}$},\ 
{\large Hee-Cheol Kim${}^{2,3}$},\  and \ 
{\large Marcus Sperling${}^{1}$} 
\\[5mm]
\noindent ${}^{1}${\em Yau Mathematical Sciences Center, Tsinghua University}\\
{\em Haidian District, Beijing, 100084, China}\\
{Email: {\tt jinchen@mail.tsinghua.edu.cn},} \\  
{{\tt babakhaghighat@tsinghua.edu.cn},} \\  
{{\tt msperling@mail.tsinghua.edu.cn }}
\\[5mm]
\noindent ${}^{2}${\em Department of Physics, POSTECH}\\
{\em  Pohang 790-784, Korea}\\
{Email: {\tt heecheol@postech.ac.kr}}
\\[5mm]
\noindent ${}^{3}${\em Asia Pacific Center for 
Theoretical Physics, POSTECH}\\
{\em Pohang 37673, Korea}
\\[5mm]
\vspace{15mm}

\begin{abstract}
Quantum curves arise from Seiberg-Witten curves associated to 4d $\Ncal=2$ gauge theories by promoting coordinates to non-commutative operators. In this way the algebraic equation of the curve is interpreted as an operator equation where a Hamiltonian acts on a wave-function with zero eigenvalue. We find that this structure generalises when one considers torus-compactified 6d $\Ncal=(1,0)$ SCFTs. The corresponding quantum curves are elliptic in nature and hence the associated eigenvectors/eigenvalues can be expressed in terms of Jacobi forms. In this paper we focus on the class of 6d SCFTs arising from M5 branes transverse to a $\C^2/\Z_k$ singularity. In the limit where the compactified 2-torus has zero size, the corresponding 4d $\Ncal=2$
theories are known as class $\Scal_k$. We explicitly show that the eigenvectors associated to the quantum curve are expectation values of codimension 2 surface operators, while the corresponding eigenvalues are codimension 4 Wilson surface expectation values.  
\end{abstract}

\end{center}

\end{titlepage}
{\baselineskip=12pt
{\footnotesize
\tableofcontents
}
}
%
%
\section{Introduction}
\label{sec:intro}
Since their classification \cite{Heckman:2013pva,Heckman:2015bfa}, 6d 
superconformal field theories (SCFTs) with 8 supercharges have played a 
prominent role in constructing lower dimensional quantum field theories. In 
particular, it appears that 5d SCFTs arise as compactifications of such 6d 
theories with Wilson line expectation values for background flavour fields 
turned on\cite{Jefferson:2017ahm,Jefferson:2018irk}, while 5d theories of KK type admitting 
an affine quiver description can be understood as twisted compactifications of 
6d SCFTs \cite{Bhardwaj:2018yhy,Bhardwaj:2018vuu,Bhardwaj:2019fzv}. 
Moreover, 4d $\Ncal=1$ SCFTs can be understood as compactifications on 
Riemann surfaces with fluxes \cite{Gaiotto:2015usa,Kim:2017toz,Razamat:2016dpl,Kim:2018lfo,Razamat:2018gro,Kim:2018bpg,Chen:2019njf}. 

In this paper we focus on 6d SCFTs arising from $N$ M5 branes probing 
$\C^2/\Z_k$ singularities. When compactified on a 2-torus 
$\T^2$, BPS partition functions of such theories have been computed in 
\cite{Haghighat:2013gba,Kim:2012qf} ($k=1$) and \cite{Haghighat:2013tka} ($k > 
1$). As it turns out, a crucial property of these partition functions is that 
they can be expressed in terms of an infinite sum over elliptic genera of BPS 
strings wrapping the torus. These elliptic genera are Jacobi forms with modular 
parameter $\tau$, being the complex structure of the torus, and several 
elliptic 
parameters arising from gauge, flavour, and R-symmetry chemical potentials. 
Using the correspondence described in the first paragraph, the 
torus-compactified theory can be equally understood as a circle 
compactification 
of a 5d gauge theory whose moduli space of vacua also carries this elliptic 
structure \cite{Haghighat:2017vch}. In particular, the corresponding 
Seiberg-Witten curve can be expressed in terms of a polynomial in a variable 
$t$ 
whose coefficients are Jacobi forms $v_l$ of an elliptic parameter $z$:
\begin{equation} \label{eq:algcurve}
	H(w,z) = t^{N} + v_1(z) t^{N-1} + \ldots + v_l(z) t^{N-l} + \ldots v_N(z) = 
0, \quad t = e^{2\pi i w}. 
\end{equation}
 A central question is about the interpretation of this curve as a 
\emph{quantum curve}. To this end, the variables $w$ and $z$ are promoted to 
operators satisfying a non-trivial commutation relation
 \begin{equation}
 	\left[\hat{w},\hat{z}\right] \sim \hbar. 
 \end{equation}
Interpreting $\hat{z}$ as a position operator, by the above commutation 
relation 
$\hat{w}$ becomes a momentum operator and $Y \equiv e^{-\hat{w}}$ will be a 
shift operator. In this framework the algebraic curve equation 
(\ref{eq:algcurve}) becomes a difference equation in the sense that the 
operator 
$\widehat{H}(\hat{w},\hat{z})$ acts on a wave-function with zero eigenvalue. 
This notion of a quantum curve is intimately related to partition functions 
arising from surface defects in gauge theories \cite{Gaiotto:2014ina}. In this 
interpretation the wave-function annihilated by the operator 
$\widehat{H}(\hat{w},\hat{z})$ is the expectation value of a codimension 2 
defect operator. In the context of our 6d SCFT such defect operators arise from 
half BPS operators extended over $\T^2 \times \R^2$ and 
localised at a point on the remaining $\R^2$. 	Localisation is done by 
turning on the Omega-background $\R^4_{\epsilon_1,\epsilon_2} \times 
\T^2$ \cite{Nekrasov:2010ka,Haghighat:2013gba} and $\hbar$ is 
identified with $\epsilon_1$, while $\epsilon_2$ is sent to zero in the 
Nekrasov-Shatashvili limit \cite{Nekrasov:2009rc}. The theory living on the 
defect flows in the IR to a 4d $\mathcal{N}=1$ SCFT and in some instances the 
defect partition function in the NS-limit can be understood as the 
superconformal index of this SCFT on $S^1 \times S^3$ \cite{Nazzal:2018brc}. In 
this correspondence, the $S^3$ is understood as a Hopf-fibration of a circle 
over a two-sphere such that the two circles are identified with $\mathbb{T}^2$ 
and the two-sphere is identified with a compactification of $\mathbb{R}^2$.

From a more geometric point of view, 6d SCFTs can be engineered by 
compactifying 
F-theory on an elliptic Calabi-Yau manifold. Performing F-theory/M-theory duality, one observes that the BPS 
partition function of the theory on $\T^2 \times \R^4_{\epsilon_1,\epsilon_2}$ 
corresponds to the refined topological string partition function of the 
Calabi-Yau manifold \cite{Haghighat:2013gba,Haghighat:2013tka}. In this picture, the surface defect arises from an 
M5 brane wrapping a Lagrangian cycle inside the Calabi-Yau threefold and 
extended over $S^1 \times \R^2$ transverse to the Calabi-Yau. The theory living 
on such a defect is expected to flow to a 3d SCFT with four supercharges 
coupled 
to the parent 5d gauge theory. Using the 3d/3d correspondence of 
\cite{Dimofte:2011ju}, the partition function of the 3d SCFT is equivalent to 
the partition function of $\mathrm{SL}(2,\C)$ Chern-Simons theory on a 
three-manifold which is a knot complement. As is well-known, the moduli space 
of flat $\mathrm{SL}(2,\C)$ connections on the knot complement is characterised 
by the so-called A-polynomial $A(z,w)$ where $z$ and $w$ characterise 
holonomies around the two cycles of the boundary torus. The equation $A(z,w) = 
0$ then describes the subspace of those holonomies which can be extended to the 
entire three-manifold. The partition function of $\mathrm{SL}(2,\C)$ 
Chern-Simons theory on the knot complement satisfies a difference equation 
which arises from the quantisation of the A-polynomial 
\cite{Gukov:2003na,garoufalidis2003characteristic,Garoufalidis_2005}. By the 
3d/3d 
correspondence, the partition function of the 3d SCFT then satisfies the same 
difference equation. In the case of our 3d defect, the 3d SCFT is coupled to a 
5d gauge theory and the A-polynomial receives a $Q$-deformation 
\cite{Gukov:2011qp,Aganagic:2012jb,Aganagic:2013jpa,Kashaev:2015kha} where by $Q$ we collectively denote the 
moduli of the 5d theory. The quantised $Q$-deformed A-polynomial can then be 
identified with our difference operator $H(\hat{w},\hat{z})$. As our 5d theory 
arises from a 6d SCFT, we find that the difference operator is elliptic with 
elliptic modulus $Q= e^{2\pi i\tau}$.

The concrete example, on which we focus in this paper, is the 6d SCFT 
arising from $2$ M5 branes probing a $\Z_k$ singularity. In this case, 
compactification on a two-torus leads to the following Seiberg-Witten curve 
\cite{Haghighat:2017vch}
\begin{equation}
	t  + q_{\phi} \prod_{l=1}^{2k}\vartheta_1(z-\mu_l)~t^{-1} - (1 + q_{\phi}) 
\prod_{l=1}^k \vartheta_1(z - z_l)= 0,
\end{equation}
where $q_{\phi} \equiv e^{2\pi i \phi}$ with $\phi$ being the tensor branch 
parameter of the 6d theory, the $\mu_l$ denote collectively the flavour 
chemical potentials, and $z_l$ are complicated functions of gauge chemical 
potentials. For the definition of the theta functions $\vartheta_1$ we refer to 
 Appendix \ref{app:theta_fct}.
A central result of the present paper is that the defect partition 
function $\Psi$ of the torus-compactified (or equivalently the 
circle-compactified 5d 
affine quiver gauge theory) satisfies the following difference equation 
corresponding to the quantisation of the above algebraic curve
\begin{equation}
		\left[Y^{-1} + q_{\phi} \prod_{l=1}^{2k} \vartheta_1(z-\mu_l) \cdot Y - 
\langle \mathcal{W} \rangle \right] \Psi = 0,
\end{equation}
where we have identified 
\begin{equation}
	\langle \mathcal{W} \rangle \equiv (1+q_{\phi}) \prod_{l=1}^k \vartheta_1(z 
+ \epsilon_1 - z_l),
\end{equation}
with $\langle \mathcal{W} \rangle$ the Wilson surface expectation value of a 
codimension 4 operator wrapping the torus to be further specified in the 
main text.

The remainder of this paper is organised as follows:
After reviewing the 6d $\Ncal=(1,0)$ theory and its partition function, Section 
\ref{sec:partition_fcts} details the inclusion of codimension 2 and 4 defects. 
For both cases, the partition functions are derived and evaluated up to order 
$q_\phi^2$.
Thereafter, the difference equation is derived in Section 
\ref{sec:difference_eq}. In detail, starting from a path integral 
representation for the  partition function of the codimension 2 defect, the 
corresponding saddle point equation naturally leads to a difference equation. 
Crucially, one contribution of the difference equation is identified with the 
partition function of the codimension 4 defect.
The 6d theories originating from 2 M5 branes on a $\C^2\slash \Z_k$ 
family have 8 supercharges for $k>1$, but 16 supercharged for $k=1$. The 
analysis of this enhanced $\Ncal=(2,0)$ case is presented in Section 
\ref{sec:enhanced_susy}, and compared to the dual 5d $\Ncal=2$ theory.
Finally, Section \ref{sec:conclusion} provides a conclusion and outlook.
Appendix \ref{app:details_part_fct} contains definitions and conventions used 
in the evaluation of the various partition functions as well as computational 
results. As a remark, most computational details are delayed to Appendix 
\ref{app:details_part_fct} in order to ease the readability of the main text.
%
%
\section{Defects for M5 branes on A-type singularity}
\label{sec:partition_fcts}
The $6$d $\Ncal=(1,0)$ SCFTs originating from $N$ M5 branes on a $A$-type 
singularity $\C^2\slash \Z_k$ are naturally labeled by two integers $(N,k)$. 
For $k=1$, the 6d world-volume theories have enhanced supersymmetry and are 
known as the $A_{N-1}$ $\Ncal=(2,0)$ theories 
\cite{Witten:1995zh,Strominger:1995ac}, whose 4d descendants are the $A_{N-1}$ 
$\Ncal=2$ theories of class $\Scal$ \cite{Gaiotto:2009we}.
For $k>1$, the resulting $\Ncal=(1,0)$ world-volume theories are well-studied 
\cite{Hanany:1997gh,Brunner:1997gf,Gaiotto:2014lca,DelZotto:2014hpa} and 
their 4d descendants are the $\Ncal=1$ theories of class $\Scal_k$ 
\cite{Gaiotto:2015usa}.
In this section, the set-up is reviewed and, thereafter, defects of codimension 2 and 4 are introduced. 

\begin{figure}
\centering
\begin{tikzpicture}
\hspace{-0.5cm}
 \node (MP) at (0,0) {
 \begin{tabular}{c|cc|cccc|c|cccc}
 \toprule
\multirow{2}{*}{M-theory}  & \multicolumn{2}{c|}{$\T^2$} & 
\multicolumn{4}{c|}{$\R^4_{\epsilon_1,\epsilon_2}$} & & \multicolumn{4}{c}{$ 
\mathrm{TN}_{k}$}\\
  & 0 & 1 & 2 & 3& 4 & 5 & 6 & 7 & 8 & 9 & 10 \\ \midrule
 $2$ M5 & $\bullet$ & $\bullet$  & $\bullet$ & $\bullet$& $\bullet$ & 
$\bullet$& 
 \\
 $l$ M2 & $\bullet$ & $\bullet$ & & & & & $\bullet$ & \\
 $1$ $\widetilde{\text{M5}}$ & $\bullet$ & $\bullet$ & &  & $\bullet$ & 
$\bullet$ & & 
$\bullet$  & & & $\bullet$  \\
 $1$ M$5^\prime$ & $\bullet$ & $\bullet$ & & & & & & 
$\bullet$ & $\bullet$ & $\bullet$& $\bullet$
 \\ \bottomrule 
\end{tabular}
 };
\node (L1) at (-4.5,-4.5) {
\begin{tabular}{c|c|cccc|c|cccc}
 \toprule
\multirow{2}{*}{IIA} & $S^1$ & 
\multicolumn{4}{c|}{$\R^4_{\epsilon_1,\epsilon_2}$} & & \multicolumn{4}{c}{$ 
\mathrm{TN}_{k}$}\\
& 0  & 2 & 3& 4 & 5 & 6 & 7 & 8 & 9 & 10 \\ \midrule
 $2$ D4 & $\bullet$   & $\bullet$ & $\bullet$& $\bullet$ & $\bullet$& 
 \\
 $l$ F1 & $\bullet$  & & & & & $\bullet$ & \\
 $1$ $\widetilde{\text{D4}}$ & $\bullet$ & &  & $\bullet$ & $\bullet$ 
& & $\bullet$ & & & $\bullet$  \\
 $1$ D$4^\prime$ & $\bullet$  & & & & & & 
$\bullet$ & $\bullet$ & $\bullet$& $\bullet$
 \\ \bottomrule 
\end{tabular}
};
\node (R1) at (4.5,-4.75) {
\begin{tabular}{c|cc|cccc|c|cccc}
 \toprule
\multirow{2}{*}{IIA} & \multicolumn{2}{c|}{$\T^2$} & 
\multicolumn{4}{c|}{$\R^4_{\epsilon_1,\epsilon_2}$} & & \\
  & 0 & 1 & 2 & 3& 4 & 5 & 6  & 7 & 8 & 9 \\ \midrule
 $2$ NS5 & $\bullet$ & $\bullet$  & $\bullet$ & $\bullet$& $\bullet$ & 
$\bullet$& 
 \\
 $l$ D2 & $\bullet$ & $\bullet$ & & & & & $\bullet$ & \\
 $1$ $\widetilde{\text{D4}}$ & $\bullet$ & $\bullet$ & &  & $\bullet$ 
& $\bullet$ & &  $\bullet$  &\\
 $1$ D$4^\prime$ & $\bullet$ & $\bullet$ & & & & & & 
 $\bullet$ & $\bullet$& $\bullet$ \\
  $k$ D6 & $\bullet$ & $\bullet$ & $\bullet$ & $\bullet$ & $\bullet$ & 
$\bullet$ & $\bullet$
 \\ \bottomrule 
\end{tabular}
};
\node (L2) at (-4.5,-9.5) {
\begin{tabular}{c|c|cccc|c|cccc}
 \toprule
\multirow{2}{*}{IIB} & $S^1$ & 
\multicolumn{4}{c|}{$\R^4_{\epsilon_1,\epsilon_2}$} & & $S^1$ & \\
 & 0  & 2 & 3& 4 & 5 & 6 & 7 & 8 & 9 & 10 \\ \midrule
 $2$ D5 & $\bullet$   & $\bullet$ & $\bullet$& $\bullet$ & $\bullet$& & & & & 
$\bullet$
 \\
 $l$ F1 & $\bullet$  & & & & & $\bullet$ & \\
 $1$ $\widetilde{\text{D3}}$ & $\bullet$ & &  & $\bullet$ & $\bullet$ & &  
$\bullet$  &\\
 $1$ D$3^\prime$ & $\bullet$  & & & & &   & $\bullet$ & 
$\bullet$& 
$\bullet$ \\
$k$ NS5 & $\bullet$ & $\bullet$ & $\bullet$ & $\bullet$ & $\bullet$ & $\bullet$
 \\ \bottomrule 
\end{tabular}
};
\node (R2) at (4.5,-9.5) {
\begin{tabular}{c|cc|cccc|c|cccc}
 \toprule
 \multirow{2}{*}{IIB} & $S^1$ & $S^1$ & 
\multicolumn{4}{c|}{$\R^4_{\epsilon_1,\epsilon_2}$} & & \\
  & 0 & 1 & 2 & 3& 4 & 5 & 6  & 7 & 8 & 9 \\ \midrule
 $2$ NS5 & $\bullet$ & $\bullet$  & $\bullet$ & $\bullet$& $\bullet$ & 
$\bullet$& 
 \\
 $l$ D1 & $\bullet$ &  & & & & & $\bullet$ & \\
 $1$ $\widetilde{\text{D3}}$ & $\bullet$ & & &  & $\bullet$ & 
$\bullet$ & &  $\bullet$  &\\
 $1$ D$3^\prime$ & $\bullet$ &  & & & & & & 
 $\bullet$ & $\bullet$& $\bullet$ \\
  $k$ D5 & $\bullet$ &  & $\bullet$ & $\bullet$ & $\bullet$ & 
$\bullet$ & $\bullet$
 \\ \bottomrule 
\end{tabular}
};
\draw[thick,->] (MP)-- node[left] {$\substack{\text{reduction} \\ S^1\subset 
\R_1}$} (L1);
\draw[thick,->] (MP)-- node[right] {$\substack{\text{reduction} \\ S^1\subset 
\R_{10}}$} (R1);
\draw[thick,<->] (L1)--node[left] {$\substack{\text{T-dual} \\ 
\text{on } \R_{10}}$} (L2);
\draw[thick,<->] (R1)--node[right] {$\substack{\text{T-dual} \\ 
\text{on } \R_1}$} (R2);
\draw[thick,<->] (L2) .. controls (0,-12) .. (R2);
\node at (0,-12.25) {$\substack{ \text{S-dual}\\\text{and }x^{10} 
\leftrightarrow x^1  }$};
\end{tikzpicture}
\caption{Brane set-up for codimension 2 and codimension 4 defect. The theory 
without defect is realised via 2 M5 branes transverse to a Taub-NUT space 
$\mathrm{TN}_k$, which is a resolution of the $\C^2\slash \Z_k$ singularity. The isometries of 
$\mathrm{TN}_k$ can be identified with the admissible $\epsilon_{1,2}$ twists  
defining the Omega background and the $\urm(1)_b$ symmetry. The codimension 2 
defect is realised via an additional $\widetilde{\text{M5}}$ brane, while the 
codimension 4 defect corresponds to an M$5^\prime$ brane. Assuming the M-theory 
circle is either along $x^1$ or $x^{10}$ direction, one arrives at two 
different Type IIA realisation. The reduction on $x^1$ leads to branes on a 
non-trivial $\mathrm{TN}_k$ background, while the reduction on $x^{10}$ results 
in an intersecting brane configuration on a flat background. Further employing 
T-duality on either $x^{10}$ or $x^1$, respectively, leads to two 5-brane web 
configurations, which are S-dual to one another. }
\label{fig:branes}
\end{figure}
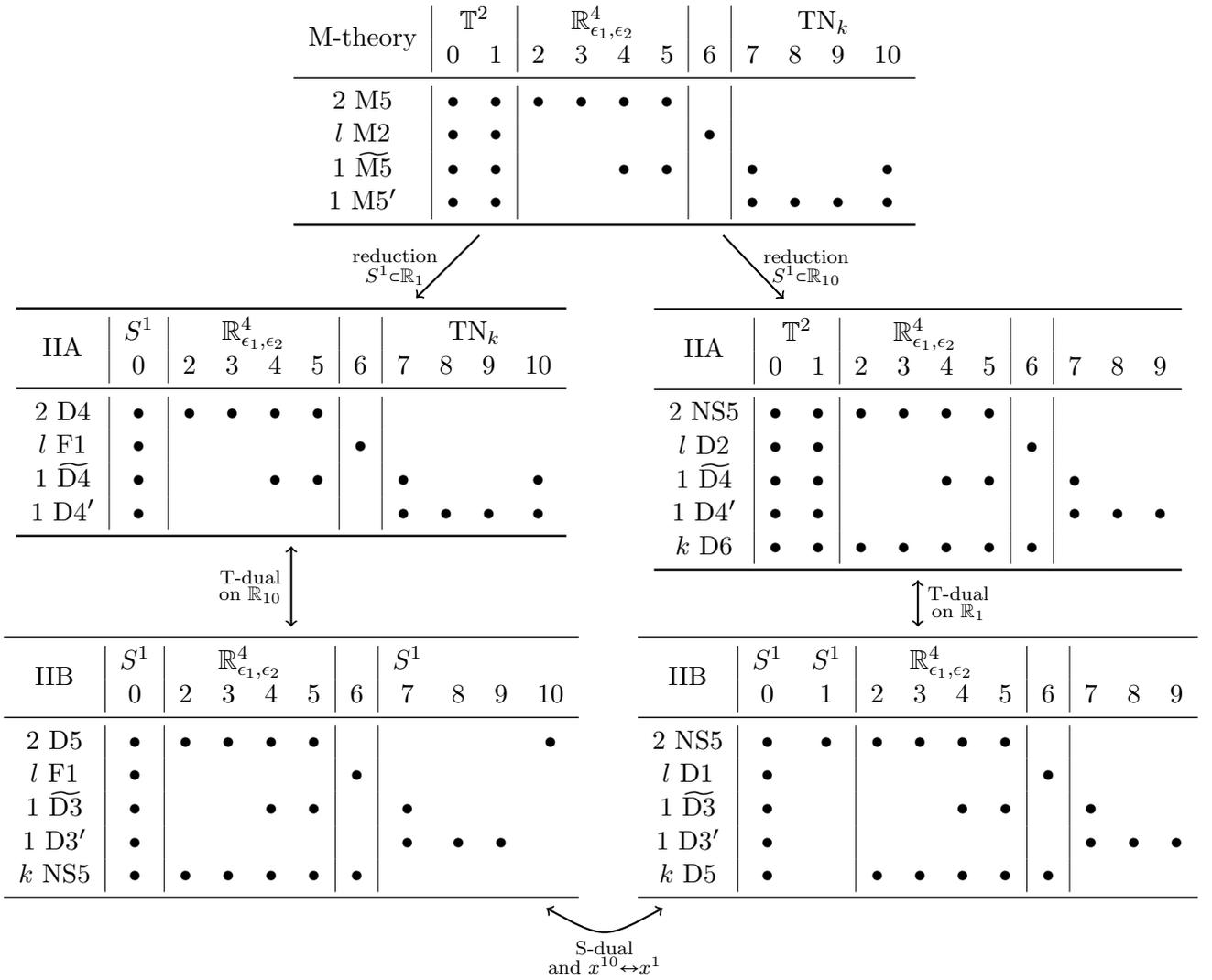
\subsection{2 M5 branes on A-type singularity}
\label{sec:6d_no_defect}
In this work, the focus is placed on $6$d $\Ncal=(1, 0)$ SCFTs for $N=2$. 
The M-theory set-up admits a dual realisation in Type IIA superstring 
theory. 
The 2 M5 branes become NS5 branes filling the space-time directions 
$x^{0},x^1,\ldots,x^5$ and being points in the transverse directions. The 
A-type ALE space $\C^2\slash \Z_k$ dualises into a stack of $k$ D6 branes 
filling space-time directions $x^{0},x^1,\ldots,x^6$, which are 
transverse to the original singularity. The set-up is summarised in Figure 
\ref{fig:branes}.
The 6d $\Ncal=(1,0)$ low-energy effective theory living on 
the world-volume of the D6 branes is composed of hypermultiplets and vector 
multiplets encoded in the following quiver diagram for 8 supercharges:
\begin{align}
	\raisebox{-.5\height}{
 	\begin{tikzpicture}
	\tikzstyle{gauge} = [circle, draw,inner sep=3pt];
	\tikzstyle{flavour} = [regular polygon,regular polygon sides=4,inner 
sep=3pt, draw];
	\node (g1) [gauge,label=below:{$\surm (k)_a$}] {};
	\node (f1) [flavour,left of=g1,label=left:{$\surm(k)_m$}] {};
	\node (f2) [flavour,right of=g1,label=right:{$\surm(k)_n$}] {};
	\draw (g1)--(f1) (g1)--(f2);
	\node at (-0.5,0.25) {$\scriptstyle{-b}$};
	\node at (0.5,0.25) {$\scriptstyle{b}$};
	\node at (0,0.75) {$\urm(1)_b$};
	\end{tikzpicture}
	}
	\qquad \cong
	\qquad
	\raisebox{-.5\height}{
	\begin{tikzpicture}
	\tikzstyle{gauge} = [circle, draw,inner sep=3pt];
	\tikzstyle{flavour} = [regular polygon,regular polygon sides=4,inner 
sep=3pt, draw];
	\node (g1) [gauge,label=below:{$\surm (k)_a$}] {};
	\node (f1) [flavour,above of=g1,label=above:{$\surm(2k)_y$}] {};
	\draw (g1)--(f1);
	\end{tikzpicture}
	}
	\label{eq:6d_quiver}
\end{align}
and one tensor multiplet. The global symmetry $\surm(2k)_y$ can be decomposed into $\surm(k)_{m,n}$, from the two stacks of semi-infinite D6 branes for $x^6 \to \pm \infty$, and $\urm(1)_b$, which is the $\C^2\slash \Z_k$ isometry.

As a remark, the general family, i.e.\ $N$ M5 branes on a $\mathbb{C}^2\slash 
\mathbb{Z}_{k}$ or $N$ NS5s intersected by $k$ D6 branes in Type IIA, 
leads to a 6d $\Ncal=(1,0)$ quiver gauge theory on the tensor branch with 
global symmetry 
$\surm(k)_m \times \urm(1) \times \surm(k)_n$. For $N=2$ there exists an 
accidental 
enhancement $\surm(k)_m \times \urm(1) \times \surm(k)_n\subset \surm(2k)$ as indicated in \eqref{eq:6d_quiver}.
\paragraph{Partition function.}
In order to evaluate the partition function, the $6$d theory is placed on 
$\T^2\times \R^4_{\epsilon_1,\epsilon_2}$, where the 2-torus is along $x^{0,1}$ 
and the 4d Omega background $\R^4_{\epsilon_1,\epsilon_2}$ fills directions 
$x^2,\ldots,x^4$, see Figure \ref{fig:branes}. The two 
parameters $\epsilon_1$ and $\epsilon_2$ denote rotations in the $x^{2,3}$ and 
$x^{4,5}$ planes, respectively. The full partition function is composed of two 
contributions 
\begin{align}
 Z_{\mathrm{6d}} = Z_{\mathrm{pert}} \cdot Z_{\str}
\end{align}
denoting the perturbative contributions $Z_{\mathrm{pert}}$ and the 
non-perturbative contributions $Z_{\str}$. The perturbative part is 
fully determined by the 6d supermultiplets in \eqref{eq:6d_quiver} plus a single tensor multiplet. In contrast, the non-perturbative 
parts originate from the $2$d $\Ncal=(0,4)$ world-volume theories of D2 branes 
filling $x^0,x^1,x^6$ directions, see Figure \ref{fig:branes}.
The instanton string partition function can be written as sum of 
elliptic genera of the 2d theories:
\begin{align}
\label{eq:full_inst_partition}
Z_{\str}
=
1+ 
\sum_{l=1}^{\infty} e^{2 \pi \im\, \cdot l \phi } Z_l 
\equiv 
\sum_{l=0}^{\infty} q_{\phi}^l Z_l 
\quad 
\text{with}
\quad 
q_\phi= e^{2 \pi \im\, \cdot \phi }
\,,
\end{align}
where $\phi$ is the vacuum expectation value of the scalar field in the tensor multiplet.
The BPS partition functions have been computed for $k=1$ in 
\cite{Kim:2012qf,Haghighat:2013gba} and for $k>1$ in \cite{Haghighat:2013tka}.
In this work, the partition function of the 6d $\Ncal=(1,0)$ without defect is 
required for the computation of the normalised partition function in the 
presence of defects, see Appendix \ref{app:normalised_part_fct}.
For completeness and concreteness, the details of $Z_{\mathrm{pert}}$ and $ Z_{\str}$ 
are discussed in turn in the following subsections.

\subsubsection{Perturbative contribution}
Following \cite{Hayashi:2016abm}, the perturbative single-letter contribution 
of the 6d supermultiplets are given as follows:
\begin{subequations}
\label{eq:single_letter_6d}
\begin{align}
 I_{\mathrm{tensor}} &= 
 -\frac{p+q}{(1-p)(1-q)}
\\
 I_{\mathrm{vector}} &=
 -\frac{(1+p\cdot q)}{(1-p)(1-q)} \left(\sum_{i,j=1}^k e^{a_i - a_j} -1 \right)
 \notag \\
 &=
 -\frac{(1+p\cdot q)}{(1-p)(1-q)} \left( (k-1) + \sum_{1\leq j < i \leq k} 
\left( e^{a_i - a_j} + e^{a_j- a_i} \right) \right)
 \\
  I_{\mathrm{hyper}} &= \frac{\sqrt{p\cdot q}}{(1-p)(1-q)} 
  \sum_{i=1}^k \left\{ 
  \sum_{l=1}^{k}  \left( e^{a_i - m_l+b} +  e^{m_l-b-a_i} \right) 
  +\sum_{l=1}^{k}  \left( e^{a_i - n_l-b} +  e^{n_l+b-a_i} \right) 
  \right\}
\end{align}
\end{subequations}
where the $\{a_i\}$ gauge as well as the $\{m_l\}$, $\{n_l\}$, and $b$ flavour charges of the hypermultiplets are derived from \eqref{eq:6d_quiver}. 
The $\surm(k)_a$ gauge as well as the $\surm(k)_{n,m}$ flavour fugacities need to satisfy 
\begin{align}
 \prod_{i=1}^{k} e^{a_i} = 
  \prod_{l=1}^{k} e^{m_l} =
   \prod_{l'=1}^{k} e^{n_{l'}} =1 
   \quad \Leftrightarrow \quad
   \sum_{i=1}^k a_i = 
   \sum_{l=1}^k m_l = 
   \sum_{l'=1}^k n_{l'} =0 
   \,.
   \label{eq:SU_fugacities}
\end{align}
Moreover, $p=e^{2\pi \im\, \epsilon_1}$, $q=e^{2\pi \im\, \epsilon_2}$ denote the Cartan generators of the rotation symmetries of the Omega background $\R^4_{\epsilon_1,\epsilon_2}$.
The total perturbative contribution becomes
\begin{align}
\label{eq:def_pert_part}
 Z_{\mathrm{pert}} = 
\PE \left[ \left( I_{\mathrm{tensor}}  + I_{\mathrm{vector}}  + 
I_{\mathrm{hyper}}\right) \cdot \left( \frac{Q}{1-Q} +\frac{1}{2} \right)\right]
\end{align}
which includes the contributions of the KK-modes generated by $\sum_{n=1}^\infty Q^n  = \tfrac{Q}{1-Q}$, with $Q=e^{2\pi \im\, \tau}$.
\subsubsection{Elliptic genus}
To compute the $l$-th instanton string partition function $Z_l$, one can add 
$l$ D$2$ branes along the  $x^0, x^1, x^2$ directions, see Figure 
\ref{fig:branes}. The D2 world-volume theory is a $2$d $\Ncal=(0, 4)$ 
effective theory, whose elliptic genera encode the $Z_l$ partition functions.
 
Considering the NS5-D6-D2 brane system in 
Figure \ref{fig:branes}, the space-time symmetry is broken to
\begin{align}
\label{eq:space-time_sym}
\begin{aligned}
 &\sorm(1,9)\to \sorm(1,1) \times \sorm(4)_{2345} \times \sorm(3)_{789}  \,,\\
&\text{with} \quad 
\sorm(4)_{2345} \cong 
\surm(2)_l\times \surm(2)_r
\quad
\text{and}
\quad 
\sorm(3)_{789} \cong
\surm(2)_I
\,.
\end{aligned}
\end{align}
The $16$ supersymmetries can be decomposed in 
representations of $(\surm(2)_l,\, \surm(2)_r,\, \surm(2)_I)_{{\pm}{\pm}}$, 
where the two ``$\pm$" label the chirality of world-sheet $x^0, x^1$ and  
space along $x^6$.
The supersymmetries preserved by the NS5-D6-D2 brane system transform as 
$(\mathbf{1},\, \mathbf 2,\, \mathbf 2)_{-+}$, such that the D2 world-volume 
theory is a $2$d $\Ncal=(0, 4)$ quiver theory, see for instance 
\cite{Kim:2015fxa}. 
The brane configuration allows one to read off the field content and 
the charges of the supermultiplets with respect to $\surm(2)_l\times 
\surm(2)_r\times \surm(2)_I$, labelled as 
$(\alpha,\,\dot\alpha,\, A)$. One finds: 
\begin{compactitem}
    \item The D2-D2 open strings give rise to the $\Ncal=(0, 4)$ vector 
$(A_\mu,\, \lambda^{\dot\alpha A})$ and a chiral multiplet $(\phi^{\alpha 
\dot\beta}, \chi^{\alpha A})$ in the adjoint representation of $\urm(l)$ group.
    \item The D2-D6 open strings, which do not cross a NS5, provide a 
$\Ncal=(0, 
4)$ hypermultiplet $(q^{\dot\alpha}, \psi^A)$ in the bi-fundamental 
representation of $\urm(l)\times \surm(k)$.  
    \item The D2-D6 open strings, which cross a NS$5$ brane, provide two 
additional $\Ncal=(0, 4)$ Fermi multiplets $\Psi$ and $\Psi^\prime$ in the 
bi-fundamental representation of $\urm(l)\times \surm(k)$.
\end{compactitem}
 All these $\Ncal=(0, 4)$ multiplets can be decomposed into $\Ncal=(0, 2)$ 
multiplets as follows:
\begin{subequations}
\label{eq:2d_multiplets}
\begin{align}
{\rm vector} \ (A_\mu,\, \lambda^{\dot\alpha A}) &\longrightarrow {\rm vector}\ 
V\ (A_\mu,\, \lambda^{\dot 1 1},\, \lambda^{\dot 2 2})\ +\ {\rm Fermi} \ 
\Lambda\ (\lambda^{\dot 1 2}) 
\;,\\
{\rm hyper} \ (\varphi^{\alpha \dot\beta}, \chi^{\alpha A}) &\longrightarrow 
{\rm chiral}\ B\ (\varphi^{1 \dot 1}, \chi^{1 2})\ +\ {\rm chiral}\ \tilde 
B^\dagger\ (\varphi^{1 \dot 2}, \chi^{1 1})
\;,\\
{\rm hyper} \ (q^{\dot\alpha}, \psi^A) &\longrightarrow {\rm chiral}\ q\ 
(q^{\dot 1}, \psi^{2})\ +\ {\rm chiral}\ \tilde q^\dagger\ (q^{\dot 2}, 
\psi^{1})
\;,\\
{\rm Fermi} \ \Psi,\ \Psi^\prime &\longrightarrow {\rm Fermi}\ \Psi,\ 
\Psi^\prime \;.
\end{align}
\end{subequations}
From the decomposition, one can read off the charges of these $\Ncal=(0, 2)$ 
multiplets as summarised in Figure \ref{fig:fields_charges}. This 2d 
quiver gauge theory is known from \cite{Kim:2015fxa} and reduces to the 
$\Ncal=(0,4)$ gauge theory description for M-strings introduced in 
\cite{Haghighat:2013tka} for case $k=1$. For completeness, the 2d quiver gauge theory with multiplets \eqref{eq:2d_multiplets} can be written as
\begin{align}
\label{eq:2d_quiver}
	\raisebox{-.5\height}{
 	\begin{tikzpicture}
 		\tikzset{node distance = 2cm}
	\tikzstyle{gauge} = [circle, draw,inner sep=3pt];
	\tikzstyle{flavour} = [regular polygon,regular polygon sides=4,inner 
sep=-2pt, draw];
	\node at (0,0) (g1) [gauge] {\tiny{$\urm (l)$}};
	\node (f1) [flavour,left of=g1] {\tiny{$\urm(k)_m$}};
	\node (f2) [flavour,above of=g1]  {\tiny{$\urm(k)_a$}};
	\node (f3) [flavour,right of=g1] {\tiny{$\urm(k)_n$}};
	\draw  (g1)--(f2) ;
	\draw [-] (0+0.11,-0.4) arc (70:-90:10pt);
	\draw [-] (0+-0.11,-0.4) arc (110:270:10pt);
	\draw[dashed] (g1)--(f1) (g1)--(f3);
    \node at (4,0) {$\longrightarrow$};
	\node at (8,0) (h1) [gauge] {\tiny{$\urm (l)$}};
	\node (ff1) [flavour,left of=h1] {\tiny{$\urm(k)_m$}};
	\node (ff2) [flavour,above of=h1] {\tiny{$\urm(k)_a$}};
	\node (ff3) [flavour,right of=h1]  {\tiny{$\urm(k)_n$}};
	\draw[->] (h1) .. controls (0.25+8,1) .. (ff2);
	\draw[<-] (h1) .. controls (-0.25+8,1) .. (ff2);
	\draw [-,->] (0+0.11+8,-0.4) arc (70:-90:10pt);
	\draw [-] (0+-0.11+8,-0.4) arc (110:270:10pt);
	\draw [-] (0+0.11+8,-0.4) arc (80:-95:15pt);
	\draw [-,->] (0+-0.11+8,-0.4) arc (100:275:15pt);
    \draw [dashed,->] (0+0.11+8,-0.4) arc (85:-95:25pt);
	\draw [dashed] (0+-0.11+8,-0.4) arc (95:275:25pt);
	\draw[dashed,->] (ff1)--(-1+8,0);
	\draw[dashed,->] (h1)--(1+8,0);
	\draw[dashed] (-1+8,0)--(h1) (1+8,0)--(ff3);
%
    \node at (-2.5,2) {\small{$\Ncal=(0,4)$ quiver}};
    \node at (-2.5+8,2) {\small{$\Ncal=(0,2)$ quiver}};
	\end{tikzpicture}
	}
\end{align}
with the conventions: circles $\circ$ denote $\Ncal=(0,4)$ or $\Ncal=(0,2)$ vector multiplets, and squares $\Box$ are flavour nodes. In addition, for lines without/with arrows: solid lines denote hypermultiplets / chiral multiplets, and dashes lines denote Fermi multiplets, respectively. The arrow in $\Ncal=(0,2)$ bifundamental matter fields points towards that node under which the field transforms in the fundamental representation.
\begin{figure}[t]
\centering
\renewcommand{\arraystretch}{1.25}
\begin{tabular}{c|cc|ccc|ccccc|cc}
\toprule 
\multicolumn{3}{c|}{$\Ncal=(0,2)$ multiplets}
&$J_l$ & $J_r$ & $J_I$ & $\urm(l)$ & $\urm(k)_a$ & 
$\urm(k)_m$ & 
$\urm(k)_n$ 
&$\urm(1)_b$ 
&$\urm(1)_x$
&$\urm(1)_z$\\ \midrule
\multirow{4}{*}{D2-D2} & 
vector&
$V$ & $0$ & $0$ & $0$ & ${\rm adj.}$ & $\mathbf 1$  & $\mathbf 1$ & $\mathbf 1$ 
& $0$ 
& $0$ & $0$ \\ 
 & Fermi& 
$\Lambda$ & $0$ & $\frac{1}{2}$ & $-\frac{1}{2}$ & ${\rm adj.}$ & $\mathbf 1$  
& $\mathbf 1$ & $\mathbf 1$ & $0$ 
& $0$ & $0$ \\ 
 & chiral&
$B$ & $\frac{1}{2}$ & $\frac{1}{2}$ & $0$ & ${\rm adj.}$ & $\mathbf 1$  & 
$\mathbf 1$ & $\mathbf 1$ & $0$
& $0$ & $0$ \\ 
 & chiral&
$\tilde B$ & $-\frac{1}{2}$ & $\frac{1}{2}$ & $0$ & ${\rm adj.}$ & $\mathbf 1$  
& $\mathbf 1$ & $\mathbf 1$ & $0$
& $0$ & $0$ \\ \midrule
\multirow{4}{*}{D2-D6} & 
chiral&
$q$ & $0$ & $\frac{1}{2}$ & $0$ & $\mathbf{ l}$ & $\mathbf {\overline k}$  & 
$\mathbf 1$ & $\mathbf 1$ & $0$
& $0$ & $0$ \\
 & chiral&
$\tilde q$ & $0$ & $\frac{1}{2}$ & $0$ & $\mathbf{\overline l}$ & $\mathbf { 
k}$  & $\mathbf 1$ & $\mathbf 1$ & $0$
& $0$ & $0$ \\ 
 & Fermi&
$\Psi$ & $0$ & $0$ & $0$ & $\mathbf{ l}$ & $\mathbf { 1}$  & $\mathbf{\overline 
k}$ & $\mathbf 1$ & $1$
& $0$ & $0$ \\ 
 & Fermi&
$\Psi^\prime$ & $0$ & $0$ & $0$ & $\mathbf{\overline l}$ & $\mathbf { 1}$  & 
$\mathbf 1$ & $\mathbf{k}$ & $1$
& $0$ & $0$ \\ \midrule
\multirow{2}{*}{D2-$\widetilde{\mathrm{D4}}$} & chiral& $\sigma$ & $0$ & $0$ &  $0$ & $\mathbf{l}$ & 
$\mathbf { 1}$ & $\mathbf { 1}$ & $\mathbf { 1}$ & 0 & $-1 $ & $ 0$
\\
 & Fermi& $\Xi$ & $-\tfrac{1}{2}$ & $\tfrac{1}{2}$ &  $0$ & $\mathbf{l}$ & 
$\mathbf { 1}$ & $\mathbf { 1}$ & $\mathbf { 1}$ & 0 & $-1$ & $0$
\\ \midrule
\multirow{4}{*}{D2-D$4^\prime$} & 
chiral&
$\phi$ & $0$ & $0$ & $\frac{1}{2}$ & $\mathbf{l}$ & $\mathbf 1$  & $\mathbf 1$ 
& 
$\mathbf 1$ & $0$
& $0$ & $-1$ \\ 
 & chiral&
$\tilde\phi$ & $0$ & $0$ & $\frac{1}{2}$ & $\mathbf{\overline l}$ & $\mathbf 1$
& $\mathbf 1$ & $\mathbf 1$ & $0$
& $0$ & $1$  \\ 
 & Fermi&
$\Gamma_1$ & $\frac{1}{2}$ & $0$ & $0$ & $\mathbf{ l}$ & $\mathbf { 1}$  & 
$\mathbf { 1}$ & $\mathbf 1$ & $0$
& $0$ & $-1$ \\ 
 & Fermi&
$\Gamma^\dagger_2$ & $\frac{1}{2}$ & $0$ & $0$ & $\mathbf{\overline l}$ & 
$\mathbf { 1}$  & $\mathbf { 1}$ & $\mathbf 1$ & $0$
 & $0$ & $1$  \\ \midrule
D6-D$4^\prime$&
Fermi&
$\rho$ & $0$ & $0$ & $0$ & $\mathbf{1}$ & 
$\mathbf{\overline{k}}$  & $\mathbf{ 1}$ & $\mathbf{1}$ & $0$ & $0$ & $1$ \\ 
\bottomrule
\end{tabular}
\caption{Charge assignments of the fields in the $2$d world-volume theory from 
the D2-D6-NS5 system with or without the presence of a $\widetilde{\mathrm{D4}}$ or D$4^\prime$ 
defect, see Figure 
\ref{fig:branes}. Here 
$J_l$, $J_r$, and $J_I$ denote the Cartans of $\surm(2)_l$, $\surm(2)_r$, and 
$\surm(2)_I$ respectively. $\urm(l)$ is the 2d gauge group on the D2 
world-volume. The $\urm(k)_{a,m,n}$ denote the 6d gauge and flavour symmetries, 
whose fugacities need to be subjected to the constraint 
\eqref{eq:SU_fugacities} in order to reduce to $\surm(k)_{a,m,n}$. $\urm(1)_b$ 
is part of the 6d global symmetry. The $\urm(1)_{x,z}$ denote the defect groups 
for the $\widetilde{\mathrm{D4}}$ and D$4^\prime$ defects, respectively.}
\label{fig:fields_charges}
\end{figure}

For a fixed number $l$ of D2 branes, the partition function of the 2d 
$\Ncal=(0,4)$ theory placed on a torus $\T^2$, with complex structure $\tau$, 
is known to coincide with the elliptic genus 
\cite{Benini:2013nda,Benini:2013xpa}.
The non-perturbative contributions are then encoded in the elliptic genera for 
all $l\geq1$. 
The elliptic genus $Z_l$ for the $2$d theory with gauge group $\urm(l)$ on 
torus $\T^2$ is computed by picking up $\Ncal=(0, 2)$ supercharges $Q\equiv Q^{1\dot 
1 }_-$ and 
$Q^\dagger\equiv Q^{2\dot 2 }_-$, and evaluating
\begin{align}
Z_l
=\Tr \left[ 
(-1)^F Q^{H_L} 
\bar{Q}^{H_R} 
e^{2\pi \im\, \epsilon_-(2 J_l)} 
e^{2\pi \im\, \epsilon_+2(J_{r}-J_I)}
e^{2\pi \im\, b F}
\prod_l^{k} e^{2\pi \im\, m_l F_l}e^{2\pi \im\, n_l F^\prime_l} 
\prod_i^{k} e^{2\pi \im\, a_i G_i}\right]\,.
\end{align}
Here  $Q=e^{2\pi \im\, \tau}$, and $\epsilon_\pm\equiv 
\frac{1}{2}(\epsilon_1\pm\epsilon_2)$, such that $2\epsilon_- J_l+2\epsilon_+ 
J_r=\epsilon_1 J_{23}+\epsilon_2 J_{45}$ with $J_{r, l}=\frac{1}{2}(J_{23}\pm 
J_{45} )$ are the Cartan generators of $\surm(2)_l\times \surm(2)_r\simeq 
\sorm(4)_{2345}$. 

Based on a path integral representation for the elliptic genus, 
a generic prescription for the elliptic 
genera via supersymmetric localisation has 
been derived \cite{Benini:2013nda,Benini:2013xpa}. To briefly summarise, the 
first step involves identifying compact zero modes $\{u_p\}$ originating from 
flat connections on $\T^2$. Keeping the zero modes fixed, the next step 
requires 
an integration over massive fluctuations, which results in a 1-loop 
determinant for each multiplet.
According to \cite{Benini:2013nda,Benini:2013xpa}, the contributions of the 
different multiplets in \eqref{eq:2d_multiplets} can be summarised as 
follows:
\begin{subequations}
\label{eq:1-loop_no_defect}
\begin{align}
Z_{\rm vec } &= \left( \frac{2\pi \eta^2}{\im} \right)^l
\prod_{\alpha\in 
{\rm 
root}}\frac{\theta_1\left({\alpha(u)}\right)}{\im \eta}
= 
\left( \frac{2\pi \eta^2}{\im} \right)^l
\prod_{1\leq p<q\leq k}\frac{\theta_1\left(\pm 
(u_p-u_q )\right)}{(\im \eta)^2}  
\,,\\
Z_{\rm 
chiral} 
&= \left(\prod_{p,q=1}^l\frac{(\im \eta)^2}{\theta_1({\epsilon_{1,2}}+u_p-u_q) 
}\right) 
\left(\prod_{p=1}^{l}\prod_{i=1}^{k}\frac{( \im \eta)^2}{
\theta_1(\epsilon_+\pm(u_p-a_i))}\right) 
\,, \\
Z_{\rm 
Fermi} &= 
\left(
\prod_{p,q=1}^l\frac{\theta_1({2\epsilon_{+}}+u_{pq})}{ \im \eta}
\right) 
\left(
\prod_{p=1}^{l}
\prod_{l=1}^{k}\frac{\theta_1(u_p-m_l+b)\,
\theta_1(-u_p+n_l+b) } {( \im \eta)^2}\right)\,,
\end{align}
\end{subequations}
where the definitions of the Dedekind eta function $\eta$  and the Theta function $\theta_1(z)\equiv \theta_1(\tau|z)$ are recalled in \eqref{eq:def_Dedekind} and \eqref{eq:theta_defs}, respectively.
As customary in the literature, the convention
\begin{align}
 \theta_1(\epsilon_+\pm(u_p-a_i)) \equiv \theta_1(\epsilon_+ +(u_p-a_i)) \cdot \theta_1(\epsilon_+ -(u_p-a_i))   
\end{align}
etc. is used. Note that the $\surm(k)_{m,n,a}$ fugacities need to satisfy 
\eqref{eq:SU_fugacities}.
Collecting all the individual contributions leads to the expression
\begin{align}
Z_{\mathrm{1-loop}}(k,l) 
&\coloneqq 
Z_{\rm vec} \cdot Z_{\rm chiral} \cdot Z_{\rm 
Fermi}
\equiv 
\left(\frac{2\pi 
\,\eta^3\theta_1(2\epsilon_+)}{  
\theta_1(\epsilon_1)\,\theta_1(\epsilon_2)}
\right)^l 
\prod_{\substack{p,q=1 \\ p\neq q}}^l
D(u_p-u_q)
\cdot
\prod_{p=1}^l
Q(u_p)
\,,
\label{eq:1-loop-part}
\end{align}
where, inspired from \cite{Nekrasov:2002qd,Fucito:2011pn}, the following conventions have been 
used:
\begin{subequations}
\label{eq:def_path_int}
\begin{align}
D(u_p -u_q)  &\coloneqq
\frac{\theta_1(u_p-u_q)\theta_1(u_p-u_q+\epsilon_1+\epsilon_2)}{
\theta_1(u_p-u_q+\epsilon_1) \theta_1(u_p-u_q+\epsilon_2)}
=
\frac{\vartheta_1(u_p-u_q)\vartheta_1(u_p-u_q+\epsilon_1+\epsilon_2)}{
\vartheta_1(u_p-u_q+\epsilon_1) \vartheta_1(u_p-u_q+\epsilon_2)}
\,,
\\
Q(u)  &\coloneqq
\frac{\prod_{l=1}^{k} \theta_1(u-m_l+b) \theta_1(-u+n_l+b)}{
\prod_{i=1}^k \theta_1(\epsilon_{+}+ (u-a_i)) \theta_1(\epsilon_{+}-( 
u-a_i))} \notag \\
&=
\frac{\prod_{l=1}^{k} \vartheta_1(u-m_l+b) \vartheta_1(u-n_l-b)}{
\prod_{i=1}^k \vartheta_1(u-a_i +\epsilon_{+} ) \vartheta_1( 
u-a_i -\epsilon_{+})}
\eqqcolon \frac{M(u)}{P_0(u) P_0(u+\epsilon_1+\epsilon_2)} 
\,,\\
\text{with} \qquad 
M(u)&\coloneqq
\prod_{l=1}^{k} \vartheta_1(u-m_l+b) \vartheta_1(u-n_l-b)
\,,\\
P_0(u) &\coloneqq
\prod_{i=1}^k \vartheta_1( u-a_i -\epsilon_{+})
\quad \text{such that} \quad 
P_0(u+\epsilon_1+\epsilon_2) = 
\prod_{i=1}^k \vartheta_1( u-a_i +\epsilon_{+})
\,.
\label{def1}
\end{align}
\end{subequations}
Note, in particular, the change to $\vartheta_1(\tau|z)$ defined in 
\eqref{eq:my_theta}, which is more convenient than the Theta function $\theta_1(\tau|z)$.
Lastly, one needs to integrate the several 1-loop determinants 
\eqref{eq:1-loop-part} over the zero modes $\{u_p\}$. As shown in 
\cite{Benini:2013nda,Benini:2013xpa}, this integral becomes a contour integral.
The contour integration needs to be performed with care, as the choice of 
integration contour determines whether the results yields the partition 
function or not. A consistent choice of contour is given by the Jeffrey-Kirwan 
residue prescription \cite{Jeffrey1995}. The expression becomes
\begin{align}
Z_l &=
\frac{1}{l!}\oint \frac{\diff^l u}{(2\pi \im)^l} Z_{\mathrm{1-loop}}(k,l) 
= \frac{1}{l!} \sum_{u_\star} \mathrm{JK-Res}_{u_\star} 
Z_{\mathrm{1-loop}}(k,l)
\label{eq:ell_genus_k-string}
\end{align}
where the sum is taken over existing poles $u_\star$ in the integrand 
$Z_{\mathrm{1-loop}}$. For details on the computational aspects of the JK 
residue, the reader is referred to \cite{Benini:2013nda,Benini:2013xpa}.
The following conventions are useful for the residue calculus of the elliptic 
genera:
\begin{subequations}
\label{eq:residue_calc}
\begin{align}
  P_0^\vee(a_i\pm \epsilon_+) &\coloneqq \prod_{\substack{j=1 \\ j\neq i }}^k 
\vartheta_1 
(u-a_j-\epsilon_+) \Big|_{u=a_i\pm \epsilon_+}  \,,
\\
Q^\vee(a_i - \epsilon_+) &\coloneqq 
\frac{M(a_i - \epsilon_+)
}{
P_0(a_i - \epsilon_+)
P_0^\vee(a_i + \epsilon_+)
}
\,.
 \end{align}
 \end{subequations}
For the Nekrasov-Shatashvili limit, the following abbreviations are used:
\begin{alignat}{2}
\label{eq:def_L_and_K}
 L(u)&\coloneqq \frac{\vartheta_1^\prime(u)}{\vartheta_1(u)}
 \;, \qquad & 
 K(u)&\coloneqq \frac{\vartheta_1^{\prime\prime}(u)}{\vartheta_1(u)} 
 \; , 
\end{alignat}
where $\vartheta_1^\prime(u) \equiv \tfrac{\partial}{\partial u} \vartheta_1(u)$ and $\vartheta_1^{\prime\prime}(u) \equiv \tfrac{\partial^2}{\partial u^2} \vartheta_1(u)$.
For later purposes, the $l=1,2$ genera are computed.
\paragraph{1-string.}
The $l=1$ elliptic genus reads
\begin{align}
Z_1 &=\frac{ \vartheta_1(2\epsilon_+)}{  
\vartheta_1(\epsilon_1)\,\vartheta_1(\epsilon_2)}
\sum_{i=1}^k Q^\vee(a_i-\epsilon_+)
\,,
\label{eq:elliptic_genus_k=1_pure}
\end{align}
and the details are presented in Appendix \ref{app:details_pure_k=1}.
\paragraph{2-string.}
The $l=2$ elliptic genus reads
\begin{align}
 Z_2&=
\left(
 \frac{  \vartheta_1(2\epsilon_+)
 }{\vartheta_1(\epsilon_1)\vartheta_1(\epsilon_2)}
\right)^2
\sum_{1 \leq i< j\leq k}
D(a_i-a_j)D(a_j-a_i)
Q^\vee(a_i-\epsilon_+)
Q^\vee(a_j-\epsilon_+)
 \\
&+
\frac{  \vartheta_1(2\epsilon_+)
 }{ \vartheta_1(2\epsilon_-)  }
 \sum_{m=1}^k
 Q^\vee(a_m-\epsilon_+)
\bigg[ 
\frac{ \vartheta_1(\epsilon_1+2\epsilon_+)
 }{ \vartheta_1(\epsilon_2) \vartheta_1(2\epsilon_1)  }
 Q(a_m-\epsilon_+-\epsilon_1)
-
\frac{ \vartheta_1(\epsilon_2+2\epsilon_+)
 }{ \vartheta_1(\epsilon_1) \vartheta_1(2\epsilon_2)}
 Q(a_m-\epsilon_+-\epsilon_2)
\bigg] \notag
\end{align}
and the derivation is summarised in Appendix \ref{app:details_pure_k=2}.
\subsubsection{Enhancement of global symmetry}
For the case of two NS5 branes, one needs to recover the global 
symmetry enhancement to $\surm(2k)$, as indicated in \eqref{eq:6d_quiver}.
\paragraph{Perturbative part.}
The perturbative contribution of the 6d $\Ncal=(1,0)$ hypermultiplets can be 
rewritten as 
\begin{align}
  I_{\mathrm{hyper}} &= \frac{\sqrt{p\cdot q}}{(1-p)(1-q)} 
  \sum_{i=1}^k \left\{ 
  \sum_{l=1}^{k}  \left( e^{a_i - m_l+b} + e^{a_i - n_l-b}  \right) 
  +\sum_{l=1}^{k}  \left(  e^{m_l-b-a_i}+  e^{n_l+b-a_i} 
\right) 
  \right\}
  \notag \\
&= \frac{\sqrt{p\cdot q}}{(1-p)(1-q)} 
  \sum_{i=1}^k 
  \sum_{l=1}^{2k}  \left( e^{a_i - y_l}+   e^{y_l-a_i} \right)
\\
&\text{with} \qquad y_l = 
\begin{cases}
 m_l-b  & , l=1,\ldots ,k \\
 n_l+b & , l=k+1,\ldots, 2k
\end{cases}
\label{eq:SU(2N)_fugacities}
\end{align}
and one verifies that $y_l$ are $\surm(2k)$ fugacities via
\begin{align}
 \prod_{l=1}^{2k} e^{y_l}=  \prod_{l=1}^k e^{m_l} \cdot \prod_{l'=1}^k e^{n_l'} 
\cdot \prod_{l''=1}^k e^{b-b} =1
\end{align}
using \eqref{eq:SU_fugacities}.
\paragraph{Non-perturbative part.}
For the 2d elliptic genus \eqref{eq:ell_genus_k-string}, the $\Psi$, $\Psi'$ 
Fermi multiplet contributions can also be rearranged
\begin{align}
 Z_{\rm Fermi} &\supset
\prod_{p=1}^{l}\prod_{l=1}^{k}\frac{\theta_1(u_p-m_l+b)\,
\theta_1(u_p-n_l-b) } {(\im\,\eta)^2} \notag \\
&=\prod_{p=1}^{l}\prod_{l=1}^{k}\frac{ \theta_1(u_p-(m_l-b))\,
\theta_1(u_p-(n_l+b) ) } {(\im\, \eta)^2} \notag \\
&=\prod_{p=1}^{l} \prod_{l=1}^{2k}\frac{  \theta_1(u_p-y_l)} 
{\im\, \eta}
\end{align}
with $y_l$ fugacities as defined in \eqref{eq:SU(2N)_fugacities}.
%
%
\subsection{Higgs mechanism in partition functions}
\label{sec:Higgs_mech}
For later purposes, in which a codimension 2 defect is introduced via a position dependent 
vacuum expectation value (VEV), this section reviews the standard Higgs 
mechanism.
To begin with, consider the Higgsing of the 6d gauge theory on the tensor 
branch: 
\begin{align}
 \surm(k+1) \, , \; N_f=2k+2 
 \quad \longrightarrow \quad 
 \surm(k) \, , \; N_f=2k \,.
 \label{eq:Higgs}
\end{align}
The first task is to find a suitable VEV assignment for a gauge invariant 
operator and then derive a condition in terms of fugacities for the 
gauge invariant operator that realises the Higgs mechanism on the level of 
partition functions.

\subsubsection{Standard Higgsing}
Consider the field theoretical description of the mesonic Higgs branch 
deformation \eqref{eq:Higgs}, but seen as $\surm(k+1)_a$ gauge theory with 
$\surm(k+1)_m \times \urm(1)_b \times \surm(k+1)_n$ global symmetry. In other 
words, there are 
the flavour 
hypermultiplets $(Q,\widetilde{Q})$ of $\surm(k+1)_m\times \urm(1)_b \times 
\surm(k+1)_a$ and 
$(Q',\widetilde{Q}')$ of $\surm(k+1)_n\times \urm(1)_b \times \surm(k+1)_a$.
Since each hypermultiplet in \eqref{eq:6d_quiver} has charge $(\tfrac{1}{2}, \tfrac{1}{2})$ under the Cartan generators $J_{r,l}$ of $\surm(2)_l\times\surm(2)_r \cong \sorm(4)_{3456}$, the fugacity contributions for each chiral are
\begin{subequations}
\begin{alignat}{2}
 &Q^i_l \in 
 \overline{\mathbf{(k{+}1)}}_a \otimes 
 \left( 
\mathbf{(k{+}1)}_m 
\otimes \mathbf{1}_n
\right)^{-1}
&\qquad \rightarrow \qquad 
 &\sqrt{pq} e^{-a_i +m_l -b}
 \,,\\
 &\widetilde{Q}^l_i \in 
 \mathbf{(k{+}1)}_a \otimes
 \left( 
 \overline{\mathbf{(k{+}1)}}_m
 \otimes \mathbf{1}_n
\right)^{+1}
 &\qquad \rightarrow \qquad  
 &\sqrt{pq} e^{a_i -m_l +b}
 \,, \\
&{Q'}^i_l  \in 
\overline{\mathbf{(k{+}1)}}_a \otimes
\left(
\mathbf{1}_m \otimes
\mathbf{(k{+}1)}_n
\right)^{+1} 
&\qquad \rightarrow \qquad  
 &\sqrt{pq} e^{-a_i +n_l +b}
 \,,\\
 &\widetilde{Q'}^l_i  \in 
 \mathbf{(k{+}1)}_a \otimes 
 \left(
 \mathbf{1}_m \otimes
 \overline{\mathbf{(k{+}1)}}_n
 \right)^{-1}
&\qquad \rightarrow \qquad  
 &\sqrt{pq} e^{a_i -n_l -b}
 \,,
\end{alignat}
\end{subequations}
with $i =1,\ldots, k+1$ for $\surm(k+1)_a$ and $l,l'=1,\ldots,k+1$ for 
$\surm(k+1)_{m,n}$ respectively. The exponent $(\ldots)^{\pm1}$ denotes the $\urm(1)_b$ charge.
There are two possibilities for meson operators
\begin{subequations}
\begin{alignat}{2}
 \mathcal{M}_{l}^{l'} &= \sum_{i=1}^{k+1} Q_{l}^{i} \widetilde{Q'}_{i}^{l'}
 &\qquad \rightarrow \qquad  
 \left( \sqrt{pq} e^{m_{l}-b} \right)  
\cdot\left( \sqrt{pq} e^{-n_{l'} -b} \right) &= pq e^{m_l -n_{l'} -2b}
\,,\\
 \widetilde{\mathcal{M}}_{l'}^{l} &= \sum_{i=1}^{k+1} \widetilde{Q}_{i}^{l} 
{Q'}_{l'}^{i}
&\qquad \rightarrow \qquad
\left( \sqrt{pq} e^{-m_{l}+b} \right)  
\cdot\left( \sqrt{pq} e^{n_{l'} +b} \right) &= pq e^{-m_l +n_{l'} +2b}
\,,
\end{alignat}
\end{subequations}
and one can consider assigning a VEV to the $(k+1,k+1)$ meson components. 
A gauge 
transformation is sufficient to see that one only needs to assign VEVs to the 
following components
\begin{align}
 \mathcal{M}_{k+1}^{k+1} = 
 \sum_{i=1}^{k+1} Q_{k+1}^{i} \widetilde{Q'}_{i}^{k+1} 
\cong Q_{k+1}^{k+1} \widetilde{Q'}_{k+1}^{k+1} 
\qquad \text{and}
\qquad 
\widetilde{\mathcal{M}}_{k+1}^{k+1} 
\cong \widetilde{Q}_{k+1}^{k+1} {Q'}_{k+1}^{k+1} \,.
\end{align}
Following the prescription of \cite{Gaiotto:2012xa}, see also \cite[Sec.\ 
2]{Gaiotto:2014ina}, Higgsing is achieved in a partition function via choosing 
the pole corresponding to the operator acquiring a VEV, i.e.\
\begin{subequations}
\label{eq:Higgs_no_defect_partFct}
\begin{alignat}{3}
 \langle \mathcal{M}_{k+1}^{k+1} \rangle &\neq 0 &
 \quad \Leftrightarrow \quad 
 pq e^{m_{k+1} -n_{k+1} -2b} &= 1 &
 \quad \Leftrightarrow \quad 
 &\begin{cases}
  n_{k+1} &= m_{k+1} -2b+ 2\epsilon_+ \\
  a_{k+1} &= m_{k+1} -b +\epsilon_+
 \end{cases}
 \,,
 \\
\langle \widetilde{\mathcal{M}}_{k+1}^{k+1} \rangle &\neq 0 &
 \quad \Leftrightarrow \quad 
 pq e^{-m_{k+1} +n_{k+1} +2b} &= 1 &
 \quad \Leftrightarrow \quad 
 &\begin{cases}
  n_{k+1}  &= m_{k+1} -2b - 2\epsilon_+ \\
  a_{k+1} &= m_{k+1} -b - \epsilon_+
 \end{cases} 
 \,,
 \label{eq:Higgs_no_defect}
\end{alignat}
\end{subequations}
and eliminating the contributions of the flat directions as well as any 
appearing Goldstone modes. Note that the condition for $a_{k+1}$ in 
\eqref{eq:Higgs_no_defect_partFct} is derived by requiring that the fugacity of 
the chiral $Q^{k+1}_{k+1}$ or $\widetilde{Q}^{k+1}_{k+1}$  equals unity, 
respectively.

In the Type IIA brane configuration, the mesonic Higgsing is realised via aligning a semi-infinite flavour D6 brane on the left and right hand side with a gauge D6 such that a single D6 is free to move along the Higgs branch directions $x^{7,8,9}$, see Figure \ref{fig:Higgsing_branes}. The codimension 2 defect is introduced via a $\widetilde{\mathrm{D4}}$ brane that connects the remaining brane configuration with the single D6 on the Higgs branch. Moving the D6 to infinity in Figure \ref{fig:Higgsing_branes}, leads to the natural connection between \emph{defect via Higgsing} and \emph{defect via additional branes}, see also Section \ref{sec:codim_2}.  
\paragraph{Perturbative contribution.}
Consider the perturbative partition function for 6d $\surm(k+1)$ theory with 
$N_f=2k+2$ flavours
\begin{align}
 Z_{\mathrm{pert}}^{k+1} &= 
 \PE \Bigg[ \frac{1}{(1-p)(1-q)} \left( \frac{Q}{1-Q} +\frac{1}{2}\right) 
\notag
 \bigg\{
 -(p+q) -(1+pq) \left(\sum_{i,j=1}^{k+1} e^{a_i -a_j} -1 \right) \notag \\
 &\qquad \qquad 
 +\sqrt{pq} \sum_{i=1}^{k+1} \sum_{l=1}^{k+1} 
 \left( e^{a_i} ( e^{-m_l+b} + e^{-n_l-b}) 
+e^{-a_i}(e^{m_l-b}+e^{n_l+b})  \right)
 \bigg\}
 \Bigg]  \,.
\end{align}
The Higgsing \eqref{eq:Higgs_no_defect} takes the form
\begin{align}
 e^{a_{k+1}} = \frac{1}{\sqrt{pq}} \cdot e^{m_{k+1}-b} 
 \qquad
 e^{n_{k+1}} = \frac{1}{pq} \cdot e^{m_{k+1}-2b} = \frac{1}{\sqrt{pq}}\cdot 
e^{a_{k+1}-b} \,.
\end{align}
A straightforward computation, see Appendix \ref{app:const_Higgs_pert}, shows 
that the Higgsing \eqref{eq:Higgs_no_defect} leads to the expected result
\begin{align}
  Z_{\mathrm{pert}}^{k+1}  =   Z_{\mathrm{pert}}^{k} \cdot   Z_{\mathrm{G}}
\end{align}
where the Goldstone modes for the breaking of the global symmetry
\begin{align}
 \surm(k+1)_m 
\times \urm(1) \times \surm(k+1)_n \to \surm(k)_m 
\times \urm(1) \times \surm(k)_n
\end{align}
contribute as 
\begin{align}
 Z_{\mathrm{G}} &= 
 \PE \bigg[ \frac{\sqrt{pq}}{(1-p)(1-q)} \left( \frac{Q}{1-Q} 
+\frac{1}{2}\right) 
  \bigg\{ \left(\frac{1}{\sqrt{pq}} + \sqrt{pq}\right) 
 +  \sum_{l=1}^{k} \left( \frac{1}{\sqrt{pq}}e^{m_{k+1}-m_l} 
+ \sqrt{pq} e^{m_l-m_{k+1}}  \right) \notag \\
&\qquad \qquad \qquad 
+ \sum_{l=1}^{k} \left( \frac{1}{\sqrt{pq}}e^{m_{k+1}-n_l-2b}
+ \sqrt{pq} e^{n_l-m_{k+1}+2b}  \right)
  \bigg\}
  \bigg]\,,
  \label{eq:Goldstone_part}
\end{align}
such that there are $4k+2$ massless chiral fields. Considering the Higgsing 
\eqref{eq:Higgs}, one computes that the sub-space of the Higgs branch, where the 
theory is broken to $\surm(k)$, has complex dimension $4k+2$, which matches the 
degrees of freedom in \eqref{eq:Goldstone_part}. Taking the closure of this sub-space, the $2k+1$ quaternionic degrees of freedom parametrise the closure of the minimal nilpotent orbit of $\surm(2k+2)$, see \cite{Bourget:2019aer}. 
\paragraph{Elliptic genus.}
Consider the elliptic genus \eqref{eq:ell_genus_k-string} for the theory 
without 
defect. Suppose one aims to realise the Higgs mechanism \eqref{eq:Higgs} on the 
level of the elliptic genus, then starting from $k+1$ one factorises 
\eqref{eq:ell_genus_k-string} as follows: 
\begin{align}
 Z_l^{k+1} &= 
 \frac{1}{l!}\oint 
 \frac{\diff^l u}{(2\pi \im)^l} 
 \,
 Z_{\mathrm{1-loop}}({k+1},l)  \notag\\
 &= \frac{1}{l!}\oint \frac{\diff^l u}{(2\pi \im)^l}
 \left(\frac{2\pi \,\eta^3\theta_1(2\epsilon_+)}{   
\theta_1(\epsilon_1)\,\theta_1(\epsilon_2)}
\right)^l 
\cdot
\prod_{\substack{p,q=1\\p\neq q}}^{l}
D(u_p-u_q)
\cdot
\prod_{p=1}^l
\left(
\frac{\prod_{l=1}^{k+1}\theta_1(u_p-m_l+b)\,
\theta_1(u_p-n_l-b) }{\prod_{i=1}^{k+1} \theta_1(u_p-a_i + \epsilon_+) \, 
\theta_1(u_p-a_i-\epsilon_+) }\right) \notag \\
&= \frac{1}{l!}\oint \frac{\diff^l u}{(2\pi \im)^l} 
 Z_{\mathrm{1-loop}}(k,l) 
\cdot \left(\prod_{p=1}^l
\frac{ \theta_1(u_p-m_{k+1}+b) \theta_1(u_p-n_{k+1}-b) }{ 
\theta_1( u_p-a_{k+1} + \epsilon_+)\, 
\theta_1( u_p-a_{k+1} -\epsilon_+ )}\right) 
\,.
\label{eq:ell_genus_Higgs_intermediate}
\end{align}
Since the Higgsing process should reduce $Z_l^{k+1} \to Z_l^{k}$, the last 
fraction is expected to be equal to one upon any of the fugacity assignments of 
\eqref{eq:Higgs_no_defect_partFct}. 
Explicitly, for \eqref{eq:Higgs_no_defect} one verifies that
\begin{align}
 &\prod_{p=1}^l
\frac{ \theta_1(u_p-m_{N+1}+b) \theta_1(u_p-n_{N+1}-b) }{ 
\theta_1( u_p-a_{N+1} + \epsilon_+) \theta_1(u_p-a_{N+1} -\epsilon_+)}
\bigg|_{\eqref{eq:Higgs_no_defect}}
\notag \\
&\qquad \qquad =
\prod_{p=1}^l
\frac{ \theta_1(u_p-m_{N+1}+b) \theta_1(u_p-m_{N+1}+2b+2\epsilon_+-b) }{ 
\theta_1( u_p-m_{N+1} + b +\epsilon_+ + \epsilon_+)
\theta_1(u_p-m_{N+1} +b+\epsilon_+ -\epsilon_+)}
=1 
\end{align}
holds.
Therefore, the elliptic genus is compatible with the fugacity assignment 
\eqref{eq:Higgs_no_defect_partFct} derived for the Higgs mechanism.
\subsubsection{Higgsing to defects}
Building on \eqref{eq:Higgs_no_defect_partFct}, a surface defect of type 
$(r,s)$ 
can be introduced via a position dependent VEV 
\cite{Gaiotto:2012xa,Gaiotto:2014ina,Ito:2016fpl} which is related to a pole 
at 
\begin{subequations}
\label{eq:Higgs_with_defect_partFct}
\begin{alignat}{5}
 \langle \mathcal{M}_{k+1}^{k+1} \rangle &= \mathrm{fct.} &
 \; &\Leftrightarrow \; &
 p^r q^s\cdot pq e^{m_{k+1} -n_{k+1} -2b} &= 1 &
\; &\Leftrightarrow \; &
 &\begin{cases}
  n_{k+1} &= m_{k+1} -2b+ 2\epsilon_+ +r \epsilon_1 + s \epsilon_2\\
  a_{k+1} &= m_{k+1} -b +\epsilon_+
 \end{cases}
 \,,
 \\
\langle \widetilde{\mathcal{M}}_{k+1}^{k+1} \rangle &= \mathrm{fct.} &
 \; &\Leftrightarrow \; &
 p^r q^s\cdot pq e^{-m_{k+1} +n_{k+1} +2b} &= 1 &
 \; &\Leftrightarrow \;  &
 &\begin{cases}
  n_{k+1}  &= m_{k+1} -2b - 2\epsilon_+ -r \epsilon_1 - s \epsilon_2 \\
  a_{k+1} &= m_{k+1} -b - \epsilon_+
 \end{cases}
 \,,
 \label{eq:Higgs_with_defect}
\end{alignat}
\end{subequations}
such that the condition for the 6d gauge fugacity remains unchanged. 

Without loss of generality, one can restrict to one choice of mesonic VEV. For 
this note, consider $\langle \widetilde{\mathcal{M}}_{k+1}^{k+1} \rangle$ such 
that \eqref{eq:Higgs_no_defect} and 
\eqref{eq:Higgs_with_defect} are relevant. 
If the defect is of type $(r,0)$ then the codimension 2 defect occupies $\R^2_{\epsilon_1}$ while being a point on $\R^2_{\epsilon_2}$ inside the 4d Omega background; whereas an $(0,s)$ defect occupies $\R^2_{\epsilon_2}$ inside $\R^4_{\epsilon_1,\epsilon_2}$ and is point-like in $\R^2_{\epsilon_1}$.
%
\subsection{Codimension 2 defect}
\label{sec:codim_2}
There are multiple ways to introduce a codimension 2 defect. For instance, one may either employ a position dependent vacuum 
expectation value \eqref{eq:Higgs_with_defect_partFct} as in 
\cite{Gaiotto:2012xa,Gaiotto:2014ina,Ito:2016fpl} or
one may include an additional $\widetilde{\mathrm{D4}}$ brane in the Type IIA brane configuration as 
in Figure \ref{fig:branes}, see also \cite{Alday:2009fs,Gadde:2013dda} for 
surface defects in 4d theories. In the original M-theory setting of Figure \ref{fig:branes}, the defect introduced via the $\widetilde{\mathrm{D4}}$ brane corresponds to another $\widetilde{\mathrm{M5}}$ brane filling $(x^0,x^1,x^3,x^4,x^7,x^{10})$, as studied in \cite{Mori:2016qof}. Further, codimension 2 defects in 6d $\Ncal=(1,0)$ $\surm(N)$ theories with adjoint matter are studied in \cite{Koroteev:2019gqi}.

\begin{figure}
\centering
\begin{tikzpicture}
  \node (6dNH) at (0,0) {
  \begin{tikzpicture}
 \draw[thick] (0,0)--(0,4) (2,0)--(2,4);
 \foreach \i in {0,...,2}
 {
    \draw (-2,{2.75+0.\i})--(0,{2.75+0.\i});
    \draw (2,{3+0.\i})--(4,{3+0.\i});
    \draw (0,{2.25+0.\i})--(2,{2.25+0.\i});    
    }
    \node at (-1,2.6) {$\cdots$}; 
    \node at (1,2.1) {$\cdots$};
    \node at (3,2.85) {$\cdots$};
    \draw (-2,2.55)--(0,2.55) (0,2.05)--(2,2.05) (2,2.8)--(4,2.8);
\draw[white,line width=1.50mm] (-0.25,0.5)--(2.25,0.5);
\draw (-2.5,0.5)--(3.5,0.5);
\draw[dotted] (-0.5,0.5)--(0,1) (-0.5+2,0.5)--(0+2,1);
\draw[black,->] (3,1.5)--(3.5,1.5);
\draw[->] (3,1.5)--(3,2);
\draw[->] (3,1.5)--(2.75,1.25);
\node at (3.75,1.5) {\footnotesize{$x^6$}};
\node at (3.25,2.25) {\footnotesize{$x^{1}$}};
\node at (3,1) {\footnotesize{$x^{7,8,9}$}};
\node at (1,2.7) {\footnotesize{$k$ D6}};
\node at (-1,3.25) {\footnotesize{$k$ D6}};
\node at (3,3.5) {\footnotesize{$k$ D6}};
\node at (0.35,4) {\footnotesize{NS5}};
\end{tikzpicture}
  };
   \node (6dNHS1) at (0,-6) {
   \begin{tikzpicture}
 \draw[thick] (0,0)--(0,4) (2,0)--(2,4);
 \draw[dashed] (0,0) .. controls (0.4,0.25) .. (0.5,2);
 \draw[dashed] (0.5,2) .. controls (0.4,4+0.25) .. (0,4);
 \draw[dashed] (0+2,0) .. controls (0.4+2,0.25) .. (0.5+2,2);
 \draw[dashed] (0.5+2,2) .. controls (0.4+2,4+0.25) .. (0+2,4);
 \foreach \i in {0,...,2}
 {
    \draw (-2,{2.75+0.\i})--(0,{2.75+0.\i});
    \draw (2,{3+0.\i})--(4,{3+0.\i});
    \draw (0,{2.25+0.\i})--(2,{2.25+0.\i});    
    }
    \node at (-1,2.6) {$\cdots$}; 
    \node at (1,2.1) {$\cdots$};
    \node at (3,2.85) {$\cdots$};
    \draw (-2,2.55)--(0,2.55) (0,2.05)--(2,2.05) (2,2.8)--(4,2.8);
\draw[white,line width=1.50mm] (-0.25,0.5)--(2.25,0.5);
\draw (-2.5,0.5)--(3.5,0.5);
\draw[dotted] (-0.5,0.5)--(0,1) (-0.5+2,0.5)--(0+2,1);
\draw[black,->] (3,1.5)--(3.5,1.5);
\draw[->] (3,1.5)--(3,2);
\draw[->] (3,1.5)--(2.75,1.25);
\node at (3.75,1.5) {\footnotesize{$x^6$}};
\node at (3.25,2.25) {\footnotesize{$x^{1}$}};
\node at (3,1) {\footnotesize{$x^{7,8,9}$}};
\node at (1,2.7) {\footnotesize{$k$ D5}};
\node at (-1,3.25) {\footnotesize{$k$ D5}};
\node at (3,3.5) {\footnotesize{$k$ D5}};
\node at (0.35+1.25,4) {\footnotesize{NS5}};
\end{tikzpicture}
   };
   \node (5dNH) at (0,-12) {
   \begin{tikzpicture}
   \node at (-2.25,0) {};
 \draw[thick] (0,0)--(0,3) (0.5,0)--(0.5,3) (1,0)--(1,3)   (2,0)--(2,3);
 \node at (1.5,1.75) {$\cdots$};
 \node at (1.5,0.75) {$\cdots$};
 \draw (-1.5,1) -- (3,1) (-1.5,2) -- (3,2); 
 \draw[dashed] (3,1) .. controls (3+0.25,1.4) .. (1,1.5);
 \draw[dashed] (1,1.5) .. controls (-1.5+0.25,1.4) .. (-1.5,1);
\draw[dashed] (3,1+1) .. controls (3+0.25,1.4+1) .. (1,1.5+1);
 \draw[dashed] (1,1.5+1) .. controls (-1.5+0.25,1.4+1) .. (-1.5,1+1);
\draw[white,line width=1.50mm] (-0.75,-0.5)--(-0.75,2.5);
 \draw[thick] (-0.75,-0.5)--(-0.75,2.5);
\draw[dotted] (-0.75,0.5)--(-0.25,1) (-0.75,0.5+1)--(-0.25,1+1);
\draw[->] (3,0)--(3.5,0);
\draw[->] (3,0)--(3,0.5);
\draw[->] (3,0)--(2.75,-0.25);
\node at (3.75,0) {\footnotesize{$x^1$}};
\node at (3.25,0.75) {\footnotesize{$x^{6}$}};
\node at (3,-0.5) {\footnotesize{$x^{7,8,9}$}};
\node at (2.8,1.8) {\footnotesize{D5}};
\draw[decoration={brace,raise=3pt,mirror},decorate,thick](-0.15,-0.15) -- 
node[below=6pt] {\footnotesize{$k$ NS5} } (2.15,-0.15);
\end{tikzpicture}
   };
  \node (6dDH) at (7,0) {
  \begin{tikzpicture}
 \draw[thick] (0,0)--(0,4) (2,0)--(2,4);
 \foreach \i in {0,...,2}
 {
    \draw (-2,{2.75+0.\i})--(0,{2.75+0.\i});
    \draw (2,{3+0.\i})--(4,{3+0.\i});
    \draw (0,{2.25+0.\i})--(2,{2.25+0.\i});    
    }
    \node at (-1,2.6) {$\cdots$}; 
    \node at (1,2.1) {$\cdots$};
    \node at (3,2.85) {$\cdots$};
    \draw (-2,2.55)--(0,2.55) (0,2.05)--(2,2.05) (2,2.8)--(4,2.8);
\draw[white,line width=1.50mm] (-0.25,0.5)--(2.25,0.5);
\draw (-2.5,0.5)--(3.5,0.5);
\draw[red,thick] (-0.5+2,0.5)--(0+2,1);
\draw[black,->] (3,1.5)--(3.5,1.5);
\draw[->] (3,1.5)--(3,2);
\draw[->] (3,1.5)--(2.75,1.25);
\node at (3.75,1.5) {\footnotesize{$x^6$}};
\node at (3.25,2.25) {\footnotesize{$x^{1}$}};
\node at (3,1) {\footnotesize{$x^{7,8,9}$}};
\node at (1,2.7) {\footnotesize{$k$ D6}};
\node at (-1,3.25) {\footnotesize{$k$ D6}};
\node at (3,3.5) {\footnotesize{$k$ D6}};
\node at (0.35,4) {\footnotesize{NS5}};
\node[red] at (1.4,0.8) {\footnotesize{$\widetilde{\mathrm{D4}}$}};
\end{tikzpicture}
  };
   \node (6dDHS1) at (7,-6) {
   \begin{tikzpicture}
 \draw[thick] (0,0)--(0,4) (2,0)--(2,4);
 \draw[dashed] (0,0) .. controls (0.4,0.25) .. (0.5,2);
 \draw[dashed] (0.5,2) .. controls (0.4,4+0.25) .. (0,4);
 \draw[dashed] (0+2,0) .. controls (0.4+2,0.25) .. (0.5+2,2);
 \draw[dashed] (0.5+2,2) .. controls (0.4+2,4+0.25) .. (0+2,4);
 \foreach \i in {0,...,2}
 {
    \draw (-2,{2.75+0.\i})--(0,{2.75+0.\i});
    \draw (2,{3+0.\i})--(4,{3+0.\i});
    \draw (0,{2.25+0.\i})--(2,{2.25+0.\i});    
    }
    \node at (-1,2.6) {$\cdots$}; 
    \node at (1,2.1) {$\cdots$};
    \node at (3,2.85) {$\cdots$};
    \draw (-2,2.55)--(0,2.55) (0,2.05)--(2,2.05) (2,2.8)--(4,2.8);
\draw[white,line width=1.50mm] (-0.25,0.5)--(2.25,0.5);
\draw (-2.5,0.5)--(3.5,0.5);
\draw[red,thick] (-0.5+2,0.5)--(0+2,1);
\draw[black,->] (3,1.5)--(3.5,1.5);
\draw[->] (3,1.5)--(3,2);
\draw[->] (3,1.5)--(2.75,1.25);
\node at (3.75,1.5) {\footnotesize{$x^6$}};
\node at (3.25,2.25) {\footnotesize{$x^{1}$}};
\node at (3,1) {\footnotesize{$x^{7,8,9}$}};
\node at (1,2.7) {\footnotesize{$k$ D5}};
\node at (-1,3.25) {\footnotesize{$k$ D5}};
\node at (3,3.5) {\footnotesize{$k$ D5}};
\node at (0.35+1.25,4) {\footnotesize{NS5}};
\node[red] at (1.4,0.8) {\footnotesize{$\widetilde{\mathrm{D3}}$}};
\end{tikzpicture}
   };
   \node (5dDH) at (7,-12) {
   \begin{tikzpicture}
   \node at (-2.25,0) {};
 \draw[thick] (0,0)--(0,3) (0.5,0)--(0.5,3) (1,0)--(1,3)   (2,0)--(2,3);
 \node at (1.5,1.75) {$\cdots$};
 \node at (1.5,0.75) {$\cdots$};
 \draw (-1.5,1) -- (3,1) (-1.5,2) -- (3,2); 
 \draw[dashed] (3,1) .. controls (3+0.25,1.4) .. (1,1.5);
 \draw[dashed] (1,1.5) .. controls (-1.5+0.25,1.4) .. (-1.5,1);
\draw[dashed] (3,1+1) .. controls (3+0.25,1.4+1) .. (1,1.5+1);
 \draw[dashed] (1,1.5+1) .. controls (-1.5+0.25,1.4+1) .. (-1.5,1+1);
\draw[white,line width=1.50mm] (-0.75,-0.5)--(-0.75,2.5);
 \draw[thick] (-0.75,-0.5)--(-0.75,2.5);
\draw[red,thick] (-0.75,0.5)--(-0.25,1);
\node[red] at (-0.3,0.5) {\footnotesize{$\widetilde{\mathrm{D3}}$}};
\draw[->] (3,0)--(3.5,0);
\draw[->] (3,0)--(3,0.5);
\draw[->] (3,0)--(2.75,-0.25);
\node at (3.75,0) {\footnotesize{$x^1$}};
\node at (3.25,0.75) {\footnotesize{$x^{6}$}};
\node at (3,-0.5) {\footnotesize{$x^{7,8,9}$}};
\node at (2.8,1.8) {\footnotesize{D5}};
\draw[decoration={brace,raise=3pt,mirror},decorate,thick](-0.15,-0.15) -- 
node[below=6pt] {\footnotesize{$k$ NS5} } (2.15,-0.15);
\end{tikzpicture}
   };
\node (5dNHaux) at (0,-12) {};
\node (5dDHaux) at (7,-12) {};
\node at (0.25,2.75) {\textbf{normal Higgsing}};
\node at (7.25,2.75) {\textbf{defect Higgsing}};
\draw[thick,<->] (6dNH)-- node[right] {$\substack{\text{T-dual} \\ S^1\subset 
\R_1}$} (6dNHS1);
\draw[thick,<->] (6dDH)-- node[right] {$\substack{\text{T-dual} \\ S^1\subset 
\R_1}$} (6dDHS1);
\draw[thick,<->] (6dNHS1)-- node[right] {\footnotesize{S-dual}} (5dNH);
\draw[thick,<->] (6dDHS1)-- node[right] {\footnotesize{S-dual}} (5dDH);
\end{tikzpicture}
\caption{Higgsing in the brane configuration of Figure \ref{fig:branes}. The mesonic Higgsing of the 6d 
theory $\surm(k{+}1)$ with $N_f=2k{+}2$ to $\surm(k)$ with $N_f=2k$, is 
realised by moving one D6 away along the $x^{7,8,9}$ direction. For the dual 5d 
theory, the corresponding baryonic Higgsing of the affine $A_{k}$ quiver to 
the affine $A_{k{-}1}$ quiver is realised by moving one NS5 brane along 
$x^{7,8,9}$. A codimension 2 defect for the 6d brane configuration is introduced 
via a $\widetilde{\mathrm{D4}}$ brane that is attached to the D6 brane which is moved along 
$x^{7,8,9}$. In the dual 5d system this becomes a $\widetilde{\mathrm{D3}}$ brane suspended between 
the 5-brane web and the NS5 that is displaced in $x^{7,8,9}$ direction.}
\label{fig:Higgsing_branes}
\end{figure}
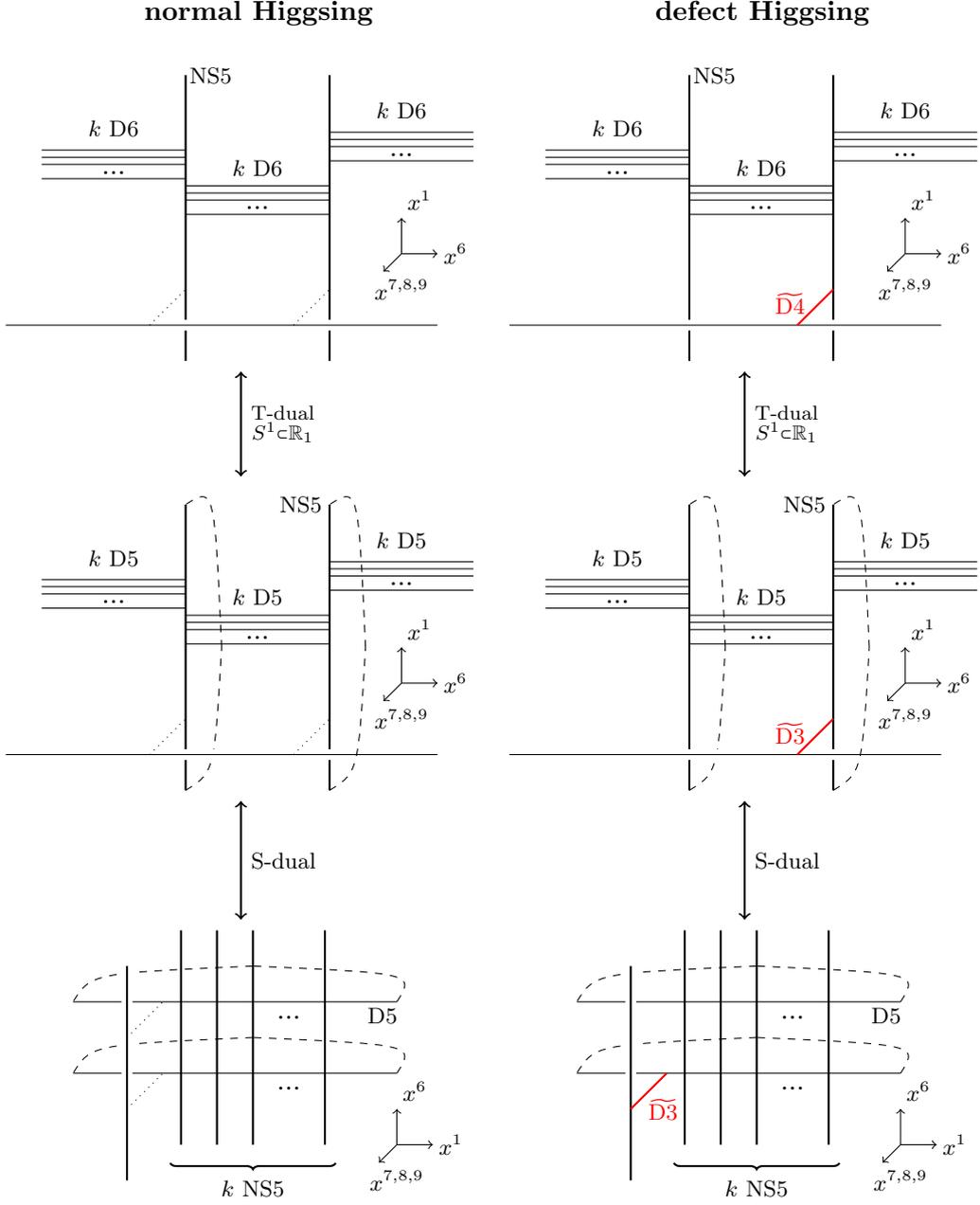

\subsubsection{Defect via D4 brane}
One way to add a codimension 2 defect into the $6$d theory is given by 
including additional D-branes. In the Type IIA brane configuration, this can be 
realised by introducing an additional $\widetilde{\mathrm{D4}}$ brane with world-volume $(x^0,x^1, 
x^4,x^5, x^7)$ ending on a NS$5$ brane, see Figure \ref{fig:branes}. 
This $\widetilde{\mathrm{D4}}$ is, indeed, of codimension 2 for the 6d world-volume theory on the D6 
branes. 
One notes that this space-time occupancy of the branes breaks supersymmetry 
further to 4 supercharges.
Moreover, the $\widetilde{\mathrm{D4}}$ brane breaks the space-time symmetry 
\eqref{eq:space-time_sym} to
\begin{align}
\label{eq:space-time_sym_D4}
\begin{aligned}
 \sorm(1,9)&\to \sorm(1,1) \times \sorm(4)_{2345} \times \sorm(3)_{789}  \\
 &\to \sorm(1,1) 
 \times  \sorm(2)_{23} 
 \times  \sorm(2)_{45} 
 \times  \sorm(3)_{789}
\,.
\end{aligned}
\end{align}
The world-volume theory on the D2 branes, which now only has $\Ncal=(0,2)$ 
supersymmetry, is read off from the open string modes as above. The open 
strings between the D2-D6-NS5 branes induce the multiplets 
\eqref{eq:2d_multiplets} from the original set-up. 
In addition, the D2-$\widetilde{\mathrm{D4}}$ open strings give rise to a pair $(\sigma,\Xi)$ of 
$\Ncal=(0, 2)$ 
bosonic and fermionic multiplets charged under gauge group 
$\urm(l)$ of the world-sheet theory, such that the supersymmetry is broken from 
$\Ncal=(0,4)$  to $\Ncal=(0, 2)$. The charges are summarised in Figure
\ref{fig:fields_charges} and the resulting 2d quiver gauge theory is given by
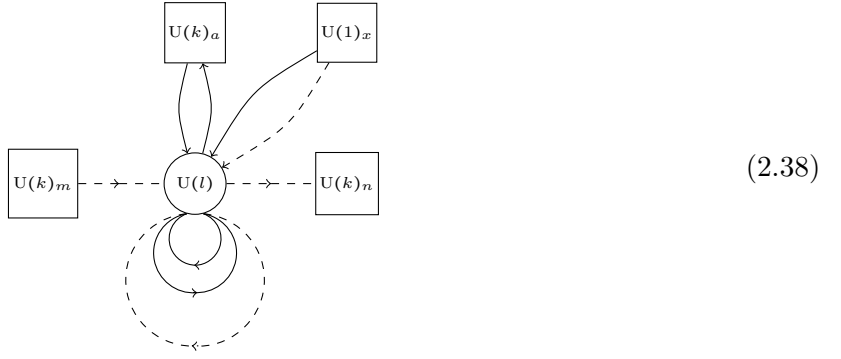
\begin{align}
\label{eq:2d_quiver_codim2}
	\raisebox{-.5\height}{
 	\begin{tikzpicture}
 		\tikzset{node distance = 2cm}
	\tikzstyle{gauge} = [circle, draw,inner sep=3pt];
	\tikzstyle{flavour} = [regular polygon,regular polygon sides=4,inner 
sep=-2pt, draw];
	\node (g1) [gauge] {\tiny{$\urm(l)$}};
	\node (f1) [flavour,left of=g1] {\tiny{$\urm(k)_m$}};
	\node (f2) [flavour,above of=g1] {\tiny{$\urm(k)_a$}};
	\node (f3) [flavour,right of=g1] {\tiny{$\urm(k)_n$}};
	\node (cd2) [flavour,above of=f3] {\tiny{$\urm(1)_x$}};
	\draw[->] (g1) .. controls (0.25,1) .. (f2);
	\draw[<-] (g1) .. controls (-0.25,1) .. (f2);
	\draw [-,->] (0+0.11,-0.4) arc (70:-90:10pt);
	\draw [-] (0+-0.11,-0.4) arc (110:270:10pt);
	\draw [-] (0+0.11,-0.4) arc (80:-95:15pt);
	\draw [-,->] (0+-0.11,-0.4) arc (100:275:15pt);
    \draw [dashed,->] (0+0.11,-0.4) arc (85:-95:25pt);
	\draw [dashed] (0+-0.11,-0.4) arc (95:275:25pt);
	\draw[dashed,->] (f1)--(-1,0);
	\draw[dashed,->] (g1)--(1,0);
	\draw[dashed] (-1,0)--(g1) (1,0)--(f3);
    \draw[->] (cd2) .. controls (1-0.25,1+0.25) .. (g1);
    \draw[dashed,->] (cd2) .. controls (1+0.25,1-0.25) .. (g1);
	\end{tikzpicture}
	}
\end{align}
where the difference compare to \eqref{eq:2d_quiver} is given by the $\urm(1)_x$ defect flavour node of the additional 
bosonic and fermionic multiplets.
Considering the elliptic genus, the 2d multiplets from the theory without 
defect contribute the 1-loop determinants \eqref{eq:1-loop-part}, while the 
additional multiplets $\sigma$ and $\Xi$ have determinants 
\begin{subequations}
\label{eq:1-loop_codim=2}
\begin{align}
 Z_{\mathrm{chiral}}^{\widetilde{\mathrm{D4}}} &=
 \prod_{p=1}^l\frac{\im\, \eta}{
\theta_1(u_p-x)} \,, \\
 Z_{\mathrm{Fermi}}^{\widetilde{\mathrm{D4}}} &=
 \prod_{p=1}^l
 \frac{\theta_1(u_p-x+\epsilon_2)}{\im\, \eta} \,,
\end{align}
\end{subequations}
where $x$ denotes the fugacity of the $\urm(1)$ symmetry.
The $\epsilon_2$ charge of the new multiplets follows because the D4 occupies $\R^2_{45}$, whose rotation parameter is $\epsilon_2$.
Collecting the determinants \eqref{eq:1-loop_no_defect} and 
\eqref{eq:1-loop_codim=2}, one finds
\begin{align}
 Z_{\mathrm{1-loop}}^{\widetilde{\mathrm{D4}}}(k,l)
 \coloneqq 
 Z_{\mathrm{vec}}
 Z_{\mathrm{chiral}}
 Z_{\mathrm{Fermi}}
 \cdot
 Z_{\mathrm{chiral}}^{\widetilde{\mathrm{D4}}}
 Z_{\mathrm{Fermi}}^{\widetilde{\mathrm{D4}}} 
 \equiv
 Z_{\mathrm{1-loop}}(k,l)
 \cdot
 Z_{\mathrm{chiral}}^{\widetilde{\mathrm{D4}}}
 Z_{\mathrm{Fermi}}^{\widetilde{\mathrm{D4}}} 
 \,.
\end{align}
The claim is that the additional $\widetilde{\mathrm{D4}}$ brane induces a $(r,s)=(0,1)$ defect in 
the sense of \eqref{eq:Higgs_with_defect_partFct}, see also 
\cite{Gaiotto:2012xa,Gaiotto:2014ina,Ito:2016fpl}. 
As a remark, a $(r,s)=(1,0)$ defect can be constructed via a $\widetilde{\mathrm{D4}}$ brane that extends along $(x^0,x^1,x^2,x^3,x^7)$, such that the 2d defect multiplets are charged under $\epsilon_1$ instead.

Consequently, one may label the resulting $2$d elliptic genera as 
follows:
\begin{align}
Z^{(0,1)\Deff}_l
=
\frac{1}{l!}\oint \frac{\diff^l u}{(2\pi \im)^l} 
Z_{\mathrm{1-loop}}^{\widetilde{\mathrm{D4}}}(k,l) 
\,.
\end{align}
As a next step, the result \eqref{eq:1-loop_codim=2} is re-derived and 
generalised to a $(r,s)$ defect via a position-dependent vacuum expectation 
values, as in Section \ref{sec:Higgs_mech}.
%
%
%
\subsubsection{Defect via Higgsing: Perturbative contribution}
Here, the chosen approach is to modify the standard Higgsing 
\eqref{eq:Higgs_no_defect_partFct} such that the VEV becomes dependent on one 
$\R^2$ plane of the $\R^4_{\epsilon_1,\epsilon_2}$ as in 
\eqref{eq:Higgs_with_defect_partFct}. 
For later purposes, one defines 
the defect fugacity $x$ in \eqref{eq:Higgs_with_defect} as follows:
\begin{align}
\label{eq:defect_fugacity}
\begin{aligned}
 a_{k+1} &= m_{k+1} - \epsilon_+ -b = x+\epsilon_+  \,,\\
 n_{k+1} &= m_{k+1} - 2 \epsilon_+ -2b -r \epsilon_1 -s \epsilon_2  \equiv x -b
 -r \epsilon_1 -s \epsilon_2 \,,\\
 &\text{with} \qquad 
 x \equiv  m_{k+1} - 2 \epsilon_+ -b\,.
\end{aligned}
\end{align}
For the exponentiated fugacities, the Higgsing \eqref{eq:Higgs_with_defect} 
takes the form
\begin{align}
 e^{a_{k+1}} =  \sqrt{pq} X\;,
 \quad
 e^{n_{k+1}} = \frac{X}{B p^r q^s}  \;,
 \quad
 e^{m_{k+1}} \equiv pq X B \;,
\quad   \text{with} \quad 
 X\equiv e^x\;, \, 
 B\equiv e^b
\end{align}
using the definition of the defect fugacity \eqref{eq:defect_fugacity}. As 
detailed in Appendix \ref{app:defect_Higgs_pert}, this Higgsing results in 
\begin{align}
   Z_{\mathrm{pert}}^{k+1}\bigg|_{\eqref{eq:Higgs_with_defect}} &=   
Z_{\mathrm{pert}}^{k} \cdot Z_G
   \cdot 
   Z_{\mathrm{pert}}^{(r,s)\Deff}\\
   Z_{\mathrm{pert}}^{(r,s)\Deff} 
 &=
   \PE \left[ \frac{( 1 - p^rq^s )}{(1-p)(1-q)} \left( \frac{Q}{1-Q} 
+\frac{1}{2}\right)
   \left\{  \frac{(1-p^{r+1} q^{s+1})}{p^r q^s} 
+ \sqrt{pq} \sum_{i=1}^{k}  \left(-\frac{e^{a_i}}{X}  + 
\frac{1}{p^r q^s} \frac{X }{e^{a_i}}   \right)
   \right\} \right]
   \label{eq:defect_pert}
\end{align}
and \eqref{eq:defect_pert} contains the additional contributions from the 
codimension 2 defect, i.e.\
\begin{align}
 Z_{\mathrm{pert}}^{k+(r,s)\Deff} =  Z_{\mathrm{pert}}^{k} \cdot  
Z_{\mathrm{pert}}^{(r,s)\Deff} \,,
\end{align}
where the Goldstone mode contribution have been removed. Note that 
$Z_{\mathrm{pert}}^{(r,s)\Deff} =1$ for $(r,s)=(0,0)$.
To be specific, specialising 
\eqref{eq:defect_pert} to $(r,s)=(0,s)$ yields
\begin{align}
 Z_{\mathrm{pert}}^{(0,s)\Deff} &=
   \PE \left[ \frac{( 1 - q^s )}{(1-p)(1-q)} \left( \frac{Q}{1-Q} 
+\frac{1}{2}\right)
   \left\{  \frac{(1-p q^{s+1})}{ q^s} 
+ \sqrt{pq} \sum_{i=1}^{k}  \left(
\frac{1}{ q^s} \frac{X }{e^{a_i}} -\frac{e^{a_i}}{X }
\right)
   \right\} \right]
   \notag \\
   &=\PE \left[ \frac{\sum_{l=0}^{s-1}q^{l}}{(1-p)} \left( 
\frac{Q}{1-Q} 
+\frac{1}{2}\right)
   \left\{  \frac{(1-p q^{s+1})}{ q^s} 
+ \sqrt{pq} \sum_{i=1}^{k}  \left(
\frac{1}{ q^s} \frac{X }{e^{a_i}}  -\frac{e^{a_i}}{X }
\right)
   \right\} \right]\,.
\end{align}
In the NS 
limit $q\rightarrow 1$, one obtains
\begin{align}
\lim_{\epsilon_2\to0} Z_{\mathrm{pert}}^{(0,s)\Deff} &=
   \PE \left[ \frac{s}{(1-p)} \left( 
\frac{Q}{1-Q} 
+\frac{1}{2}\right)
   \left\{ (1-p) 
+ \sqrt{p} \sum_{i=1}^{k}  \left(
 \frac{X }{e^{a_i}}  -\frac{e^{a_i}}{X }
\right)
   \right\} \right]
   \label{eq:pert_part_NS_limit}\\
 &= \left( \lim_{\epsilon_2\to0} Z_{\mathrm{pert}}^{(0,1)\Deff}  
\right)^s 
\,.
\notag
\end{align}
Thus, the contribution of a $(0,s)$ defect factorises into $s$ copies of a $(0,1)$ defect in the NS limit.
%
%
\subsubsection{Defect via Higgsing: Elliptic genus}
The fugacity assignment for a  Higgsing with a position dependent VEV  has 
been derived in \eqref{eq:Higgs_with_defect}. 
Inserting the fugacity assignment into \eqref{eq:ell_genus_Higgs_intermediate} 
yields the following:
\begin{align}
  Z_l^{k+1} \bigg|_{\eqref{eq:Higgs_with_defect}}
&= \frac{1}{l!}\oint \frac{\diff^l u}{(2\pi \im)^l} 
\,
Z_{\mathrm{1-loop}}(k,l)
\cdot 
\prod_{p=1}^l
\frac{ \theta_1(u_p-x-2\epsilon_+ ) 
\theta_1(u_p-x +r \epsilon_1 +s \epsilon_2) }{ 
\theta_1( u_p-x )\, 
\theta_1( u_p-x -2\epsilon_+ )}
\notag \\
&= \frac{1}{l!}\oint \frac{\diff^l u}{(2\pi \im)^l}
\,
Z_{\mathrm{1-loop}}(k,l)
\cdot
\prod_{p=1}^l
\frac{  \theta_1(u_p-x +r \epsilon_1 +s \epsilon_2 ) }{ 
 \theta_1(u_p-x)}
 \label{eq:elliptic_genus_codum2}
 \\
&\equiv 
\frac{1}{l!}\oint \frac{\diff^l u}{(2\pi \im)^l} 
\,
Z_{\mathrm{1-loop}}(k,l)
\cdot
\prod_{p=1}^l
 V_{(r,s)}(u_p)
 \notag
\end{align}
with the definition
\begin{align}
\label{eq:defect_term}
 V_{(r,s)}(u)  &\coloneqq
\frac{\theta_1( u-x +r \epsilon_1 +s \epsilon_2 
)}{\theta_1( u-x)}
=
\frac{\vartheta_1( u-x +r \epsilon_1 +s \epsilon_2 
)}{\vartheta_1( u-x)} 
\,,
\end{align}
which corresponds to the contribution of an $(r,s)$ defect. In other words, 
\eqref{eq:defect_term} are the 1-loop determinants of the Fermi and chiral 
multiplet that define the defect.  In particular, for $(r,s)=(0,1)$ the defect 
contribution reduces to the results \eqref{eq:1-loop_codim=2} of the defect 
introduced by the $\widetilde{\mathrm{D4}}$ brane. 

The resulting 1-loop determinant and  elliptic genus are then defined as follows:
\begin{align}
Z_{\mathrm{1-loop}}^{(r,s)\Deff}(k,l)&\coloneqq
Z_{\mathrm{1-loop}}(k,l)
\cdot \prod_{p=1}^l V_{(r,s)}(u_p) \,, \\
Z^{(r,s)\Deff}_l
&=
\frac{1}{l!}\oint \frac{\diff^l u}{(2\pi \im)^l} 
Z_{\mathrm{1-loop}}^{(r,s)\Deff}(k,l)
\,,
\end{align}
employing the definitions \eqref{eq:1-loop-part} and \eqref{eq:defect_term}.
\paragraph{1-string.}
Performing the integration for $l=1$ yields:
\begin{align}
  Z_{1}^{(0,s)\Deff} 
&=
\frac{ \vartheta_1(2\epsilon_+)}{  
\vartheta_1(\epsilon_1)\,\vartheta_1(\epsilon_2)}
\left[
\sum_{i=1}^k \bigg(
Q^\vee(a_i-\epsilon_+)
\cdot
V_{(0,s)}(a_i-\epsilon_+)
\bigg) 
+
\vartheta_1(s\epsilon_2 )
\cdot 
Q(x)
\right] \label{eq:elliptic_genus_k=1}
\,.
\end{align}
The normalised 1-string contribution in the NS-limit \cite{Nekrasov:2009rc} 
reads
\begin{align}
 \widetilde{Z}_{1}^{(0,s)\Deff}  &= 
 Z_{1}^{(0,s)\Deff}  - Z_{1}  \notag \\
\lim_{\epsilon_2\to0} \widetilde{Z}_{1}^{(0,s)\Deff}
&=
\frac{ s}{  
\vartheta_1^\prime(0)}
\sum_{i=1}^k 
Q_{(0)}^\vee(a_i-\tfrac{1}{2}\epsilon_1)
\cdot
L(a_i-x-\tfrac{1}{2}\epsilon_1)
+s \cdot
Q_{(0)}(x)
\label{eq:elliptic_genus_k=1_normalise_NS}  \\
&= s \cdot \lim_{\epsilon_2\to0} \widetilde{Z}_{1}^{(0,1)\Deff}
\notag
\end{align}
and the $(0,s)$ defect part is the product of $s$ copies of the $(0,1)$ defect contribution. $L(\cdot)$ is defined in \eqref{eq:def_L_and_K}.
The detailed derivation of \eqref{eq:elliptic_genus_k=1} and \eqref{eq:elliptic_genus_k=1_normalise_NS} is provided in Appendix \ref{app:details_codim=2_k=1}. 
\paragraph{2-string.}
The $l=2$ case yields the following elliptic genus:
\begin{align}
 Z_2^{(0,s)\Deff}
&=
\left(
 \frac{  \vartheta_1(2\epsilon_+)
 }{\vartheta_1(\epsilon_1)\vartheta_1(\epsilon_2)}
\right)^2
\sum_{1\leq i< j \leq k}
D(a_i-a_j)D(a_j-a_i)  
\label{eq:codim2_result_2-string}\\
&\qquad \qquad \qquad \qquad \qquad  \cdot
Q^\vee(a_i-\epsilon_+)
Q^\vee(a_j-\epsilon_+)
V_{(0,s)}(a_i-\epsilon_+)
V_{(0,s)}(a_j-\epsilon_+) 
\notag \\
&+
 \frac{  \vartheta_1(2\epsilon_+) 
 }{ \vartheta_1(2\epsilon_-) }
 \sum_{m=1}^k
 Q^\vee(a_m-\epsilon_+)
V_{(0,s)}(a_m-\epsilon_+)
\notag \\
&\qquad\qquad \qquad  \cdot
\bigg[
\frac{\vartheta_1(\epsilon_1+2\epsilon_+)}{
\vartheta_1(\epsilon_2)
\vartheta_1(2\epsilon_1)
}
Q(a_m-\epsilon_+-\epsilon_1)
V_{(0,s)}(a_m-\epsilon_+ -\epsilon_{1})
\notag
\\
&\qquad \qquad \qquad \qquad \qquad -
\frac{\vartheta_1(\epsilon_2+2\epsilon_+)}{
\vartheta_1(\epsilon_1)
\vartheta_1(2\epsilon_2)
}
Q(a_m-\epsilon_+-\epsilon_2)
V_{(0,s)}(a_m-\epsilon_+ -\epsilon_{2})
\bigg]
\notag \\
&+
\left(
\frac{\vartheta_1(2\epsilon_+)}{
\vartheta_1(\epsilon_1)
\vartheta_1(\epsilon_2)}
\right)^2
\vartheta_1(s \epsilon_2)
\sum_{m=1}^k
D(a_m-x-\epsilon_+)
D(x+\epsilon_+-a_m)
\notag \\
&\qquad \qquad \qquad \qquad  \cdot
Q^\vee(a_m-\epsilon_+)
Q(x)
V_{(0,s)}(a_m-\epsilon+)
\notag \\
&+
\frac{
\vartheta_1(2\epsilon_+)}{
\vartheta_1(2\epsilon_-)
}
\cdot 
Q(x)
\vartheta_1(s\epsilon_2)
\cdot 
\bigg[
 \frac{
\vartheta_1(\epsilon_1+2\epsilon_+)}{
\vartheta_1(\epsilon_2)
 \vartheta_1(2\epsilon_1)}
 Q(x-\epsilon_1)
V_{(0,s)}(x-\epsilon_1) 
\notag \\
&\qquad \qquad \qquad \qquad \qquad \qquad 
-
 \frac{
\vartheta_1(\epsilon_2+2\epsilon_+)}{
\vartheta_1(\epsilon_1)
 \vartheta_1(2\epsilon_2)}
 Q(x-\epsilon_2)
V_{(0,s)}(x-\epsilon_2)
\bigg]
\notag
\,.
\end{align}
Consider the normalised 2-string elliptic genus
\begin{align}
 \widetilde{Z}_{2}^{(0,s)\Deff}  &= 
 Z_{2}^{(0,s)\Deff}  - Z_{2} 
 -  Z_{1} \left(  Z_{1}^{(0,s)\Deff}  - Z_{1} \right)\,,
 \end{align}
 see Appendix \ref{app:normalised_part_fct}.
The full normalised 2-string elliptic genus for the codimension 2 defect in the 
NS-limit is given by
\begin{align}
  \widetilde{Z}_{l=2}^{(0,s)\Deff}  &= 
  -\frac{s}{2}
  \sum_{j=1}^k 
  \left( 
  \frac{Q_{(0)}^\vee(a_j - \tfrac{\epsilon_1}{2})}{
  \vartheta_1^\prime(0)}
    \right)^2
   K(a_j -x - \tfrac{\epsilon_1}{2}) 
  \label{eq:codim2_result_2-string_normalised} \\
 &\quad + \sum_{j=1}^k 
  \left( 
  \frac{Q_{(0)}^\vee(a_j - \tfrac{\epsilon_1}{2})}{
  \vartheta_1^\prime(0)}
    \right)^2
  \Bigg\{
  \frac{s(s+1)}{2}     
  L(a_j -x - \tfrac{\epsilon_1}{2})^2 
+2 s \cdot L(\epsilon_1)
L(a_j -x- \tfrac{\epsilon_1}{2})
  \notag \\
&\qquad \qquad \qquad 
+s 
L(a_j -x - \tfrac{\epsilon_1}{2})
\bigg[
\sum_{i=1}^k L(a_j -a_i -\epsilon_1)
+ \sum_{\substack{i=1\\ i\neq j}}^k L(a_j -a_i) 
\notag \\
  &\qquad \qquad \qquad \qquad \qquad \qquad \qquad \qquad 
-\sum_{i=1}^k \left( 
L(a_j -\tfrac{\epsilon_1}{2} -m_i +b)
+L(a_j -\tfrac{\epsilon_1}{2} -n_i -b) 
\right)
\bigg]
\Bigg\}
 \notag\\
&\qquad 
+
\sum_{\substack{i,j=1\\ i\neq j}}^k
\frac{  Q_{(0)}^\vee(a_i - \tfrac{\epsilon_1}{2})}{
\vartheta_1^\prime(0)}
\frac{ Q_{(0)}^\vee(a_j - \tfrac{\epsilon_1}{2})}{
\vartheta_1^\prime(0)}
\Bigg\{
\frac{s^2}{2}
  L(a_i -x - \tfrac{\epsilon_1}{2})
  L(a_j -x - \tfrac{\epsilon_1}{2}) 
\notag \\
  &\qquad \qquad 
+s L(a_i -x - \tfrac{\epsilon_1}{2})
  \big[ 
  L(a_i - a_j +\epsilon_1)
  -L(a_i - a_j )
  +L(a_j - a_i +\epsilon_1)
  -L(a_j - a_i )
  \bigg]
  \Bigg\}
  \notag \\
&\qquad
+s
\sum_{j=1}^k  
\frac{Q_{(0)}^\vee(a_j - \tfrac{\epsilon_1}{2})}{
\vartheta_1^\prime(0)}
Q_{(0)}(a_j - \tfrac{3\epsilon_1}{2})
\left[
L(a_j -x - \tfrac{\epsilon_1}{2})
+
L(a_j -x - \tfrac{3\epsilon_1}{2})
\right]
\notag \\
&\qquad 
+s\cdot Q_{(0)}(x)
\sum_{j=1}^k 
\frac{Q_{(0)}^\vee(a_j - \tfrac{\epsilon_1}{2})}{
\vartheta_1^\prime(0)}
\bigg[
L(a_j -x +\tfrac{\epsilon_1}{2} )
- L(a_j -x -\tfrac{\epsilon_1}{2} )
\notag \\
  &\qquad \qquad \qquad \qquad \qquad \qquad \qquad \qquad 
+L(x- a_j +\tfrac{3\epsilon_1}{2} )
- L(x-a_j  +\tfrac{\epsilon_1}{2} )
+ s L(a_j -x -\tfrac{\epsilon_1}{2} )
\bigg] 
\notag \\
&\qquad 
+s\cdot Q_{(0)}(x) 
\left( 
Q_{(0)}(x-\epsilon_1)
-\frac{1-s}{2}
Q_{(0)}(x)
\right)
\,,
\notag
\end{align}
with $L(\cdot)$ and $K(\cdot)$ as defined in \eqref{eq:def_L_and_K}.
The computational details of \eqref{eq:codim2_result_2-string} and \eqref{eq:codim2_result_2-string_normalised} are presented in Appendix \ref{app:details_codim=2_k=2}.%
\paragraph{Full defect partition function.}
The 6d partition function in the presence of the codimension 2  defect is then denoted as 
\begin{align}
    Z_{\mathrm{6d}}^{(r,s)\Deff} \coloneqq Z_{\mathrm{pert}}^{(r,s)\Deff} \cdot Z_{\str}^{(r,s)\Deff}
\end{align}
in the rest of this paper.
%
%
\subsection{Codimension 4 defect}
\label{sec:Wilson_surface}
A natural candidate for a  codimension 4 defect is a Wilson surface $\Sigma$
\cite{Ganor:1996nf,Chen:2007ir} that acquires a vacuum expectation value.
The VEV of a Wilson surface in representation $\mathcal{R}$ can formally be 
expressed in terms of the two-form 
potential $B_{\mu\nu}$ and the associated supersymmetric strings as,
\begin{align}\label{eq:Wilson-tensor}
\mathcal W_{\mathcal{ R}}[\Sigma]=\Tr_{\mathcal{R}}\left(\mathcal P  \ e^{
i\int_\Sigma 
d\sigma^{\mu\nu}(B_{\mu\nu}+\ldots) }\right)\,,
\end{align}
where $\ldots$ denotes the necessary supersymmetric partners of $B_{\mu\nu}$. There exists another type of codimension 4 BPS defects that couples to the 6d gauge symmetry. One can consider a 2d chiral fermion field $\psi$ localised at the origin of $\R^4$ that couples to the bulk 6d gauge group through the following action:
\begin{align}\label{eq:Wilson-vector}
	S^{\mathrm{2d}} = \int \diff^2x\, \bar\psi_- (D_0+D_1)\psi_- \ ,
\end{align}
where $D_i = \partial_i +iA_i$ with $i=0,1$ and $A_i$ is the bulk $\surm(k)$ gauge field. Adding this action to the path integral introduces a codimension 4 defect preserving half the supersymmetries. This defect is a 6d generalisation of the Wilson loop generating function in a 5d gauge theory that can also be called the 6d qq-character \cite{Nekrasov:2015wsu,Kim:2016qqs}. The codimension 4 defect that is discussed below is a product of these two types \eqref{eq:Wilson-tensor} and \eqref{eq:Wilson-vector} which is called the Wilson surface defect from now on. Consequently, the Wilson surface defect carries both tensor and gauge charges.

In practice, because of the lack of a field theoretical formulation of $6$d 
SCFTs, one has to resort to string theory to formulate the Wilson surface 
defect and compute it. Wilson surface defects have, for example, been 
considered on the $\Omega$-deformed $\R^4 \times \T^2$ in 
\cite{Bullimore:2014upa,Nekrasov:2015wsu,Kim:2016qqs,Agarwal:2018tso}, see also \cite{Tong:2014yna}.
Following \cite{Agarwal:2018tso}, a Wilson surface defect in 
the 
$6$d $\Ncal=(1, 0)$ $A_{1}$ SCFTs can be realised in the Type IIA 
brane construction via an additional D$4^\prime$ brane filling the 
$x^0,x^1,x^7,x^8,x^9$ space-time directions, 
see Figure \ref{fig:branes}.
As the D$4^\prime$ occupies different space-time directions as the $\widetilde{\text{D4}}$ 
brane of Section \ref{sec:codim_2}, the codimension $4$ defect differs from 
the codimension $2$ defect. 
In contrast to the $\widetilde{\text{D4}}$ brane, the addition of the D$4^\prime$ brane to the 
D2-D6-NS5 brane preserves the broken space-time symmetry 
\eqref{eq:space-time_sym} of the original set-up.
As a consequence, the 2d world-volume theory is composed of the multiplets 
\eqref{eq:2d_multiplets} of the D2-D6-NS5 system which are then supplemented by 
additional multiplets that originate from the presence of the D$4^\prime$ brane.
These new  multiplets originate from the following:
\begin{compactitem}
 \item The D2-D$4^\prime$ open string modes give rise to an additional 
$\Ncal=(0,4)$ twisted hyper $\phi^A$ 
and a Fermi multiplet $\Gamma_\alpha$, which do not break the $\Ncal=(0, 
4)$ supersymmetry of the resulting $2$d quiver theory.
 \item The D6-D$4^\prime$ open strings introduce an additional Fermi multiplet 
$\rho$, which is a singlet under the 2d gauge group as well as the $\sorm(4)_R$ 
R-symmetry.
\end{compactitem}
Decomposing $\Ncal=(0,4)$ multiplets into $\Ncal=(0,2)$ multiplets, yields the 
field content from the original theory \eqref{eq:2d_multiplets} plus the 
additional $\Ncal=(0,2)$ multiplets due to the additional D$4^\prime$ brane. 
For the latter, one finds \cite{Agarwal:2018tso}
\begin{subequations}
\begin{align}
{\rm twist\ hyper} \ (\phi^{A}, \eta^{\dot\alpha}) &\longrightarrow {\rm 
chiral}\ \phi\ (\phi^{1}, \eta^{\dot1})\ +\ {\rm chiral}\ \tilde \phi^\dagger\ 
(\phi^{\dot 2}, \eta^{2}) \\
{\rm Fermi} \  
\Gamma_\alpha, \rho &\longrightarrow {\rm 
Fermi}\  
\Gamma_\alpha,\rho \,.
\end{align}	
\end{subequations}
and the charges are detailed in Figure 
\ref{fig:fields_charges}.
The resulting 2d quiver gauge theory can be encoded in
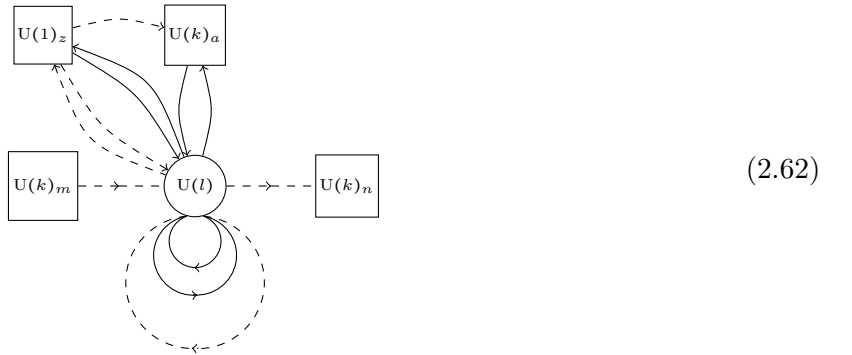
\begin{align}
\label{eq:2d_quiver_codim4}
	\raisebox{-.5\height}{
 	\begin{tikzpicture}
 		\tikzset{node distance = 2cm}
	\tikzstyle{gauge} = [circle, draw,inner sep=3pt];
	\tikzstyle{flavour} = [regular polygon,regular polygon sides=4,inner 
sep=-2pt, draw];
	\node (g1) [gauge] {\tiny{$\urm (l)$}};
	\node (f1) [flavour,left of=g1] {\tiny{$\urm(k)_m$}};
	\node (f2) [flavour,above of=g1] {\tiny{$\urm(k)_a$}};
	\node (f3) [flavour,right of=g1] {\tiny{$\urm(k)_n$}};
	\node (cd4) [flavour,above of=f1] {\tiny{$\urm(1)_z$}};
	\draw[->] (g1) .. controls (0.25,1) .. (f2);
	\draw[<-] (g1) .. controls (-0.25,1) .. (f2);
	\draw [-,->] (0+0.11,-0.4) arc (70:-90:10pt);
	\draw [-] (0+-0.11,-0.4) arc (110:270:10pt);
	\draw [-] (0+0.11,-0.4) arc (80:-95:15pt);
	\draw [-,->] (0+-0.11,-0.4) arc (100:275:15pt);
    \draw [dashed,->] (0+0.11,-0.4) arc (85:-95:25pt);
	\draw [dashed] (0+-0.11,-0.4) arc (95:275:25pt);
	\draw[dashed,->] (f1)--(-1,0);
	\draw[dashed,->] (g1)--(1,0);
	\draw[dashed] (-1,0)--(g1) (1,0)--(f3);
    \draw[<-] (cd4) .. controls (-1+0.45,1+0.45) .. (g1);
    \draw[->] (cd4) .. controls (-1+0.25,1+0.25) .. (g1);
    \draw[dashed,<-] (cd4) .. controls (-1-0.45,1-0.45) .. (g1);
    \draw[dashed,->] (cd4) .. controls (-1-0.25,1-0.25) .. (g1);
    \draw[dashed,->] (cd4) .. controls (-1,2+0.25) .. (f2);
	\end{tikzpicture}
	}
\end{align}
where the changes due to the D$4^\prime$ brane are manifest in the additional $\urm(1)_z$ defect flavour node compared to \eqref{eq:2d_quiver}.
Analogously to the elliptic genus \eqref{eq:ell_genus_k-string} of the theory 
without defect, the 1-loop determinant contributions from the 2d multiplets 
include the terms \eqref{eq:1-loop_no_defect} form the original theory plus the 
following defect parts:
\begin{subequations}
\label{eq:1-loop_codim=4}
\begin{align}
Z_{\rm chiral}^{D4^\prime} 
&=
\prod_{p=1}^l\frac{(\im\, \eta)^2}{
\theta_1(-\epsilon_+\pm (u_p-z))}
\,,
\\
Z_{\rm fermi}^{D4^\prime} 
&=
\prod_{p=1}^l\frac{\theta_1(\epsilon_-\pm 
(u_p-z))}{(\im\,\eta)^2}
\cdot
\prod_{j=1}^k \frac{\theta_1(z-a_j)}{\im\, \eta }
\,,
\end{align}
\end{subequations}
where the $z$-fugacity labels the $\urm(1)$ charge of the additional twisted 
hyper 
multiplet $\phi^A$ and Fermi multiplets  $\Gamma_\alpha$, $\rho$ due to the 
D$4^\prime$ brane. 
Collecting all the contributions from \eqref{eq:1-loop_no_defect} and 
\eqref{eq:1-loop_codim=4}, one obtains
\begin{align}
 Z_{\mathrm{1-loop}}^{\mathrm{D}4^\prime}(k,l)  
 &\coloneqq
 Z_{\mathrm{1-loop}}(k,l) 
\cdot
Z_{\rm chiral}^{D4^\prime} 
Z_{\rm fermi}^{D4^\prime} 
\notag\\
&= W_{\mathrm{pert}} \cdot
\left(\frac{2\pi 
\,\eta^3\theta_1(2\epsilon_+)}{\theta_1(\epsilon_1)\,\theta_1(\epsilon_2)}
\right)^l 
\cdot
\prod_{\substack{p,q=1 \\ p\neq q}}^{l} D(u_p-u_q)
\cdot 
\prod_{p=1}^l Q(u_p)
W(u_p)
\label{eq:1-loop_det_codim4}
\\
&\equiv W_{\mathrm{pert}}\cdot Z_{\mathrm{1-loop}}^{\rm Wilson}(k,l) \notag
\,,
\end{align}
where the following definitions have been used
\begin{subequations}
\label{eq:Ell_genus_Wilson}
\begin{align}
 W(u)&\coloneqq 
 \frac{ \theta_1(\epsilon_-\pm (u-z))
 }{
 \theta_1(-\epsilon_+\pm (u-z)) }
=
\frac{ \theta_1( u-z \pm \epsilon_-)
 }{
 \theta_1(u-z \pm \epsilon_+ ) } 
=
\frac{ \vartheta_1( u-z \pm \epsilon_-)
 }{
 \vartheta_1(u-z \pm \epsilon_+ ) } \,,\\
W_\mathrm{pert} &\coloneqq 
\prod_{j=1}^k \frac{\theta_1(z-a_j)}{\im\, \eta }
=
\prod_{j=1}^k \frac{\vartheta_1(z-a_j)}{\im\, Q^{-\frac{1}{12}} } \,.
\end{align}
\end{subequations}
Note that $W_\mathrm{pert}$ is independent of the 2d gauge fugacities such that 
its contribution in the contour integral reduces to an identical prefactor 
for all elliptic genera. Hence, one may define
\begin{align}
Z^{\rm Wilson}_l &=
\frac{1}{l!}\oint \frac{\diff^l u}{(2\pi \im)^l}  Z_{\mathrm{1-loop}}^{\rm 
Wilson}(k,l) 
\label{eq:ell_genus_k-string_w/_Wilson_surface} \,,
\end{align}
using the definitions \eqref{eq:1-loop_det_codim4}--\eqref{eq:Ell_genus_Wilson}.
Therefore, the partition 
function of the theory in the presence of a Wilson surface 
is given by
\begin{align}
    Z_{\mathrm{6d}}^{\mathrm{Wilson}}
=
Z_{\mathrm{pert}}\cdot W_\mathrm{pert} \cdot \left(1+\sum_{l=1}^\infty 
q_\phi^l\, Z_l^{\rm Wilson}\right)
\,,
\end{align}
where $Z_{\rm pert}$ is the perturbative contribution 
\eqref{eq:def_pert_part} of the theory without defect, see also 
\cite[Sec.\ 3.3]{Agarwal:2018tso}.
Since the interest is placed on the Wilson surface expectation value, one has 
to normalise the partition function with respect to the partition function of 
the theory without codimension 4 defect. Therefore, the expectation value of 
Wilson 
surface is given by
\begin{align}
 \langle \mathcal{W}\rangle
 = W_\mathrm{pert} \cdot \frac{
 \left(1+\sum_{l=1}^\infty q_\phi^l \ Z_l^{\rm Wilson}\right)
 }{ \left(1+\sum_{l'=1}^\infty q_\phi^{l'}\ Z_{l'}\right)}
 =
  W_\mathrm{pert} \cdot \left[ 1 + \left(Z_{1}^{\rm Wilson} - Z_{1} 
\right)  q_\phi +\mathcal{O}(q_\phi^2) \right] 
\,,
\end{align}
see also Appendix \ref{app:normalised_part_fct}.
Before turning to the computation details, one may wonder in which 
representation $\mathcal{R}$ the Wilson surface transforms. As argued in 
\cite{Agarwal:2018tso}, the codimension 4 defect of a single D$4^\prime$ brane 
introduces a Wilson surface in the fundamental representation.
\subsubsection{Wilson surface: Perturbative contribution}
The perturbative contribution acts as a multiplicative factor. The explicit 
contribution is 
\begin{align}
  W_\mathrm{pert} =  (-\im)^k Q^{\frac{k}{12}}  P_0(z+\epsilon_+) \,.
\end{align}
\subsubsection{Wilson surface: Elliptic genus}
For the non-perturbative contributions of the Wilson surface expectation value, the 1-string and 2-string contributions are computed in this section.
\paragraph{1-string.}
Similar to the codimension 2 defect computation \eqref{eq:elliptic_genus_k=1}, 
one finds for the $l=1$ case of the contour integral \eqref{eq:ell_genus_k-string_w/_Wilson_surface} the following result
\begin{align}
  Z_{1}^{\mathrm{Wilson}} 
  &=
\frac{\vartheta_1(2\epsilon_+)}{\vartheta_1(\epsilon_1)\,\vartheta_1(\epsilon_2)
}
\sum_{i=1}^k 
Q^\vee(a_i -\epsilon_+)
\cdot W(a_i-\epsilon_+)
+
Q(z+\epsilon_+)
\,.
\label{eq:Wilson_1-string_pure}
\end{align}
The normalised 1-string contribution in the NS-limit becomes
\begin{align}
 \widetilde{Z}_{1}^{\mathrm{Wilson}}  &=  Z_{1}^{\mathrm{Wilson}}  -  
Z_{1} \,, \notag \\
 \lim_{\epsilon_2 \to 0} \widetilde{Z}_{1}^{\mathrm{Wilson}}
&= \frac{1}{\vartheta_1^\prime(0)}
\sum_{i=1}^k 
Q^\vee_{(0)}\left(a_i -\tfrac{1}{2}\epsilon_1\right)
\cdot \left[ 
L\left(a_i-z-\epsilon_1\right)
-
L\left(a_i-z\right)
\right]
+
Q_{(0)}\left(z+\tfrac{1}{2}\epsilon_1\right)
\,.
\label{eq:Wilson_1-string}
\end{align}
The detailed computations that lead to \eqref{eq:Wilson_1-string_pure} and \eqref{eq:Wilson_1-string} are summarised in Appendix \ref{app:details_codim=4_k=1}.
\paragraph{2-string.}
Consider the $l=2$ elliptic genus \eqref{eq:ell_genus_k-string_w/_Wilson_surface}, 
a computation yields 
\begin{align}
 Z_2^{\mathrm{Wilson}}
&=
\left(
 \frac{  \vartheta_1(2\epsilon_+)
 }{\vartheta_1(\epsilon_1)\vartheta_1(\epsilon_2)}
\right)^2
\sum_{1\leq i< j \leq k}
D(a_i-a_j)D(a_j-a_i)  
\label{eq:Wilson_2-string_pure}\\
&\qquad \qquad \qquad \qquad \qquad  \cdot
Q^\vee(a_i-\epsilon_+)
Q^\vee(a_j-\epsilon_+)
W(a_i-\epsilon_+)
W(a_j-\epsilon_+) 
\notag \\
&+
 \frac{  \vartheta_1(2\epsilon_+) 
 }{ \vartheta_1(2\epsilon_-) }
 \sum_{j=1}^k
 Q^\vee(a_j-\epsilon_+)
W(a_j-\epsilon_+)
\notag \\
&\qquad\qquad \qquad  \cdot
\bigg[
\frac{\vartheta_1(\epsilon_1+2\epsilon_+)}{
\vartheta_1(\epsilon_2)
\vartheta_1(2\epsilon_1)
}
Q(a_j-\epsilon_+-\epsilon_1)
W(a_j-\epsilon_+ -\epsilon_{1})
\notag
\\
&\qquad \qquad \qquad \qquad \qquad -
\frac{\vartheta_1(\epsilon_2+2\epsilon_+)}{
\vartheta_1(\epsilon_1)
\vartheta_1(2\epsilon_2)
}
Q(a_j-\epsilon_+-\epsilon_2)
W(a_j-\epsilon_+ -\epsilon_{2})
\bigg]
\notag \\
&+
\frac{\vartheta_1(2\epsilon_+)}{
\vartheta_1(\epsilon_1)
\vartheta_1(\epsilon_2)}
\sum_{j=1}^k
D(a_j-z-2\epsilon_+)
D(z+2\epsilon_+-a_j)
\notag \\
&\qquad \qquad \qquad \qquad  \cdot
Q^\vee(a_j-\epsilon_+)
Q(z+\epsilon_+)
W(a_j-\epsilon_+)
\,.
\notag 
\end{align}
The normalised 2-string elliptic genus for the codimension 4 defect reads
\begin{align}
 \widetilde{Z}_2^{\mathrm{Wilson}} &=
 \sum_{\substack{i,j=1\\ i\neq j}}^k  
 \frac{Q_{(0)}^\vee(a_i - \tfrac{\epsilon_1}{2}) }{\vartheta_1^\prime(0)}
 \frac{Q_{(0)}^\vee(a_j - \tfrac{\epsilon_1}{2}) }{\vartheta_1^\prime(0)}
 \bigg[
 \frac{1}{2} L(a_i-z)L(a_j-z) \notag \\
 &\qquad \qquad \qquad \qquad -L(a_i-z)L(a_j-z-\epsilon_1)
 +\frac{1}{2} L(a_i-z-\epsilon_1)L(a_j-z-\epsilon_1)
 \bigg] \notag \\
 &+ \sum_{\substack{i,j=1\\ i \neq j}}^k 
\frac{Q_{(0)}^\vee(a_i - \tfrac{\epsilon_1}{2}) }{\vartheta_1^\prime(0)} 
\frac{Q_{(0)}^\vee(a_j - \tfrac{\epsilon_1}{2}) }{\vartheta_1^\prime(0)} 
\left[ 
L(a_i-z-\epsilon_1)-L(a_i-z)
\right] 
\notag  \\
&\qquad \qquad \qquad \qquad
\cdot \bigg[
L(a_i -a_j +\epsilon_1) 
- L(a_i -a_j )
+L(a_j -a_i +\epsilon_1) 
- L(a_j -a_i )  
\bigg]
\notag  \\
&+\frac{1}{2} \sum_{j=1}^k 
\left( \frac{Q_{(0)}^\vee(a_j - \tfrac{\epsilon_1}{2}) }{\vartheta_1^\prime(0)} 
\right)^2 
\left[ 
K(a_j-z)-K(a_j-z-\epsilon_1)
\right] \notag  \\
&+ \sum_{j=1}^k 
\left( \frac{Q_{(0)}^\vee(a_j - \tfrac{\epsilon_1}{2}) }{\vartheta_1^\prime(0)} 
\right)^2 
\bigg( 
L(a_j-z-\epsilon_1)-L(a_j-z)
\bigg)
\bigg[
\frac{1}{2}
L(a_j-z-\epsilon_1)
+ 
2 L(\epsilon_1)
\notag  \\
&\qquad  + 
\sum_{i=1}^k L(a_j -a_i -\epsilon_1) 
+\sum_{\substack{i=1\\ i\neq j}}^k L(a_j -a_i -\epsilon_1)  
-\sum_{i=1}^N \left( L(a_j-\tfrac{\epsilon_1}{2} -m_i +b)
+ L(a_j-\tfrac{\epsilon_1}{2} -n_i -b)\right)
\bigg]
\notag  \\
&+\sum_{j=1}^k 
\frac{Q_{(0)}^\vee(a_j - \tfrac{\epsilon_1}{2}) }{\vartheta_1^\prime(0)} 
Q_{(0)}(a_j - \tfrac{3\epsilon_1}{2})
\left[ 
L(a_j-z-2\epsilon_1)-L(a_j-z)
\right] 
\notag \\
&+
Q_{(0)}(z + \tfrac{\epsilon_1}{2})
\sum_{j=1}^k 
\frac{Q_{(0)}^\vee(a_j - \tfrac{\epsilon_1}{2}) }{\vartheta_1^\prime(0)}
\left[ 
L(z-a_j+2\epsilon_1)-L(z-a_j+\epsilon_1)
\right] 
\,,
\label{eq:Wilson_2-string}
\end{align}
with $L(\cdot)$, $K(\cdot)$ as in \eqref{eq:def_L_and_K}.
The derivation of \eqref{eq:Wilson_2-string_pure} and \eqref{eq:Wilson_2-string} is detailed in Appendix \ref{app:details_codim=4_k=2}.
%
%
\section{Difference equation}
\label{sec:difference_eq}
In Section \ref{sec:partition_fcts}, several partition functions have been 
discussed.
Focusing on the defects introduced by a \emph{single} $\widetilde{\text{D4}}$ and \emph{single} 
D$4^\prime$, the partition functions are related as follows:
\begin{subequations}
\begin{align}
 Z_{k+1}^{\mathrm{6d}} 
 &\xrightarrow{\qquad \text{normal Higgs \eqref{eq:Higgs_no_defect_partFct}} 
\qquad }  
Z_{G} \cdot Z_{k}^{\mathrm{6d}} \,,
\\
Z_{k+1}^{\mathrm{6d}} 
&\xrightarrow{\quad \; \text{$(0,1)$-defect  Higgs 
\eqref{eq:Higgs_with_defect_partFct}}  \; \quad} 
Z_{G} \cdot Z_{k}^{\mathrm{6d/4d}}(x)  \,,
\end{align}
\end{subequations}
where $Z_{k}^{\mathrm{6d/4d}} \coloneqq Z_{\mathrm{6d}}^{(0,1)\Deff} \slash Z_{\mathrm{6d}} $ denotes the normalised partition function in the presence 
of a codimension 2 defect. Consequently, $Z_{k}^{\mathrm{6d/4d}} $  depends on 
the defect fugacity $x$.
In addition, one may introduce a codimension 4 defect to the 6d 
theory, which in terms of partition functions means
\begin{align}
 Z_{k}^{\mathrm{6d}} 
 &\xrightarrow{\quad \text{codim 4 defect} \quad }  
 Z_{k}^{\mathrm{6d/2d}}(z) 
\,,
\end{align}
where $Z_{k}^{\mathrm{6d/2d}}(z) \coloneqq Z_{\mathrm{6d}}^{\mathrm{Wilson}} \slash Z_{\mathrm{6d}} $ is the normalised partition function in the presence of the codimension 4 defect.
This codimension 4 defect is characterised by another defect fugacity $z$.

The aim of this section is to derive a difference operator $\mathcal{D}$, which 
acts via shifts on the codimension 2 defect fugacity 
$x$, and, similarly to \cite{Gaiotto:2012xa,Gaiotto:2015usa,Ito:2016fpl,Maruyoshi:2016caf,Yagi:2017hmj,Nazzal:2018brc}, is expected to generate the partition functions for the 6d 
theory in 
the presence of both, the codimension 2 and the codimension 4 defect, i.e.
\begin{align}
 \mathcal{D} Z^{\mathrm{6d/4d}}(x) = Z^{\mathrm{6d/4d/2d}}(x) \,.
 \label{eq:def_diff_op}
\end{align}
Clearly, since $Z^{\mathrm{6d/4d}}$ only depends on the defect fugacity $x$, 
and the flavour and gauge fugacities inherited from the pure 6d theory, the 
generated $Z^{\mathrm{6d/4d/2d}}$ cannot depend on $z$.
In the NS-limit \cite{Nekrasov:2009rc}, one expects a factorisation 
of the 
latter 
\begin{align}
 Z^{\mathrm{6d/4d/2d}}(x) \xrightarrow{\quad \mathrm{NS} \quad }
 \langle \mathcal{W} \rangle(x) \cdot Z^{\mathrm{6d/4d}}(x)
\end{align}
with $\langle \mathcal{W} \rangle$ being the Wilson surface expectation value 
of the 6d theory. In other words, $\langle \mathcal{W} \rangle\cong 
Z^{\mathrm{6d/2d}} $ for a \emph{suitable identification} of the defect 
fugacity $z$.
As a consequence, the $Z^{\mathrm{6d/4d}}$ partition function 
is annihilated by the following operator in the NS-limit:
\begin{align}
 \mathcal{D}- \langle \mathcal{W} \rangle 
 \quad \cong  \quad 
 \text{quantised SW-curve}
\end{align}
which, in the spirit of \cite{Gaiotto:2014ina,Bullimore:2014awa}, is expected to yield a quantisation of the Seiberg-Witten 
curve of the 
$6$d $\Ncal=(1,0)$ $A_{1}$ theory. The defect fugacity $x$ becomes the 
coordinate of the SW-curve.
\subsection{Path integral representation}
\label{sec:path_integral_rep}
As a first step towards the quantised SW-curve, one may try to express the 
non-perturbative parts of 
the partition function with codimension 2 defect as a path 
integral.
Following the approach of \cite{Ferrari:2012dy}, one may write the elliptic 
genus contributions via \eqref{eq:def_path_int} and \eqref{eq:defect_term} 
as follows:
\begin{align}
Z_{\str}^{(r,s)\Deff}
&=\sum_{l=0}^{\infty}
 \frac{1}{l!}q_\phi^l
 \oint  \left( \prod_{p=1}^l \frac{\diff u_p}{2\pi \im} \right) 
 \left( \frac{2\pi  \eta^3 
\theta_1(\epsilon_1+\epsilon_2)}{ 
\theta_1(\epsilon_1) \theta_1(\epsilon_2)}
\right)^l 
\prod_{\substack{p,q=1\\p\neq q}}^l D(u_p -u_q) 
\prod_{p=1}^l Q(u_p) 
\prod_{p=1}^l V_{(r,s)}(u_p)  
\,.
\end{align}
For all specific considerations, the defect is specialised to $(r,s)=(0,s)$.
Next, introduce the density
\begin{align}
 \bar{\rho}(u) =  \sum_{p=1}^l (\#)^{-1} \cdot \delta(u-u_p) 
\qquad 
\text{with}
\qquad
\#\coloneqq \frac{2\pi  \eta^3 
\theta_1(\epsilon_1+\epsilon_2)}{
\theta_1(\epsilon_1) \theta_1(\epsilon_2)}
\end{align}
and rewrite the partition function
\begin{align}
Z_{\str}^{(r,s)\Deff}&= 
\sum_{l=0}^{\infty}
 \frac{1}{l!}q_\phi^l
 \oint  \left( \prod_{p=1}^l \frac{\diff u_p}{2\pi \im} \right) \cdot 
 \left(  \# \right)^l 
 \int \mathcal{D}\rho(u) 
 \delta \left( \rho(u) - \sum_{p=1}^l (\#)^{-1} \cdot \delta(u-u_p)  \right) 
 \notag \\
&\quad \cdot \exp\bigg[\int \diff u \, \diff u' (\#)^{2} \rho(u) \log(D(u-u')) 
\rho(u')  \notag \\
&\qquad \qquad +\int \diff u\, \# \,\rho(u) (\log(Q(u))+ \log(V_{(r,s)}(u)))
\bigg] \,.
\end{align}
With the Fourier representation 
\begin{align}
 \delta \left( \rho(u) - \bar{\rho}  \right)
 = \int \mathcal{D}\lambda
 \exp \left[ \im\, \int \diff u \; \lambda(u) \left(  \rho(u) -\bar{\rho}\right)  
\right]
\end{align}
of the Delta function, one obtains
\begin{align}
Z_{\str}^{(r,s)\Deff}&= 
\sum_{l=0}^{\infty}
 \frac{1}{l!}q_\phi^l
 \int  \left( \prod_{p=1}^l \frac{\diff u_p}{2\pi \im} \right) \cdot 
 \left( \# \right)^l 
 \int \mathcal{D}\rho(u) 
 \int \mathcal{D}\lambda(u)
 \prod_{p=1}^l
 e^{-\im\, \int \diff u (\#)^{-1} \delta(u-u_p) \lambda(u)}
 \notag \\
&\quad \cdot \exp\bigg[\int \diff u \, \diff u' (\#)^{2} \rho(u) \log(D(u-u')) 
\rho(u')  \notag \\
&\qquad \qquad +\int \diff u  
\left( \im\, \lambda(u)\cdot \rho(u)  +
\, \# \, \rho(u) \left[\log(Q(u))+ \log(V_{(r,s)}(u))\right]
\right)
\bigg] \notag \\
&= 
\sum_{l=0}^{\infty}
 \frac{1}{l!}q_\phi^l
 \int  \left( \prod_{p=1}^l \frac{\diff u_p}{2\pi \im} \right) \cdot 
 \left(  \# \right)^l 
 \int \mathcal{D}\rho(u) 
 \int \mathcal{D}\lambda(u)
 \prod_{p=1}^l
 e^{-\im\,  (\#)^{-1}  \lambda(u_p)}
 \notag \\
&\quad \cdot \exp\bigg[\int \diff u \, \diff u' (\#)^{2} \rho(u) \log(D(u-u')) 
\rho(u')  \notag \\
&\qquad \qquad +\int \diff u  
\left( \im\, \lambda(u)\cdot \rho(u)  + \#\, \rho(u) \left[\log(Q(u))+ 
\log(V_{(r,s)}(u))\right]
\right)
\bigg] \,.
\end{align}
The sum over $l$ can be evaluated
\begin{align}
 \sum_{l=0}^{\infty}
 \frac{1}{l!}q_\phi^l
 \int  \left( \prod_{p=1}^l \frac{\diff u_p}{2\pi \im} \right) \cdot 
 \left( \# \right)^l
 \prod_{p=1}^l
 e^{-\im\,  (\#)^{-1}  \lambda(u_p)} 
 &=
 \sum_{l=0}^{\infty}
 \frac{1}{l!} \left( 
 q_\phi \, \# \,
 \int \frac{\diff u}{2\pi\, \im }
 e^{-\im\,  (\#)^{-1}   \lambda(u)}
 \right)^l \notag \\
&= \exp \left[ 
q_\phi \, \# \,
 \int \frac{\diff u}{2\pi\,\im  }
 e^{-\im\,  (\#)^{-1}   \lambda(u)}
\right]
\end{align}
such that 
\begin{align}
Z_{\str}^{(r,s)\Deff} &= 
 \int \mathcal{D}\rho(u) 
 \int \mathcal{D}\lambda(u)
 \, \exp\bigg[\int \diff u \, \diff u' (\#)^{2} \rho(u) \log(D(u-u')) 
\rho(u')   \\
&\qquad +\int \diff u 
\left( \im\, \lambda(u)\cdot \rho(u)  + \#\, \rho(u) \left[\log(Q(u))+ 
\log(V_{(r,s)}(u))\right]
+ \frac{\# \, q_\phi}{2\pi\, \im }  e^{-\im\,  (\#)^{-1}   \lambda(u)}
\right)
\bigg] \,. \notag
\end{align}
Analogous to \cite{Ferrari:2012dy}, one may employ a shift in the auxiliary 
variable\footnote{The shift here differs from the 4d case in 
\cite{Ferrari:2012dy} by a minus sign in front of $q_\phi$. The alteration 
seems necessary as the quantised 6d SW-curve derived in this 
way passes nontrivial consistency checks, see Section 
\ref{sec:comparison}} 
\begin{align}
 \lambda(u) &=  \lambda'(u) - \im\, \, \#\, \log(-q_\phi) 
 \qquad 
 \text{such that}
\qquad 
e^{-\im\,  (\#)^{-1}   \lambda(u)} 
=- \frac{1}{q_\phi}e^{-\im\,  (\#)^{-1}  \lambda'(u)}
\end{align}
which yields
\begin{align}
 Z_{\str}^{(r,s)\Deff} &= 
 \int \mathcal{D}\rho(u) 
 \int \mathcal{D}\lambda'(u)
 \, \exp\bigg[\int \diff u \, \diff u' (\#)^{2} \rho(u) \log(D(u-u')) 
\rho(u') 
\label{eq:path_integral} \\
&\quad +\int \diff u 
\left( \im\, \lambda'(u)\cdot \rho(u)   + \#\, \left[ \rho(u) \log\left(-q_\phi 
Q(u) V_{(r,s)}(u) \right) - \frac{1}{2\pi\, i }  e^{-\im\,  (\#)^{-1}
\lambda'(u)}\right]
\right)
\bigg] \,.
 \notag
\end{align}
This represents a path integral representation of the elliptic genera for the 
theory with codimension 2 defect. For the the theory without defect, one simply puts 
$(r,s)=(0,0)$ because $V_{(0,0)}(u)=1$.
\subsubsection{Leading and next-to-leading order}
Following \cite{Ferrari:2012dy}, consider the behaviour as $\epsilon_2\to 0$. 
One computes the following expansions:
\begin{subequations}
\begin{align}
 \# = \frac{1}{\epsilon_2} 
 + L(\epsilon_1)
+ \mathcal{O}(\epsilon_2)  
\,,
\end{align}
where the abbreviation $L(\cdot)$ is defined in \eqref{eq:def_L_and_K}.
For the $D(u-u^\prime)$-terms one considers 
\begin{align}
 &\int \diff u \, \diff u' (\#)^{2} \rho(u) \log(D(u-u')) 
\rho(u') \notag \\
&= (\#)^{2} \int \diff u \, \int_{-\infty}^u \diff u' \rho(u) 
\left[ \log(D(u-u')) + \log(D(u'-u))  \right]
\rho(u')
\,, \notag
\end{align}
such that the $\epsilon_2$-expansion leads to
\begin{align}
&\log(D(u-u')) + \log(D(u'-u)) = 
G_1(u-u')\cdot \epsilon_2 + G_2(u-u')\cdot \epsilon_2^2  +\mathcal{O}(\epsilon_2^3)\,, \\
G_1(u-u') &=
L(u-u'+\epsilon_1) - L(u-u'-\epsilon_1)
\,,
\notag \\
G_2(u-u')&= 
\bigg[
L(u-u')^2 
- K(u-u') 
+ \frac{1}{2}K(u-u'+\epsilon_1) 
-\frac{1}{2} L(u-u'+\epsilon_1)^2
\notag\\
&\qquad \qquad \qquad \qquad \qquad \quad \,
+
\frac{1}{2}
K(u-u'-\epsilon_1)
-
\frac{1}{2}
L(u-u'-\epsilon_1)^2
\bigg] \,,
\notag
\end{align}
using the $L(\cdot)$, $K(\cdot)$ notation \eqref{eq:def_L_and_K}.
Similarly, for the $Q(u)$-terms the $\epsilon_2$ expansion yields
\begin{align}
\log Q(u) &= \mathcal{Q}_0 + \mathcal{Q}_1 \cdot\epsilon_2 
+\mathcal{O}(\epsilon_2^2) \,,\\
\mathcal{Q}_0 &= \log Q(u)|_{\epsilon_2=0} \notag \,, \\
\mathcal{Q}_1 &=  - 
\sum_{i=1}^k\left[
L\left(\tfrac{\epsilon_1}{2}-(u-a_i) \right)
+
L\left(\tfrac{\epsilon_1}{2}+(u-a_i) \right)
\right]
\,,
\notag  
\end{align}
and similarly for the $(0,s)$ defect terms $V_{(0,s)}(u)$ one finds
\begin{align}
\log V_{(0,s)}(u) &= \mathcal{V}_1^{(0,s)} \cdot \epsilon_2 
+\mathcal{O}(\epsilon_2^2)
\label{eq:def_defect_terms} \,,\\
\mathcal{V}_1^{(0,s)} 
&= 
s\cdot 
L(u-x)
\equiv s \cdot \mathcal{V}_1^{(0,1)} \,.
\notag
\end{align}
\end{subequations}
The $\epsilon_2$ expansion of the path integral for $\epsilon_2\ll 1$ becomes
\begin{align}
Z_{\str}^{(0,s)\Deff}
&= 
 \int \mathcal{D}\rho(u) 
 \int \mathcal{D}\lambda'(u)
 \, \exp\Bigg[
 \frac{1}{\epsilon_2}
\int \diff u \, \diff u'
 \frac{1}{2} 
 \rho(u) G_1(u-u')  \rho(u') 
\label{eq:expansion_path_integral}\\
&\qquad +\frac{1}{\epsilon_2} \int \diff u
 \left( \rho(u) 
 \left(
\log(-q_\phi) +\mathcal{Q}_0 
\right)
-
\frac{1}{2\pi\, i  }  e^{-i  \epsilon_2   \lambda'(u)}
\right)
\notag \\
&\qquad +
\int \diff u \, \diff u'
 \rho(u)  \left(\frac{1}{2} G_2(u-u') 
 + 
 L(\epsilon_1)
 G_1(u-u')
 \right)
\rho(u')
\notag \\
&\qquad + \int \diff u 
\left(
\rho(u)  \left(  \im\, \lambda'(u) + 
\mathcal{Q}_1+ \mathcal{V}_1^{(0,s)}
+
L(\epsilon_1)
\left(\log(-q_\phi) +\mathcal{Q}_0 \right)
\right)
-
L(\epsilon_1)
\frac{1}{2\pi\, \im }  e^{-i  \epsilon_2   \lambda'(u)}
\right)
\notag \\
&\qquad +\mathcal{O}(\epsilon_2)
\Bigg] 
\notag 
\end{align}
and the expression for $Z_\str\equiv 
Z_\str^{(0,0)\Deff}$ is obtained by setting all the codimension 2 defect 
contributions $\mathcal{V}_n$ to zero, i.e.\ $r=s=0$.
\subsubsection{Saddle point analysis}
Considering \eqref{eq:expansion_path_integral}, the saddle point contribution 
comes from the leading order term
\begin{align}
 \frac{\delta}{\delta \rho(u)} Z_\str^{(0,s)\Deff} \sim 
 Z_\str^{(0,s)\Deff} 
 \cdot \frac{1}{\epsilon_2}
 \left( \int \diff u' G_1(u-u') \rho(u') + \log(-q_\phi) +\mathcal{Q}_0(u) 
\right)
\end{align}
such that the saddle point equation is 
\begin{align}
  \int \diff u' G_1(u-u') \rho(u') + \log(-q_\phi) +\mathcal{Q}_0(u) =0 \,,
  \label{eq:saddle_pt_initial}
\end{align}
which then defines a critical density $\rho_\ast$.
Inspired by \cite{Fucito:2011pn}, define the following objects:
\begin{align}
 \mathcal{Y}(u) &\coloneqq 
\exp\left[ - \int \diff u' \,  
\rho(u')\frac{\diff}{\diff u} \log( \vartheta_1(u-u'))   \right] \,,
\label{eq:def_Upsilon}
\\
\omega(u) &\coloneqq \frac{\mathcal{Y}(u-\epsilon_1)}{\mathcal{Y}(u)P_0(u)}
\,,
\label{eq:def_omega}
\end{align}
and observe that 
\begin{subequations}
\begin{align}
 \int \diff u' \rho(u') \mathcal{G}_1(u-u')
 &= \log \frac{\mathcal{Y}(u-\epsilon_1)}{\mathcal{Y}(u+\epsilon_1)} \,,\\
 \mathcal{Q}_0(u)  &=  \log(M(u))- \log(P_0(u)) -\log( P_0(u+\epsilon_1)) 
\notag \\
 &= \log \frac{\mathcal{Y}(u+\epsilon_1)}{\mathcal{Y}(u-\epsilon_1)} +  
\log\left(M(u) \omega(u) \omega(u+\epsilon_1) \right)
\,.
\end{align}
\end{subequations}
The saddle point equation \eqref{eq:saddle_pt_initial} becomes
\begin{align}
 \log\left(-q_\phi M(u_\ast) \omega(u_\ast) \omega(u_\ast+\epsilon_1) \right) 
=0 
 \qquad \Leftrightarrow \qquad 
 1+ q_\phi M(u_\ast) \omega(u_\ast) \omega(u_\ast+\epsilon_1) =0 \,,
 \label{eq:saddle_pt}
\end{align}
for some points $u_\ast$.
Next, define the following function:
\begin{align}
 f(u)\coloneqq 
 \frac{1+ q_\phi M(u-\epsilon_1) \omega(u) \omega(u-\epsilon_1)}{\omega(u) }
\end{align}
the properties of $f$ indicate that it can be written as a product of $k$ Theta 
functions
\begin{align}
 f(u) \equiv  P(u)
 =  \prod_{l=1}^k \vartheta_1(u-e_l)
 \label{eq:P_deg_N_modular}
\end{align}
with roots $e_l$ to be determined. The saddle point equation 
\eqref{eq:saddle_pt} becomes equivalent to
\begin{align}
 -q_\phi  M(u-\epsilon_1) \omega(u) \omega(u-\epsilon_1) + \omega(u) 
P(u) - 1 =0 \,.
\label{eq:saddle_pt_mod}
\end{align}
From \eqref{eq:saddle_pt_mod} one can now derive a difference equation for the defect partition function.
%
%
\subsection{Shift operator}
Having derived a path integral expression \eqref{eq:path_integral}, which is 
dominated by the contribution of the saddle point \eqref{eq:saddle_pt_mod}, the 
next step is to define a shift operator. For this, the 
(exponentiated) defect 
fugacity $X$ is promoted to a non-commutative parameter together with conjugate 
coordinate $Y$ such that 
\begin{align}
 Y X = \frac{1}{p} XY 
 \label{eq:def_shift_op}
\end{align}
i.e.\ $Y f(x) = f(x-\epsilon_1)$.
Now, one can act with the shift operator $Y$ on the two parts of the partition 
function. For the perturbative part, one proceeds with the natural expressions; 
while the $Y$-action on the non-perturbative part is greatly simplified by the 
path integral representation.
\paragraph{Perturbative contribution.}
The normalised perturbative part \eqref{eq:pert_part_NS_limit} for an 
$(0,s)$ defect can be written as  
\begin{align}
\widetilde{Z}_{\mathrm{pert}}^{(0,s)\Deff} &= 
    \PE \bigg[ 
\frac{s}{2(1-p)} \left( \frac{1+Q}{1-Q} \right)
   \left\{ (1-p)
   +\sqrt{p} \sum_{i=1}^{k}  
   \left( 
   \frac{X}{A_i }  - \frac{A_i}{X } 
\right) 
   \right\} \bigg] \,,
\end{align}
for $A_i = e^{a_i}$. 
A direction computation, see Appendix \ref{app:shift_perturbative}, shows that 
the action of $Y$ is given by
\begin{align}
\label{eq:shift_op_pert_part}
Y \widetilde{Z}_{\mathrm{pert}}^{(0,s)\Deff} &=
\left[ 
\sqrt{ 
\left(
\prod_{i=1}^k 
\frac{1 
}{\vartheta_1(a_i-x+\tfrac{1}{2}\epsilon_1,\tau)}
\right)^2
}
\right]^s
\widetilde{Z}_{\mathrm{pert}}^{(0,s)\Deff}
=
\left[
\frac{1}{P_0(x)}
\right]^s
\widetilde{Z}_{\mathrm{pert}}^{(0,s)\Deff}
\,.
\end{align}
Note that \emph{the sign of the argument of the theta function can be flipped 
without any consequence}.

\paragraph{Elliptic genus.}
Consider the defect contribution \eqref{eq:defect_term}, \eqref{eq:def_defect_terms}, which one may write as
\begin{align}
 \widetilde{Z}_{\str}^{(0,s)\Deff} \supset \exp\left[ \int \diff 
u 
\rho_{\ast} 
(u) \mathcal{V}_1^{(0,s)}  \right]
  &=
  \exp\left[s\cdot \int \diff u \rho_{\ast} (u)\partial_u \log \theta_1(u-x)  
\right] \notag \\
&=     
\exp\left[- s \cdot \int \diff u \rho_{\ast} (u)\partial_x \log \theta_1(u-x)
\right]
= \left(\mathcal{Y}(x) \right)^s \,.
\end{align}
The shift operator 
acting on the \emph{normalised} instanton-strings partition function yields
\begin{align}
 Y \widetilde{Z}_{\str}^{(0,s)\Deff} 
 &\sim Y \int \mathcal{D}\rho 
 \exp\left[ \int \diff u \rho(u) \mathcal{V}_1^{(0,s)}(u,x)  \right] \notag \\
 &\sim Y 
 \exp\left[ \int \diff u \rho_{\ast} (u) \mathcal{V}_1^{(0,s)}(u,x)  \right]
\notag \\
 &\sim  
 \exp\left[ \int \diff u \rho_{\ast} (u) \mathcal{V}_1^{(0,s)}(u,x-\epsilon_1)  
\right] \,.
\end{align}
Consequently, one arrives at
\begin{align}
 Y Z_{\str}^{(0,s)\Deff} 
 &=  
 \left(
\frac{\mathcal{Y}(x-\epsilon_1)}{\mathcal{Y}(x)} 
\right)^s
\cdot \widetilde{Z}_{\str}^{(0,s)\Deff} 
\qquad 
\text{in leading order} 
\notag\\ 
&= 
\left(
\omega(x) P_0(x)
\right)^s
\cdot \widetilde{Z}_{\str}^{(0,s)\Deff} 
\qquad 
\text{using \eqref{eq:def_omega}} \,.
\label{eq:shift_op_string_part}
\end{align}
Alternatively, a direct computation on the defect contribution 
\eqref{eq:defect_term} leads to the same conclusion as in 
\eqref{eq:shift_op_string_part}, as detailed in Appendix 
\ref{app:shift_elliptic_genus}.
\paragraph{Full partition function.}
Combining \eqref{eq:shift_op_string_part} and \eqref{eq:shift_op_pert_part} 
implies that  
\begin{align}
 Y \left( \widetilde{Z}_{\mathrm{pert}}^{(0,s)\Deff} \cdot 
\widetilde{Z}_{\str}^{\Deff} \right)
&=   Y \widetilde{Z}_{\mathrm{pert}}^{(0,s)\Deff} \cdot 
Y \widetilde{Z}_{\str}^{(0,s)\Deff} 
=
\left[
\frac{1 }{P_0(x) }
\right]^s
\widetilde{Z}_{\mathrm{pert}}^{(0,s)\Deff}
\cdot
\left[
\omega(x) P_0(x)
\right]^s
\cdot \widetilde{Z}_{\str}^{(0,s)\Deff}
\notag\\
&=
\left[
\omega(x)
\right]^s
\cdot 
\left(
\widetilde{Z}_{\mathrm{pert}}^{(0,s)\Deff} 
\cdot \widetilde{Z}_{\str}^{(0,s)\Deff} \right) 
\,,
\end{align}
where one notes the cancellation of the contribution from the perturbative part.
In particular, notice the ratio
\begin{align}
  \frac{\widetilde{Z}_\mathrm{tot}^\mathrm{(0,1)def} 
(x-\epsilon_1)}{\widetilde{Z}_\mathrm{tot}^\Deff 
(x)}
&\equiv 
\frac{Y \left( \widetilde{Z}_{\mathrm{pert}}^{(0,1)\Deff} \cdot 
\widetilde{Z}_{\str}^{(0,1)\Deff} \right) }{
\widetilde{Z}_{\mathrm{pert}}^{(0,1)\Deff} 
\cdot \widetilde{Z}_{\str}^{(0,1)\Deff}
}
=
\omega(x)
\,, 
\label{eq:ratio_omega}
\end{align}
which is reminiscent of \cite[Eq.\ (55)]{Fucito:2011pn}. 
\subsection{Difference equation}
Finally, following the logic of \cite{Poghossian:2010pn,Fucito:2011pn}, the 
saddle point equation can be used to derive a difference equation on the level 
of the normalised codimension 2 partition function.
For this, one starts from the saddle point equation 
\eqref{eq:saddle_pt_mod} and performs the following manipulations:
\begin{align}
 0&=-q_\phi M(x-\epsilon_1) \omega (x) \omega (x-\epsilon_1)
 +\omega(x) P(x) -1 
 \notag \\[5pt] 
 &=-q_\phi M(x) \omega (x+\epsilon_1) \omega (x)
 +\omega(x+\epsilon_1) P(x+\epsilon_1) -1 
 \qquad \text{by shifting $x \to x+\epsilon_1$}
 \notag \\[5pt] 
&=-q_\phi M(x)  
\frac{\widetilde{Z}^{(0,1)\Deff}(x)}{\widetilde{Z}^{(0,1)\Deff}
(x+\epsilon_1)} 
\frac{\widetilde{Z}^{(0,1)\Deff}(x-\epsilon_1)}{\widetilde{Z}^{(0,
1)\Deff}(x)}
 + P(x+\epsilon_1)  
\frac{\widetilde{Z}^{(0,1)\Deff}(x)}{\widetilde{Z}^{(0,1)\Deff}
(x+\epsilon_1)}
 -1
 \qquad \text{using \eqref{eq:ratio_omega}}
 \notag \\[5pt] 
&=-q_\phi M(x)
  \cdot 
\frac{\widetilde{Z}^{(0,1)\Deff}(x-\epsilon_1)}{\widetilde{Z}^{(0,
1)\Deff} (x+\epsilon_1)} 
 + P(x+\epsilon_1)
\frac{\widetilde{Z}^{(0,1)\Deff}(x)}{\widetilde{Z}^{(0,1)\Deff}
(x+\epsilon_1)}
 -1
 \notag \\[5pt] 
&=-q_\phi M(x)
  \cdot 
\widetilde{Z}^{(0,1)\Deff}(x-\epsilon_1)
 + P(x+\epsilon_1)
  \widetilde{Z}^{(0,1)\Deff}(x) 
- \widetilde{Z}^{(0,1)\Deff}(x+\epsilon_1) 
\notag \\[5pt] 
&=
\left[ -q_\phi M(x)   \cdot 
Y    
 + P(x+\epsilon_1)   
-Y^{-1} \right] \widetilde{Z}^{(0,1)\Deff}(x)
\,.
\label{eq:diff_equation}
\end{align}
Hence, \eqref{eq:diff_equation} shows the existence of an operator that 
annihilates the codimension 2 defect partition function. Nevertheless, the 
expression needs to be considered with care. Comparing to the results of 
\cite{Haghighat:2013tka}, the form is already suggestive of the Seiberg-Witten 
curve. In order to consolidate this further, one can equivalently rewrite 
\eqref{eq:diff_equation} as
\begin{align}
\left[ q_\phi M(x)   \cdot Y   
+ Y^{-1} \right] \widetilde{Z}^{(0,1)\Deff}(x)
=
  P(x+\epsilon_1)   \cdot \widetilde{Z}^{(0,1)\Deff}(x)
  \label{eq:SW-curve}
\end{align}
where the left-hand-side contains expressions that are fully known, while the 
right-hand-side contains the degree $k$ modular form $P(u)$ of 
\eqref{eq:P_deg_N_modular}, whose existence follows from the saddle point 
analysis. Therefore, the purpose of the remainder of this section is to 
establish a physical interpretation of $P(x+\epsilon_1)$. As it turns out, the 
codimension 4 defect in form of the VEV of a Wilson surface is a suitable 
object to consider.
%
%
\subsection{Comparison to Wilson surface}
\label{sec:comparison}
The strategy for determining the physical meaning of $P(x+\epsilon_1) $ has two 
steps:
\begin{compactenum}[(i)]
    \item Starting from \eqref{eq:SW-curve}, together with the \emph{known} 
normalised codimension 2 defect partition function 
$\widetilde{Z}^{(0,1)\Deff}(x)$, one can compute $ P(x+\epsilon_1) $ 
order by order in $q_\phi$.
    \item The \emph{predictions} for $ P(x+\epsilon_1) $ are compared to the 
normalised codimension 4 defect partition function 
$\widetilde{Z}_2^{\mathrm{Wilson}}(z) $, i.e. the Wilson surface VEV. This 
determines $z$ as a function of $x$.
\end{compactenum}
To begin with, consider the difference equation \eqref{eq:diff_equation} or \eqref{eq:SW-curve} 
together with the $q_\phi$-expansions  
\begin{align} 
\widetilde{Z}^{(0,1)\Deff}(x)=\widetilde{Z}^{(0,1)\Deff}
_0(x)\left(1 + \sum_{l=1}^\infty  q_\phi^l\, 
\widetilde{Z}^{(0,1)\Deff}_l(x) \right)
 \,,\qquad 
 P(x)= P_0(x) \left(1 + \sum_{l=1}^\infty  q_\phi^l\, P_l(x) \right)
 \,,
\label{defZP}
\end{align}
such that \eqref{eq:diff_equation} becomes
\begin{align}
 0=\bigg[ 
 P_0(x+\epsilon_1) -Y^{-1}
 &+ q_\phi \left( 
 P_0(x+\epsilon_1) P_1(x+\epsilon_1) - M(x)Y
 \right) \notag \\
 &+\sum_{l=2}^\infty q_\phi^l  P_0(x+\epsilon_1) P_l(x+\epsilon_1)
 \bigg] \widetilde{Z}^{(0,1)\Deff}_0(x) \left( 1+\sum_{j=1}^{\infty} 
q_\phi^j \widetilde{Z}^{(0,1)\Deff}_j (x) \right) \,.
\end{align}
Next, one can try to match the predictions for $P_l(x+\epsilon_1)$ with the 
results from the Wilson surface. Based the explicit computations detailed 
below, the \emph{claim} is that
\begin{align}
 P_l(x+\epsilon_1) = \widetilde{Z}_l^{\mathrm{Wilson}}(z) \qquad \forall l
 \qquad \Leftrightarrow \qquad 
 z=x+\frac{1}{2}\epsilon_1 \,,
 \label{eq:identification_z_x}
\end{align}
i.e.\ the fugacities $x$ and $z$ are suitably identified.
\subsubsection{Perturbative level}
\label{sec:compare_Wilson_perturbative}
The lowest order in the $q_\phi$ expansion reads
 \begin{align}
  0&= \left[P_0(x+\epsilon_1) - Y^{-1}\right] 
\widetilde{Z}^{(0,1)\Deff}_0(x) 
  \label{eq:new_SW_pert_level}
 \end{align}
 and one finds 
 \begin{align}
 P_0(x+\epsilon_1) =\frac{ Y^{-1} 
\widetilde{Z}^{(0,1)\Deff}_0(x)}{\widetilde{Z}^{(0,1)\Deff}_0(x)} 
\,.
 \end{align}
Comparing to $W_{\text{part}}$ in the NS-limit yields
\begin{align}
 P_0(x+\epsilon_1) = \frac{W_{\text{part}}(z)}{(-\im)^N Q^{\frac{k}{12}}} 
 \quad \Leftrightarrow \quad 
 z= x+\frac{\epsilon_1}{2} \,.
\end{align}
 \subsubsection{1-string level}
 \label{sec:compare_Wilson_1-string}
Next, the linear $q_\phi$ order reads
 \begin{align}
 \label{eq:new_SW_1_level}
 \begin{aligned}
 0= \left[P_0(x+\epsilon_1) - Y^{-1}\right] 
&\widetilde{Z}^{(0,1)\Deff}_0(x) \widetilde{Z}^{(0,1)\Deff}_1(x)
 \\
 & + 
\left[ P_0(x+\epsilon_1) P_1(x+\epsilon_1) - M(x) Y    \right] 
\widetilde{Z}^{(0,1)\Deff}_0(x)
\,,
  \end{aligned}
  \end{align}
  and, using \eqref{eq:new_SW_pert_level}, one finds
\begin{align}
 P_1(x+\epsilon_1) = 
 Q_{(0)}(x)
  + Y^{-1} \widetilde{Z}^{(0,1)\Deff}_1(x) 
-\widetilde{Z}^{(0,1)\Deff}_1(x) \,.
  \label{eq:prediction_P1}
 \end{align}
Using the results from above, one computes the prediction 
\eqref{eq:prediction_P1} to be
\begin{align}
 P_1(x+\epsilon_1) &= 
 Q_{(0)}(x+\epsilon_1)
+ \frac{1}{ 
 \vartheta_1^\prime(0)
 }
\sum_{i=1}^k
Q^\vee_{(0)}(a_i-\tfrac{1}{2}\epsilon_1)
\cdot
\left[
L(a_i-x-\tfrac{3}{2}\epsilon_1)
-
L(a_i-x-\tfrac{1}{2}\epsilon_1)
\right] \,,
  \label{eq:prediction_P1_calc}
\end{align}
see Appendix \ref{app:details_P1} for details.
Comparing to the Wilson surface result \eqref{eq:Wilson_1-string}, one finds
\begin{align}
  P_1(x+\epsilon_1) = \widetilde{Z}_1^{\mathrm{Wilson}}(z) 
  \quad \Leftrightarrow \quad
  z= x+\frac{\epsilon_1 }{2}\,.
\end{align}
\subsubsection{2-string level}
\label{sec:compare_Wilson_2-string}
Lastly, the quadratic $q_\phi$ order in the expansion reads
\begin{align}
  \label{eq:new_SW_2_level}
\begin{aligned}
 0&= \left[P_0(x+\epsilon_1) - Y^{-1}\right] 
\widetilde{Z}^{(0,1)\Deff}_0(x) \widetilde{Z}^{(0,1)\Deff}_2(x)
 \\
 &\qquad \qquad \qquad + 
\left[ P_0(x+\epsilon_1) P_1(x+\epsilon_1) - M(x) Y    \right] 
\widetilde{Z}^{(0,1)\Deff}_0(x) \widetilde{Z}^{(0,1)\Deff}_1(x) 
 \\
&\qquad \qquad \qquad + P_0(x+\epsilon_1)P_2(x+\epsilon_1) 
\widetilde{Z}^{(0,1)\Deff}_0(x)
\,,
\end{aligned}
  \end{align}
  and using \eqref{eq:new_SW_pert_level} and \eqref{eq:new_SW_1_level} one finds
  \begin{align}
  \label{eq:prediction_P2}
  \begin{aligned}
P_2(x+\epsilon) = 
Q_{(0)}(x)
\left[Y-1\right] \widetilde{Z}^{(0,1)\Deff}_1(x)
&+\left[Y^{-1}-1\right] \widetilde{Z}^{(0,1)\Deff}_2(x)  \\
&-\widetilde{Z}^{(0,1)\Deff}_1(x) \left[Y^{-1}-1\right] 
\widetilde{Z}^{(0,1)\Deff}_1(x) \,.
\end{aligned}
  \end{align}
Using the results from above, one compute the prediction 
\eqref{eq:prediction_P2} to be
\begin{align}
 P_2(x+\epsilon_1)
 &=
 -\frac{1}{2} \sum_{j=1}^k 
 \left( 
 \frac{Q_{(0)}^\vee(a_j-\tfrac{1}{2}\epsilon_1) }{\vartheta_1^\prime(0)}
 \right)^2
 \big[ 
 K(a_j -x -\tfrac{3}{2}\epsilon_1)
 -K(a_j -x -\tfrac{1}{2}\epsilon_1)
 \big] 
 \notag \\
&+ 
\sum_{j=1}^k
\left( 
 \frac{Q_{(0)}^\vee(a_j-\tfrac{1}{2}\epsilon_1) }{\vartheta_1^\prime(0)}
 \right)^2
 \bigg\{
 L(a_j-x-\tfrac{3}{2}\epsilon_1)
 \left[ 
 L(a_j-x-\tfrac{3}{2}\epsilon_1)
 - L(a_j-x-\tfrac{1}{2}\epsilon_1)
 \right]
 \notag \\
 &\qquad \qquad \qquad \qquad \qquad \qquad
 +2 L(\epsilon_1)
 \left[ 
 L(a_j-x-\tfrac{3}{2}\epsilon_1)
 - L(a_j-x-\tfrac{1}{2}\epsilon_1)
 \right]
 \bigg\} 
 \notag \\
&+\sum_{\substack{i,j=1 \\ i\neq j}}^k
\frac{Q_{(0)}^\vee(a_i-\tfrac{1}{2}\epsilon_1) }{\vartheta_1^\prime(0)}
\frac{Q_{(0)}^\vee(a_j-\tfrac{1}{2}\epsilon_1) }{\vartheta_1^\prime(0)}
 \left[ 
 L(a_i-x-\tfrac{3}{2}\epsilon_1)
 - L(a_i-x-\tfrac{1}{2}\epsilon_1)
 \right] \notag \\
 &\qquad \qquad \cdot \bigg\{
 L(a_i-a_j-\epsilon_1)
 +L(a_j-a_i-\epsilon_1)
 -L(a_i-a_j)
 -L(a_j-a_i)
 \bigg\} 
 \notag  \\
&+\sum_{\substack{i,j=1 \\ i\neq j}}^k
\frac{Q_{(0)}^\vee(a_i-\tfrac{1}{2}\epsilon_1) }{\vartheta_1^\prime(0)}
\frac{Q_{(0)}^\vee(a_j-\tfrac{1}{2}\epsilon_1) }{\vartheta_1^\prime(0)}
\bigg[
\frac{1}{2} L(a_i-x-\tfrac{3}{2}\epsilon_1) L(a_j-x-\tfrac{3}{2}\epsilon_1)
\notag \\
&\qquad \qquad \qquad 
- L(a_i-x-\tfrac{1}{2}\epsilon_1) L(a_j-x-\tfrac{3}{2}\epsilon_1)
+\frac{1}{2} L(a_i-x-\tfrac{1}{2}\epsilon_1) L(a_j-x-\tfrac{1}{2}\epsilon_1)
\bigg]
\notag \\
&+ \sum_{j=1}^k 
\frac{Q_{(0)}^\vee(a_j-\tfrac{1}{2}\epsilon_1) }{\vartheta_1^\prime(0)}
Q_{(0)}(a_j -\tfrac{3}{2}\epsilon_1)
\left[ 
 L(a_j-x-\tfrac{5}{2}\epsilon_1)
 - L(a_j-x-\tfrac{1}{2}\epsilon_1)
 \right]
\notag \\
&+ Q_{(0)}(x+\epsilon_1)
\sum_{j=1}^k 
\frac{Q_{(0)}^\vee(a_j-\tfrac{1}{2}\epsilon_1) }{\vartheta_1^\prime(0)}
\left[ 
 L(x+\tfrac{5}{2}\epsilon_1 - a_j)
 - L(x+\tfrac{3}{2}\epsilon_1 - a_j)
 \right] \,,
 \label{eq:prediction_P2_calc}
\end{align}
see Appendix \ref{app:details_P2} for details. Comparing to the Wilson surface 
result \eqref{eq:Wilson_2-string}, one finds
\begin{align}
  P_2(x+\epsilon_1) = \widetilde{Z}_2^{\mathrm{Wilson}}(z) 
  \quad \Leftrightarrow \quad
  z= x+\frac{\epsilon_1 }{2}\,.
\end{align}
\subsubsection{Implications}
The results of Sections 
\ref{sec:compare_Wilson_perturbative}--\ref{sec:compare_Wilson_2-string} 
provide evidence that the claim \eqref{eq:identification_z_x} is correct. Thus, 
the difference equation \eqref{eq:SW-curve} can be re-written 
\begin{align}
 \mathcal{D}_{\mathrm{NS}} \widetilde{Z}^{(0,1)\Deff}(x) &\equiv
 \left[ q_\phi M(x)   \cdot Y   
+ Y^{-1} \right] \widetilde{Z}^{(0,1)\Deff}(x)
\notag \\
&=
  P(x+\epsilon_1)   \cdot \widetilde{Z}^{(0,1)\Deff}(x)
\equiv \langle \mathcal{W}\rangle \left( z=x+\tfrac{\epsilon_1}{2} \right)
 \cdot \widetilde{Z}^{(0,1)\Deff}(x) \,,
\end{align}
which identifies the operator $\mathcal{D}$ of \eqref{eq:def_diff_op} in the NS 
limit. In addition, the degree $k$ modular form $P$ of 
\eqref{eq:P_deg_N_modular} has been identified with the expectation value of 
the Wilson surface defect.

As a comment, the found identification \eqref{eq:identification_z_x} is a qualitatively new feature of the $\Ncal=(1,0)$ theories in contrast to the $\Ncal=(2,0)$ case discussed in the next section. As shown in the 6d $\Ncal=(2,0)$ $A_1$ case \cite{Agarwal:2018tso}, the Wilson surface expectation value is independent of the defect fugacity $z$; similarly, the dual 5d picture has been considered in \cite{Bullimore:2014awa}, where the Wilson loop expectation values also has no dependence on the defect fugacity.
%
%
%
\section{2 M5 branes: Matching 6d and 5d with enhanced SUSY}
\label{sec:enhanced_susy}
In this section, the methods developed in the above sections are applied to 
the simplest $6$d $\Ncal=(1, 0)$ theory with $\surm(2)$ gauge group and $4$ 
flavours. 
The interest in this model comes because Higgsing the $\surm(2)$ gauge group 
as above leads to a theory with no gauge theory left. Put differently, in the 
Type IIA brane construction the Higgsing is realised by removing a D6 brane, see Figure \ref{fig:Higgsing_branes}. 
Starting from the 2 D6 branes for the $\surm(2)$ theory and removing one of 
them, leads to a single D6 which is dual to  $\C^2\slash \Z_1 \cong \C^2$, 
i.e.\ the $A_0$ singularity in the original M-theory setup. Thus, the Higgsing 
leads to a system of M5 branes which preserve $16$ supercharges instead of the 
$8$ supersymmetries of the generic case with an $A_k$ singularity.

Building on Section \ref{sec:Higgs_mech}, one can study the $\Ncal =(2,0)$ 
$A_1$ theory in the presence of a codimension 2 defect by Higgsing the $\Ncal 
=(1, 0)$ $\surm(2)$ theory with a position dependent VEV.
In addition, the path integral formalism developed in previous section allows one to derive the quantised Seiberg-Witten curve therein.
As a consistency check, it is verify in this section that the established SW-curve matches  
the result obtained from the 5d/3d perspective by compactifying the 6d $\Ncal=(2, 0)$ $A_1$ theory onto $S^1$ \cite{Bullimore:2014awa}.

\subsection{\texorpdfstring{ Defects in $6$d $\Ncal=(2, 0)$ $A_1$ theory}{Defects in 6d (2,0) 
A1 theory}}
To begin with, one computes the partition function for the 6d case. In order to find agreement with the 5d result of Section \ref{sec:5d_SYM}, the derivation is repeated in a slightly different manner compared to Sections \ref{sec:partition_fcts} and \ref{sec:difference_eq}. 
\subsubsection{Elliptic genus}
Firstly, the saddle point approach is used to derive the difference equation 
of the non-perturbative part of the partition function, analogously to Section 
\ref{sec:path_integral_rep}.
The $\surm(2)$ gauge and $\surm(4)\subset \sorm(8)$ flavour fugacities  
are labeled in terms of $\alpha$, $\mu_i$ and $t$ as
\begin{align}
\alpha=e^a\,,\qquad
e^{m_1}=t\mu_1\,,\qquad 
e^{m_2}=\frac{t}{\mu_1}\,,\qquad 
e^{m_3}=\frac{\mu_2}{t}\,,\qquad 
\text{and}\qquad 
e^{m_4}=\frac{1}{t\mu_2}\,.
\end{align}
The instanton partition function $Z_k$ for $\mathcal N=(1, 0)$ $\surm(2)$ with $4$ 
flavours is thus given by
the $l$-th elliptic genus, contributing to the non-perturbative partition 
function \eqref{eq:full_inst_partition}
\begin{align}
Z_l=\frac{1}{l!}
\oint\prod_{I=1}^{l}\frac{\diff\phi_I}{2\pi \im}
\prod_{I, J=1}^{l}
\frac{\vartheta^\vee_1(\phi_{IJ})\vartheta_1(\phi_{IJ}+2\epsilon_+)}{
\vartheta_1(\phi_{IJ}+\epsilon_{1,2})}
\cdot
\prod_{I=1}^{l}\frac{\prod_{i=1}
^4\vartheta_1(\phi_I-m_i)}{\vartheta_1(\phi_I\pm a\pm\epsilon_+)}\,,
\label{eq:A1-inst}
\end{align}
with $\phi_{IJ}\coloneqq \phi_I -\phi_J$. Here, $\vartheta^\vee_1(\phi_{IJ})$ means that those terms in 
$\vartheta_1(\phi_{IJ})$ with $\phi_I=\phi_J$ are replaced by 
$\vartheta^{\prime}_1(0)$.
Next, a $(0,1)$ codimension 2 defect is introduced via the Higgsing \eqref{eq:Higgs_with_defect}, 
which becomes
\begin{subequations}
\label{eq:codim2}
\begin{alignat}{3}
a&=m_3-\epsilon_+\,\qquad  &
&\text{and} \qquad &
m_4&=m_3-2\epsilon_+-\epsilon_2\equiv 
x-\epsilon_2\,,
\label{codim2a} 
\\
\text{or}\qquad 
\alpha&=\frac{\sqrt{q}}{t}\,\qquad &
&\text{and} \qquad  &
\mu_2&=\sqrt{p q^2}\,,
\label{codim2b}
\end{alignat}
\end{subequations}
such that the elliptic genus of the $\Ncal=(2,0)$ theory with defect is given 
by
\begin{align}
\begin{aligned}
Z^{\Deff}_l
=
\frac{1}{l!}\oint\prod_{I=1}^{l}\frac{\diff \phi_I}{2\pi \im}
&\prod_{I, J=1}^{l}
\frac{\vartheta^\vee_1(\phi_{IJ})\vartheta_1(\phi_{IJ}+2\epsilon_+)}{
\vartheta_1(\phi_{IJ}+\epsilon_{1,2})}
\\
&\cdot 
\prod_{I=1}^{l}
\frac{
\vartheta_1(\phi_I-m_1)\vartheta_1(\phi_I-m_2)\vartheta_1(\phi_I-m_3+2\epsilon_+
+\epsilon_2)}{
\vartheta_1(\phi_I+m_3)\vartheta_1(\phi_I+m_3-2\epsilon_+)\vartheta_1(\phi_I-m_3
+2\epsilon_+)}\,.
\end{aligned}
\end{align}
Further notice that
\begin{align}
e^{m_1+m_3}=\mu_1\mu_2\,\qquad 
\text{and} \qquad  
e^{m_2+m_3}=\mu_1^{-1}\mu_2\,.
\end{align}
For convenience, one defines $\mu_1\equiv e^{m-\epsilon_2/2}$, and, additionally, shifts the 2d gauge variables as
\begin{align}
\phi_I\mapsto \phi_I-m_3+\epsilon_+\,.
\end{align}
Finally, one ends up with the instanton partition function for the theory with a
codimension 2 defect of type $(0,1)$, which is given by
\begin{align}
\begin{aligned}
Z_l^{\Deff}
=
\frac{1}{l!}\oint\prod_{I=1}^{l}\frac{\diff \phi_I}{2\pi \im}
\prod_{I, J=1}^{l}
\frac{\vartheta_1(\phi_{IJ})\vartheta_1(\phi_{IJ}+2\epsilon_+)}{
\vartheta_1(\phi_{IJ}+\epsilon_{1,2})}
&\cdot
\prod_{I=1}^{l}
\frac{
\vartheta_1(\phi_I-m)\vartheta_1(\phi_I+m-\epsilon_2)}{
\vartheta_1(\phi_I\pm\epsilon_+)}
\\
&\cdot
\prod_{I=1}^{l}
\frac{\vartheta_1(\phi_I-2x-\epsilon_-)
} {
\vartheta_1(\phi_I-2x-\epsilon_+)}\,,
\end{aligned}
\end{align}
and, following the definitions \eqref{eq:def_path_int} of Section \ref{sec:partition_fcts}, one defines
\begin{subequations}
\begin{align}
D(u)&=
\frac{\vartheta_1(u)\vartheta_1(u+2\epsilon_+)
}{
\vartheta_1(u+\epsilon_{1}
)\vartheta_1(u+\epsilon_{2})}
\,,\\ 
Q(u)&=
\frac{\vartheta_1(u-m)\vartheta_1(u+m-\epsilon_2)
}{
\vartheta_1(u+\epsilon_+)
\vartheta_1(u-\epsilon_+)}
\,,\\
V(u)&=
\frac{\vartheta_1(u-2x-\epsilon_-)}{\vartheta_1(u-2x-\epsilon_+)}
\,.
\end{align}
\end{subequations}
Having set-up the notation, one recasts the instanton partition function in 
a path integral, analogous to Section \ref{sec:path_integral_rep}, as follows:
\begin{align}
Z^{\Deff}_{\str}
\sim\int 
\mathcal{D}\rho(u)\exp
\left[
\frac{1}{\epsilon_2}\int \diff u \diff u' 
\frac{1}{2}\rho(u) G_1(u-u')\rho(u')
+\frac{1}{\epsilon_2}\int \diff u
\log(-q_\phi Q_0(u))+\mathcal{O}(\epsilon_2^0)
\right]\,,
\end{align} 
with the expansion coefficients
\begin{align}
G_1(u-u')=
L(u-u'+\epsilon_1)
-
L(u-u'-\epsilon_1)
\,,\qquad 
\text{and}
\qquad 
Q_0(u)=Q(u)\vert_{\epsilon_2=0}=\frac{\vartheta_1(u\pm m)}{\vartheta_1(u\pm 
\frac{\epsilon_1}{2})}\,,
\end{align}
with $L(\cdot)$ introduced in \eqref{eq:def_L_and_K}.
As in \eqref{eq:def_Upsilon}, one may define
\begin{align}
\mathcal Y(u)=\exp\left[-\int \diff u' 
\rho(u')\frac{\vartheta_1^\prime(u-u')}{\vartheta_1(u-u')}\right]\,,
\label{Y} 
\end{align}
such that the saddle point equation can be written as 
\begin{align}
\log\left(-q_\phi\frac{\mathcal Y(u_\ast-\epsilon_1)}{\mathcal 
Y(u_\ast+\epsilon_1)}Q_0(u_\ast)\right)=0
\,,\qquad  \text{or}\qquad  
1+q_\phi\frac{\mathcal Y(u_\ast-\epsilon_1)}{\mathcal 
Y(u_\ast+\epsilon_1)}Q_0(u_\ast)=0
\label{SW}
\end{align}
for certain specified solutions $\rho_\ast$ and $u_\ast$.

On the other hand, one can apply the saddle point equation to the normalised 
$Z_{\str}^{\Deff}$ and take the NS-limit, $\epsilon_2\rightarrow 
0$ and $q\rightarrow 1$, 
\begin{align}
\widetilde Z_{\str}^{\Deff}(x)
&\equiv \lim_{\epsilon_2\rightarrow 
0}\frac{Z_{\str}^{\Deff}}{Z_{\str}}
=\exp
\left[\int \diff u \rho_\ast(u) \mathcal V_1(u)\right]\,, \\
\text{with } \qquad
\mathcal{V}_1(u)
&=\log V(u)\vert_{\mathcal{O}(\epsilon_2^1)}
=
L\left(u-2x-\tfrac{\epsilon_1}{2} \right)
=
-L\left(2x+\tfrac{\epsilon_1}{2}-u\right)
\,.
\end{align}
Therefore, by the virtue of \eqref{Y}, one finds
\begin{align}
\widetilde Z_{\str}^{\Deff}(x)
=\mathcal{Y}\left(2x+\tfrac{\epsilon_1}{2}\right)
\end{align}
For a shift operator $Y: x\mapsto x-\epsilon_1$, the action on the partition 
function is
\begin{align}
Y\cdot\widetilde Z_{\str}^{\Deff}(x)
=
\mathcal{Y}
\left(2x-\tfrac{3\epsilon_1}{2}\right)
=\frac{\mathcal 
Y\left(2x-\tfrac{3\epsilon_1}{2}\right)}{\mathcal 
Y\left(2x+\tfrac{\epsilon_1}{2}\right)}
\widetilde{Z}_{\str}^{\Deff}(x)\,.
\label{DE}
\end{align}
Next, consider the left-hand-side of \eqref{SW} for arbitrary values of $u$, 
i.e.\
\begin{align}
\mathcal L\equiv1+q_\phi\frac{\mathcal Y\left(u-{\epsilon_1}\right)}{\mathcal 
Y\left(u+{\epsilon_1}\right)}Q_0(u)\,.
\end{align}
whose purpose is clarified shortly. 
For $u=2x-\frac{\epsilon_1}{2}$, one has
\begin{align}
Q_0\left(2x-\tfrac{\epsilon_1}{2}\right)
=
\frac{\vartheta_1(2x\pm 
m-\epsilon_1/2)}{\vartheta_1(2x)\vartheta_1(2x-\epsilon_1)}
\equiv
\frac{\widetilde{\theta}_1(p^{-1}X\eta^{-1})\widetilde{\theta}_1(p^{-2}X\eta)}{
\widetilde{\theta}_1(p^{-1}X)\widetilde{\theta}_1(p^{-2}X)}\,,
\end{align}
due to \eqref{eq:theta_defs}. To compare with the results in \cite{Bullimore:2014awa}, one defines the following variables
\begin{align}
X\equiv e^{2x+\epsilon_1}=t^{-2}
\,\qquad \text{and} \qquad 
\eta\equiv  e^{m+\epsilon_1/2}=\sqrt{pq}\mu_1
\,.
\label{Xparameter}
\end{align}
Therefore, using \eqref{DE}, one finds
\begin{align}
\Lcal
&=
1+q_\phi 
\frac{\widetilde{\theta}_1(p^{-1}X\eta^{-1})\widetilde{\theta}_1(p^{-2}X\eta)}{
\widetilde{\theta}_1(p^{-1}X)\widetilde{\theta}_1(p^{-2}
X)} 
\cdot
\frac{\mathcal Y(2x-\tfrac{3\epsilon_1}{2})}{\mathcal 
Y(2x+\tfrac{\epsilon_1}{2})}
\notag \\
&=
1+q_\phi 
\frac{\widetilde{\theta}_1(p^{-1}X\eta^{-1})\widetilde{\theta}_1(p^{-2}X\eta)}{
\widetilde{\theta}_1(p^{-1}X)\widetilde{\theta}_1(p^{-2}
X)}\,\frac{Y\cdot\widetilde Z_{str}^{def}(x)}{\widetilde 
Z_{\str}^{def}(x)}\,.
\label{DEleft}
\end{align}
Notice that $YX=p^{-2} XY$, for convenience, one defines
\begin{subequations}
\begin{alignat}{3}
Y_X X &\coloneq p^{-1} XY_X &
\,\qquad  &\text{meaning} \qquad &
Y_X^2&=Y\,,\\
\widetilde Z(X) 
&\coloneq 
\widetilde{Z}_{\str}^{\Deff}(x) &
\,,\qquad  &\text{such that} \qquad  &
\widetilde{Z}(p^{-1}X)&= Y_X \widetilde{Z}_{\str}^{\Deff}(x)\,.
\end{alignat}
\end{subequations}
Now \eqref{DEleft} can be recast as 
\begin{align}
Y_X^{-1}\cdot\widetilde Z(p^{-1}X)+q_\phi 
\frac{\widetilde{\theta}_1(p^{-1}X\eta^{-1})\widetilde{\theta}_1(p^{-2}X\eta)}{
\widetilde{\theta}_1(p^{-1}X)\widetilde{\theta}_1(p^{-2}
X)}\,Y_X\cdot\widetilde Z(p^{-1}X)=\mathcal L\,\,Y_X^{-1}\cdot\widetilde 
Z(p^{-1}X)\,.
\label{DEleft2}
\end{align}
Lastly, one shifts $X\rightarrow p X$ and re-defines the right-hand-side of 
\eqref{DEleft2} to be
\begin{align}
Y_X^{-1}\cdot\widetilde Z(X)+q_\phi 
\frac{\widetilde{\theta}_1(X\eta^{-1})\widetilde{\theta}_1(p^{-1}X\eta)}{
\widetilde{\theta}_1(X)\widetilde{\theta}_1(p^{-1}X)}\,
Y_X\cdot\widetilde Z(X)
\eqqcolon \mathcal W(X)\cdot\widetilde Z(X)\,,
\label{instDE}
\end{align}
where $\mathcal{W}$ is identified with the 6d partition 
function of the codimension $4$ defect, i.e.\ the Wilson surface, in 
Section \ref{WS2}. 
Therefore, \eqref{instDE} is exactly the difference equation obtained 
from 5d/3d perspective in \cite{Bullimore:2014awa}.
\subsubsection{Perturbative part}
Next, the difference equation for the perturbative part of the 
$\mathcal N=(2, 0)$ $A_1$ theory is derived. As above, the starting 
point is the 6d $\Ncal=(1,0)$ $\surm(2)$ theory with 4 flavours, whose perturbative contributions to the 
partition function are given by
\begin{subequations}
\label{eq:A1-pert}
\begin{align}
Z_{\Ncal=(1, 0)\,\,A_1}^{\mathrm{pert}} &=  \PE[I_t+I_v+I_h]\,, \\
\text{with} \qquad 
I_t &=-\frac{p+q}{(1-p)(1-q)}\frac{Q}{1-Q} \,, \\
I_v &=-\frac{1+pq}{(1-p)(1-q)(1-Q)}\left(\alpha^2+\alpha^{-2}Q+Q\right)  \,,\\
I_h 
&=\frac{\sqrt{pq}}{(1-p)(1-q)(1-Q)}\left(\alpha+\alpha^{-1}Q\right)\left(t+t^
{-1}\right)\left(\mu_1+\mu_1^{-1}+\mu_2+\mu_2^{-1}\right)\,,
\end{align}
\end{subequations}
where the contributions of the tensor, vector, and hyper 
multiplets $I_t$, $I_v$ and $I_h$, respectively, have been \emph{flopped} 
compared to \eqref{eq:single_letter_6d}, for comparison with the $5$d result.

Before introducing the codimension 2 defect, one first computes the 
contribution of Goldstone bosons from the usual Higgsing procedure by assigning
\begin{align}
\alpha=t^{-1} 
\qquad  \text{and} \qquad  
\mu_2=\sqrt{pq} \,.
\end{align}
The Goldstone boson part is given by
\begin{align}
Z_{G}&=
\PE
\left[
\frac{\sqrt{pq}}{(1-p)(1-q)(1-Q)}
\left(\alpha+\alpha^{-1}Q\right)
\left(t+t^{-1}\right)
\left(\mu_1+\mu_1^{-1}\right)
\right]\Bigg\vert_{
\alpha=t^{-1}}
\notag\\
&=\PE
\left[
\frac{\sqrt{pq}}{(1-p)(1-q)(1-Q)}
\left(t^{-1}+tQ\right)
\left(t+t^{-1}\right)
\left(\mu_1+\mu_1^{-1}\right)
\right]\,.
\end{align}
With this preparation, one can introduce a $(0,1)$ codimension 2 defect as in 
\eqref{eq:codim2}. The partition function 
$Z_{\Ncal=(1,0)\,\,A_1}^{\mathrm{pert}}$ can be factorised as
\begin{align}
Z_{\Ncal =(1, 0)\,\,A_1}^{\mathrm{pert}}
=
Z_{\mathcal N=(2, 0)\,\,A_1}^{\mathrm{pert}}
\cdot 
Z_{G}
\cdot 
Z_{\mathrm{pert}}^{\Deff}(X)\,,
\end{align}
where only $Z_{\mathrm{pert}}^{\Deff}(X)$ is a function of the
defect parameter $X$. 
Using \eqref{eq:codim2}, computing $I_v$ and the $\mu_2$-dependent 
part of $I_h$ leads to
\begin{align}
Z_1&=
\PE
\left[
I_v+\frac{\sqrt{pq}}{(1-p)(1-q)(1-Q)}
\left(\alpha+\alpha^{-1}Q\right)
\left(t+t^{-1}\right)
\left(\mu_2+\mu_2^{-1}\right)
\right]
\Bigg\vert_{
\substack{
\alpha=\sqrt{q}t^{-1}
\\
\mu_2=\sqrt{pq^2}
}}
\notag\\
&=\PE 
\left[
\frac{1}{(1-p)(1-Q)}
\left(t^{-2}-t^2p\,Q\right)
\right]
+\mathrm{etc.}\notag\\
&=\prod_{i=0}^{\infty}\frac{1}{\widetilde{\theta}_1(Xp^i)}
+\mathrm{etc.} 
\end{align}
using \eqref{Xparameter} in the last line. Further, all irrelevant terms 
independent of the defect parameter $X$ have been omitted.

On the other hand, one also needs to extract additional 
$Z_{\mathrm{pert}}^{\mathrm{def}}(X)$ 
contributions\footnote{Different from the generic $\Ncal=(1, 0)$ case, there is an additional contribution depending on the defect parameter $X$ and flavour fugacity $\eta$. Because the $(2, 0)$ $A_1$ theory contains no vector multiplet, the new piece thus originates from the term depending on the $a_1$ gauge fugacity and the left flavour fugacity $\eta$. Since the $\surm(2)$ fugacities satisfy $a_1 +a_2=0$, both $a_1$ \emph{and} $a_2$ have been replaced by the defect parameter $X$ after Higgsing.}, which are $\mu_1$-dependent, from the Goldstone part $Z_{G}$. 
In detail
\begin{align}
Z_2=
\PE 
\left[
\frac{\sqrt{pq}}{(1-p)(1-q)(1-Q)}
\left(\alpha+\alpha^{-1}Q\right)
\left(t+t^{-1}\right)
\left(\mu_1+\mu_1^{-1}\right)
\right]
\Bigg\vert_{ 
\substack{ 
\alpha=\sqrt{q}t^{-1}
\\
\mu_2=\sqrt{pq^2}
\\
\mu_1\rightarrow 
\sqrt{q}\mu_1
}
}
\cdot 
\frac{1}{Z_{G}}
\,,
\end{align}
where one has shifted $\mu_1\rightarrow \sqrt{q}\mu_1$ in order to compare with 
the contribution of Goldstone bosons. 
With some algebra, apart from some irrelevant terms, one finds
\begin{align}
Z_2&=
\PE
\left[
\frac{1}{(1-p)(1-Q)} 
\left( -X\eta+(X\eta)^{-1}p\,Q \right)
\right]\notag\\
&=\prod_{i=0}^{\infty}\widetilde{\theta}_1(X\eta\, p^i)\,,
\end{align}
using \eqref{Xparameter}. Hence, combining the various parts, one 
arrives at
\begin{align}
Z_{\mathrm{pert}}^{\Deff}(X)
=
Z_1\cdot Z_2
=\prod_{i=0}^{\infty}\frac{\widetilde{\theta}_1(X\eta\, 
p^i)}{\widetilde{\theta}_1(Xp^i)}\,.
\end{align}
By acting with $Y_X$ on it, one finds the following difference equation
\begin{align}
Y_X\cdot 
Z_{\mathrm{pert}}^{\Deff}(X)
=\frac{\theta(p^{-1}X\eta)}{\theta(p^{-1}X)}
Z_{\mathrm{pert}}^{\Deff}(X)
\,.
\end{align}
Therefore, the full partition function
\begin{align}
Z(X)=Z^{\Deff}_{\mathrm{pert}}(X)\cdot \widetilde Z(X)\,,
\end{align}
satisfies the following difference equation in the NS-limit $q\rightarrow 1$:
\begin{align}
\begin{aligned}
\frac{\widetilde{\theta}_1(X\eta)}{\widetilde{\theta}_1(X)}Y_X^{-1}\cdot 
Z(X)+q_\phi 
\frac{\widetilde{\theta}_1(X\eta^{-1})}{\widetilde{\theta}_1(X)}\,Y_X\cdot Z(X)
&\eqqcolon \Wcal(X)\cdot 
Z(X)
\\
\Leftrightarrow \qquad
\left[ 
\frac{\widetilde{\theta}_1(X\eta)}{\widetilde{\theta}_1(X)}Y_X^{-1}+q_\phi 
\frac{\widetilde{\theta}_1(X\eta^{-1})}{\widetilde{\theta}_1(X)}\,Y_X
- \Wcal(X)
\right]
Z(X) &= 0
\,,
\end{aligned}
\label{fullDE}
\end{align}
where the last line already bears resemblance to \eqref{eq:diff_eq_5d}. As in \eqref{instDE}, one still has to 
provide an interpretation of $\Wcal$, which is the subject of the next section.

\subsubsection{Wilson surface}
\label{WS2}
In this subsection, $\Wcal$ is identified with the Wilson 
surface from $6$d perspective as discussed above for the generic $6$d 
$\Ncal=(1, 0)$ case.
As in Section \ref{sec:comparison}, the identification proceeds in  
two steps:
\begin{compactenum}[(i)]
 \item Computation of the prediction for $\Wcal$ from the difference equation 
\eqref{instDE}.
\item Direct evaluation of the Wilson surface expectation value.
\end{compactenum}
Firstly, one computes $\Wcal$ from \eqref{instDE} up to 
one-instanton order. A computation shows that
\begin{align}
\Wcal 
&=
\frac{Y_X^{-1}\widetilde Z(X)}{\widetilde Z(X)}+q_\phi 
\frac{\widetilde{\theta}_1(X\eta^{-1})\widetilde{\theta}_1(p^{-1}X\eta)}{
\widetilde{\theta}_1(X)\widetilde{\theta}_1(p^{-1}X)}\frac{
Y_X\widetilde Z(X)}{\widetilde Z(X)}
\notag \\
&=
1
+
q_\phi
\left(
\widetilde{Z}_1(p X)
-\widetilde{Z}_1(X)
+\frac{\widetilde{\theta}_1(X\eta^{-1})\widetilde{\theta}_1(p^{-1}X\eta)}{
\widetilde{\theta}_1(X)\widetilde{\theta}_1(p^{-1}X)}
\right)
+\Ocal(q_\phi^2)
\notag\\
&=
1
+
q_\phi
\left(
\frac{\widetilde{\theta}_1(pX\eta^{-1})\widetilde{\theta}_1(X\eta)}{\widetilde{
\theta}_1(X)\widetilde{\theta}_1(pX)}
+
pX 
\frac{\widetilde{\theta}_1(\eta)\widetilde{\theta}_1(p\eta^{-1})\widetilde{
\theta}_1^\prime(pX)}{\widetilde{\theta}_1^{\prime}
(1)\widetilde{\theta}_1(p)\widetilde{\theta}_1(pX)}
-X
\frac{\widetilde{\theta}_1(\eta)\widetilde{\theta}_1(p\eta^{-1})\widetilde{
\theta}_1^\prime(X)}{
\widetilde{\theta}_1^{\prime}(1)\widetilde{\theta}_1(p)\widetilde{\theta}_1(X)}
\right)
\label{W1}\\
&  \; \; \quad  +\Ocal(q_\phi^2)\,,
\notag 
\end{align}
where $\widetilde{\theta}_1^\prime(X)$ denotes the derivative of 
$\widetilde{\theta}_1(X)$. 
As it turns out, the $\Wcal$ expression is independent on $X$, as can be 
verified by expanding \eqref{W1} with respect to $Q$, i.e.\
\begin{align}
\label{eq:W_(2,0)_z-independent}
\begin{aligned}
\Wcal=
1
&+
q_\phi
\left(
1
+
\frac{(1-\eta)^2(1-p^{-1}\eta)^2}{p^{-1}\eta^2}Q
+
\frac{
(1-\eta)^2(1-p^{-1}\eta)^2(p^{-2}+4p^{-1}+1)}{p^{-2}\eta^2}Q^2
+\Ocal(Q^3)\right)
\\
&+\Ocal(q_\phi^2)\,.
\end{aligned}
\end{align}
In fact, $q_\phi^{-1/2}\Wcal$ coincides with the Wilson line 
$\Wcal_{\surm(2)}$ computed from the $5$d $\surm(2)$ SYM via compactifying the 
$6$d theory on a  circle as in \cite{Bullimore:2014awa, Agarwal:2018tso}.

Secondly, one can directly compute the expectation value of the 6d Wilson 
surface in the 6d $\Ncal=(2, 0)$ $A_1$ theory, as studied in 
\cite{Agarwal:2018tso}. For a Wilson surface in a minuscule representation, 
for instance the fundamental representation, one finds either from 
\cite{Agarwal:2018tso} or Section \ref{sec:Wilson_surface} that 
\begin{align}
\mathcal W^{(2,0)}
&=\sum_{l=0}^{\infty}q_\phi^l W_l\notag\\
W_l
&=
\frac{1}{l!}\int\prod_{I=1}^{l}\frac{\diff \phi_I}{2\pi i}
\prod_{I, J=1}^{l}
\frac{\vartheta_1(\phi_{IJ})\vartheta_1(\phi_{IJ}+2\epsilon_+)}{
\vartheta_1(\phi_{IJ}+\epsilon_{1,2})}
\prod_{I=1}^{l}
\frac{\vartheta_1(m\pm\phi_I)}{\vartheta_1(\epsilon_+\pm\phi_I)}
\prod_{I=1}^l
\frac{\vartheta_1(\epsilon_-\pm(\phi_I-z))}{
\vartheta_1(-\epsilon_+\pm(\phi_I-z))}\,,
\end{align}
where $z$ denotes the $U(1)$ fugacity from D$4^\prime$ brane, see Figure 
\ref{fig:branes}. Up to one-instanton order, one finds
\begin{align}\label{eq:Wilson-VEV}
\Wcal^{(2,0)}
&=
1
+
q_\phi
\left(\frac{\widetilde{\theta}_1(p 
Z\eta^{-1})\widetilde{\theta}_1(Z\eta)}{\widetilde{\theta}_1(p 
Z)\widetilde{\theta}_1(Z)}
+pZ 
\frac{\widetilde{\theta}_1(\eta)\widetilde{\theta}_1(p\eta^{-1})\widetilde{
\theta}_1^\prime(p Z)}{\widetilde{\theta}_1(p)\widetilde{\theta}_1(p 
Z)\widetilde{\theta}_1^\prime(1)}
-
Z\frac{\widetilde{\theta}_1(\eta)\widetilde{\theta}_1(p\eta^{-1})\widetilde{
\theta}_1^\prime(Z)}{
\widetilde{\theta}_1(q)\widetilde{\theta}_1(Z)\widetilde{\theta}_1^\prime(1)} 
\right)
\\
&\quad \;\; 
+\Ocal (q_\phi^2) \notag \,,
\end{align}
which is the same as \eqref{W1} by replacing $Z\equiv e^{z}$ with $X$. 
However, as shown in \eqref{eq:W_(2,0)_z-independent}, $\Wcal^{(2,0)}$ is 
independent of $Z$ or $X$. 
As a consequence, the direct $6$d computation of the Wilson surface, which 
coincides with 
the $5$d Wilson loop result \cite{Agarwal:2018tso}, also verifies the quantised 
SW-curve \eqref{instDE} proposed in the subsection above for the $6$d 
$\Ncal=(2, 0)$ $A_1$ case.

\subsection{\texorpdfstring{Codimension 2 defect in 5d $\Ncal=2$ $\surm(2)$ SYM}{Codimension 2 defect in 5d N=2 SU(2) SYM}}
\label{sec:5d_SYM}
A circle compactification of the 6d $\Ncal=(2, 0)$ $A_1$ theory gives rise to the 5d $\mathcal{N}=2$ maximal supersymmetric Yang-Mills theory with gauge group $\surm(2)$. In fact, the instanton states in this 5d theory capture the Kaluza-Klein momentum modes. Therefore, the 5d $\surm(2)$ maximal SYM theory at strong coupling is conjectured to be dual to the 6d $\Ncal=(2, 0)$ $A_1$ theory \cite{Lambert:2010wm,Douglas:2010iu,Lambert:2010iw}.

A codimension 2 defect preserving half of the supersymmetries in the 5d $\mathcal{N}=2$ $\surm(2)$ gauge theory has been studied in \cite{Bullimore:2014awa}. This defect was introduced as a monodromy defect. However, the same defect can also be introduced by Higgsing the $\surm(2)\times \surm(2)$ affine quiver theory with two bi-fundamental hypermultiplets with position dependent VEV of a baryonic operators formed by one of the bi-fundamental hypermultiplets.  In terms of a 8 supercharges quiver, the Higgsing is summarised as follows:
\begin{align}
	\raisebox{-.5\height}{
 	\begin{tikzpicture}
	\tikzstyle{gauge} = [circle, draw,inner sep=3pt];
	\tikzstyle{flavour} = [regular polygon,regular polygon sides=4,inner 
sep=3pt, draw];
	\node (g1) [gauge,label=left:{$\surm(2)_1$}] {};
	\node (g2) [gauge,right of=g1,label=right:{$\surm(2)_2$}] {};
	\draw (g1) .. controls (0.5,0.25) .. (g2);
	\draw (g1) .. controls (0.5,-0.25) .. (g2);
	\end{tikzpicture}
	}
	\qquad \xrightarrow[\text{Higgsing}]{\text{baryonic}}
	\qquad
	\raisebox{-.5\height}{
	\begin{tikzpicture}
	\tikzstyle{gauge} = [circle, draw,inner sep=3pt];
	\tikzstyle{flavour} = [regular polygon,regular polygon sides=4,inner 
sep=3pt, draw];
	\node (g1) [gauge,label=below:{$\surm(2)$}] {};
	\draw [-] (0.11,0.105) arc (-70:90:10pt);
	\draw [-] (-0.11,0.105) arc (250:90:10pt);
	\end{tikzpicture}
	}
	\,.
	\label{eq:5d_quiver}
\end{align}
Equivalently, the Higgsing of a 5d $\Ncal =1$ affine $A_k$ quiver gauge theory with a constant or position dependent VEV is realised in Type IIB superstring theory as shown in Figure \ref{fig:Higgsing_branes}.
From the 6d viewpoint, this corresponds to a mesonic Higgsing of the $\surm(2)$ gauge theory with $4$ flavours towards the $\Ncal=(2,0)$ $ A_1$ theory with a codimension 2 defect. The duality between the 5d and 6d description can be verified on the level of partition functions.
\paragraph{Partition function before Higgsing.}
Let us start with the partition function of the 5d $\surm(2)\times \surm(2)$ affine quiver gauge theory on $\R^4_{\epsilon_1,\epsilon_2}\times S^1$. The perturbative partition function can be written as
\begin{equation}
	Z^{\mathrm{ 5d,pert}}_{\surm(2)^2} {=}  \PE\left[-\frac{1+pq}{(1-p)(1-q)}(A_1^2+A_2^2) + \frac{\sqrt{pq}}{(1-p)(1-q)}A_1(A_2+A_2^{-1})(\mu_1+\mu_2+\mu_1^{-1}+\mu_2^{-1})\right] ,
\end{equation}
where $A_{1,2}\equiv e^{a_{1,2}}$ are the gauge fugacities for two $\surm(2)$ gauge groups in \eqref{eq:5d_quiver} and $\mu_{1}\equiv e^{M_1}$, $\mu_{2}\equiv e^{M_2}$ are the fugacities for the bi-fundamental flavours. The instanton partition function can be evaluated from a 1d gauged quantum mechanics and is given by
\begin{align}
	Z^{\mathrm{ 5d, inst}}_{\surm(2)^2} &= \sum_{k_1,k_2=0}^\infty y^{k_1}\left(\frac{Q}{y}\right)^{k_2} Z^{5d}_{k_1,k_2} \nonumber \\
	Z^{\mathrm{5d}}_{k_1,k_2} &=\frac{1}{k_1!k_2!}
	\oint\left(\prod_{I=1}^{k_1}\frac{\diff \phi_I}{2\pi \im}\right)
	\left(\prod_{J=1}^{k_2}\frac{\diff \tilde{\phi}_J}{2\pi \im}\right) 
	\frac{
	\prod_{I\neq J}^{k_1} \sh(\phi_{IJ})
	\prod_{I, J}^{k_1} \sh(\phi_{IJ}+2\epsilon_+)
	}{
	\prod_{I, J}^{k_1} \sh(\phi_{IJ}+\epsilon_{1,2})
	}
	\cdot
	\frac{
	\prod_{I\neq J}^{k_2} \sh(\tilde\phi_{IJ})
	\prod_{I, J}^{k_2} \sh(\tilde\phi_{IJ}+2\epsilon_+)
	}{
	\prod_{I, J}^{k_2} \sh(\tilde\phi_{IJ}+\epsilon_{1,2})
	} \notag \\
	& \qquad \cdot
	\prod_{I=1}^{k_1}\prod_{J=1}^{k_2}
	\frac{
 \sh(\phi_I\pm a_2+M_{1,2})
 \sh(\tilde\phi_J\pm a_1-M_{1,2})
	}{
	 \sh(\phi_I\pm a_1 \pm \epsilon_+)
	 \sh(\tilde{\phi}_J\pm a_2 \pm \epsilon_+)
	}
	\cdot 
	\frac{
	\sh(\phi_I-\tilde\phi_J+M_{1,2}\pm \epsilon_-)
	}{
	 \sh(\phi_I-\tilde\phi_J+M_{1,2}\pm \epsilon_+)
	} \ ,
    \label{eq:5d_instanton}
\end{align}
where $y$ and $Q/y$ are the instanton fugacities for the $\surm(2)$ gauge groups, respectively, and $\sh(x)\equiv 2\sinh(\tfrac{x}{2})$ as well as $\phi_{IJ} = \phi_I -\phi_J$, $\tilde{\phi}_{IJ} = \tilde{\phi}_I -\tilde{\phi}_J$. The contour integral \eqref{eq:5d_instanton} at each instanton sector can again be evaluated by using the JK-prescription \cite{Hwang:2014uwa}. 
As expected from the duality between the 5d $\surm(2){\times} \surm(2)$ affine quiver gauge theory and the 6d SCFT for 2 M5-branes on $A_1$ singularity, the full partition function for the 5d $\surm(2){\times }\surm(2)$ affine quiver theory coincides with the partition function of the 6d SCFT given in Section \ref{sec:6d_no_defect}. Namely,
\begin{align}\label{eq:equality-5d/6d}
\begin{aligned}
	Z_{\mathcal{N} =(1,0)\ A_1}^{\mathrm{6d}} &= Z^{\mathrm{5d}}_{\surm(2)^2}\cdot Z_{\rm extra} \\
	Z_{\mathcal{N} =(1,0)\ A_1}^{\mathrm{6d}} &\equiv Z_{\mathcal{N} =(1,0)\ A_1}^{\rm pert}\cdot \sum_{l=0}^\infty q_\phi^l Z_l \ , \quad Z^{5d}_{\surm(2)^2} \equiv Z^{\mathrm{5d,pert}}_{\surm(2)^2} \cdot Z^{\rm 5d,inst}_{\surm(2)^2} \ ,
	\end{aligned}
\end{align}
with the identification of the 5d/6d fugacities as
\begin{equation}
	(A_1,A_2,y, \mu_{1,2})^{5d} = (q_\phi^{1/2}\alpha^{-1} t , q_\phi^{1/2} ,t^{2},\mu_{1,2})^{6d} \ .
\end{equation}
Here, $Z_{\mathcal{N} =(1,0)\ A_1}^{\rm pert}$ and $Z_l$ are given in \eqref{eq:A1-pert} and \eqref{eq:A1-inst}, respectively; and $Z_{\rm extra}$ is an extra factor independent of dynamical fugacities defined as
\begin{equation}
	Z_{\rm extra} = \PE\left[
	-\frac{(1+p)(1+q)Q}{(1-p)(1-q)(1-Q)} 
	-\frac{
	\left( 
	\frac{t^2 \mu_1}{\mu_2}
	+\frac{\mu_2}{t^2 \mu_1}Q
	+\frac{pqt^2 \, \mu_2}{\mu_1}
	+\frac{pq\, \mu_1}{t^2\mu_2}Q
	\right)
	}{
	(1-p)(1-q)(1-Q)
	}
	\right]  \ .
\end{equation}
One can check the equality (\ref{eq:equality-5d/6d}) by expanding both sides in terms of $Q$ and $q_\phi$.
\paragraph{Higgsing.}
Higgsing \eqref{eq:5d_quiver} to the 5d $\mathcal{N}=2$ $\surm(2)$ gauge theory can be performed by tuning the fugacities in the partition function as
\begin{equation}
	A_1 \rightarrow A_2 \ , \quad \mu_2 \rightarrow \frac{1}{\sqrt{pq}} \ .
\end{equation}
This leads to the partition function of the 5d $\mathcal{N}=2$ $\surm(2)$ gauge theory as
\begin{equation}
	Z^{5d}_{\surm(2)^2}\bigg|_{\substack{ A_1 \rightarrow A_2 \\ \mu_2 \rightarrow 1\slash \sqrt{pq} }} 
	= Z^{5d}_{\mathcal{N}=2 \ \surm(2)}\cdot Z_{\rm extra'} \ , \quad 
	Z_{\rm extra'} = \PE\left[-\frac{\left(1-pq\right)\left(1-\tfrac{pq}{ \mu_1^{2}} \right) \mu_1}{\sqrt{pq}(1-p)(1-q)} y \right] \ ,
\end{equation}
up to the extra factor $Z_{\rm extra'}$ independent of the dynamical fugacity $A_2$. After the Higgsing, $A_2$ becomes the fugacity for the $\surm(2)$ gauge symmetry and $\mu_1$ becomes the fugacity for the $\surm(2)\subset \sorm(5)$ flavour symmetry. 

Next consider the Higgsing with a position dependent VEV that introduces a codimension 2 defect in the 5d $\mathcal{N}=2$ $\surm(2)$ theory. The Higgsing can achieve by the following fugacity assignment:
\begin{equation}
	A_1\rightarrow A_2 \sqrt{q} \ , \quad \mu_2 \rightarrow \frac{1}{\sqrt{pq^2}} \ , \quad \mu_1 \rightarrow\mu_1 \sqrt{q} \,.
\end{equation}
With this specialisation of the fugacities, the partition function reduces to that of the 5d $\mathcal{N}=2$ $\surm(2)$ theory in the presence of the monodromy defect, called $\mathcal{Z}_{[1,1]}$, introduced in \cite{Bullimore:2014awa}:
\begin{equation}
	 Z^{5d}_{\surm(2)^2}\bigg|_{\substack{ A_1 \rightarrow A_2\sqrt{q} \\ \mu_2 \rightarrow 1\slash \sqrt{pq^2} \\ \mu_1 \rightarrow \mu_1 \sqrt{q} } }
	 =\mathcal{Z}_{[1,1]} \cdot Z_{\rm extra'}\ .
\end{equation}
This shows that the codimension 2 defect introduced by the Higgsing is identical to the monodromy defect considered in \cite{Bullimore:2014awa}.
The instanton part of the codimension 2 defect partition function is expanded in terms of $y$ and $Q/y$, and the first few terms are given by
\begin{equation}
\mathcal{Z}_{[1,1]}^{{\rm inst}}=  1 
- \frac{(\eta-1)(\eta- A_2^{2})}{(1-p)(1-A_2^{2}/p)\eta} y 
-\frac{(\eta-1)(1- \eta q A_2^{2})p }{(1-p)(1- pqA_2^{2})\eta} \frac{Q}{y} +\ldots \ ,
\end{equation}
with $\eta$ defined in \eqref{Xparameter}.

In the NS limit $q\rightarrow1$, the codimension 2 defect partition function satisfies the following difference equation \cite{Bullimore:2014awa}:
\begin{equation}
	\left[A_2^{-1}\frac{\tilde\theta(y\eta)}{\tilde\theta_1(y)}Y_y^{-1}+A_2\frac{\tilde\theta_1(y/\eta)}{\tilde\theta_1(y)}Y_y  - \langle W_{\surm(2)} \rangle \right]
	\lim_{q\rightarrow1} \mathcal{Z}_{[1,1]}^{{\rm inst}} = 0 \ ,
	\label{eq:diff_eq_5d}
\end{equation} 
where $Y_y y = py Y_y$ and $\langle W_{\surm(2)} \rangle$ is the $\surm(2)$ fundamental Wilson loop expectation value in the 5d maximal SYM discussed in \cite{Bullimore:2014awa,Bullimore:2014upa}. This is the difference equation of the two-body elliptic Ruijsenaars-Schneider integrable system. Here, the Wilson loop expectation value $\langle W_{\surm(2)}\rangle$ of the 5d theory is related to the VEV of Wilson surface $\mathcal{W}^{(2,0)}$ in the 6d (2,0) $A_1$ theory in \cite{Agarwal:2018tso} as 
\begin{equation}
    \mathcal{W}^{(2,0)} = q_\phi^{1/2} \langle W_{\surm(2)}\rangle \ .
\end{equation}
One can verify that, by replacing
\begin{equation}
X=y^{-1}\,\ \ \ {\rm and} \ \ \ Y_X=\eta^{-1} Y_y^{-1}\,,
\end{equation}
\eqref{fullDE} becomes \eqref{eq:diff_eq_5d}. Hence, the difference equations agree.
%
%
\section{Conclusions}
\label{sec:conclusion}
In this paper we explored elliptic difference equations arising from quantisation of Seiberg-Witten curves of compactified 6d $A$-type $\Ncal=(1,0)$ SCFTs. In order to obtain a 4d $\mathcal{N}=2$ supersymmetric theory, the 6d theory is compactified on a two-torus together with an Omega-background. This allows to compute the BPS partition function of the theory together with expectation values of various defect operators using localisation. We explicitly showed, using a matrix-model approach, that the corresponding quantum curves annihilate expectation values of codimension $2$ surface defects inside the 6d theory. Moreover, we found that our difference equations can be rewritten as eigenvalue equations with eigenvectors being our codimension $2$ defects and eigenvalues corresponding to expectation values of codimension $4$ defects arising from Wilson surfaces wrapping the two-torus.

One important insight of our analysis is the fact that our difference operator equally well applies to the 5d dual of the 6d SCFT. This duality, as for example recently explored in \cite{Bhardwaj:2018yhy,Bhardwaj:2018vuu,Bhardwaj:2019fzv}, results in a 5d supersymmetric gauge theory admitting an affine quiver description. In our case, this is an affine $A$-type quiver with $\surm(N)$ gauge nodes \cite{Haghighat:2013tka}. BPS partition functions of the circle-compactified 5d theory are then equal to the torus-compactified 6d partition function. The codimension $2$ defect of the 6d theory is mapped to a codimension $2$ defect inside the 5d theory giving rise to a coupled 3d/5d system. Difference operators for such systems are not easy to obtain, but our approach via the dual 6d theory gives a recipe to construct such operators from first principles. 

Another direction, particularly interesting for future research, is the realisation of 4d $\Ncal=1$ SCFTs as surface defects inside a 6d SCFT. Indeed, our codimension $2$ defect is itself such a 4d theory extended over $\T^2 \times_{\epsilon_2} \R^2$. The expectation value of the defect operator on such a geometry is related to the supersymmetric index of the corresponding 4d $\Ncal=1$ SCFT and, thus, it is expected that such indices satisfy similar difference equation. From this point of view, it would be interesting to ask whether the knowledge of the difference operator is enough to reconstruct the index of the corresponding 4d SCFT. First steps in this direction have been taken in \cite{Gaiotto:2012xa,Gaiotto:2015usa,Nazzal:2018brc}. In particular, in \cite{Nazzal:2018brc} the authors give a detailed derivation of the difference operator associated to $\Ncal=1$ compactifications of E-string theory. It would be interesting to extend these results by applying our techniques to the torus-compactified E-string theory. The corresponding difference operator should in this case arise from the quantisation of the SW-curve derived in \cite{Haghighat:2018dwe}. We leave this and the derivation of quantum curves for a wider class of 6d SCFTs for future work.
Likewise, the difference equations of other 6d SCFTs and their relation to integrable models, as for example considered in \cite{Koroteev:2019gqi}, are interesting future directions.

\paragraph{Acknowledgments.}
B.H. would like to thank the Bethe Center for Theoretical Physics at the University of Bonn, where part of this work was carried out, for hospitality. The work of J.C., B.H., and M.S. is supported by the National Thousand-Young-Talents Program of China. 
H.K. is supported by the POSCO Science Fellowship of POSCO TJ Park Foundation and the National Research Foundation of Korea (NRF) Grant 2018R1D1A1B07042934.
M.S. is further supported by the National Natural Science Foundation of China (grant no.\ 11950410497), and the China Postdoctoral Science Foundation (grant no.\ 2019M650616).
%
%
\appendix
%
%
\section{Details of partition functions}
\label{app:details_part_fct}
The computational details of the various partition functions are provided in this appendix.
\subsection{Perturbative contribution}
The perturbtative part of the partition function is composed of the single letter contributions \eqref{eq:single_letter_6d} for the 6d $\Ncal=(1,0)$ multiplets.
\subsubsection{Higgsing: constant VEV}
\label{app:const_Higgs_pert}
The perturbative part can be written as
\begin{align}
  Z_{\mathrm{pert}}^{k+1} &=   Z_{\mathrm{pert}}^{k}
  \cdot 
  \PE \Big[ \frac{1}{(1-p)(1-q)} \left( \frac{Q}{1-Q} +\frac{1}{2}\right) 
\notag 
\\
  &\qquad \bigg\{
  -(1+pq) \sum_{i=1}^{k}  \left( e^{a_i -a_{k+1} }+ e^{a_{k+1} -a_i } \right) 
-(1+pq)
  \notag \\
  &\qquad \qquad 
  +\sqrt{pq}  \sum_{l=1}^{k} \left( e^{a_{k+1}} (e^{-m_l+b}+e^{-n_l-b}) 
+ e^{-a_{k+1}} (e^{m_l-b}  +e^{n_l+b}) \right)
  \notag \\
  &\qquad \qquad 
  +\sqrt{pq} \sum_{i=1}^{k}  \left(e^{a_i} (e^{-m_{k+1}+b} + e^{-n_{k+1}-b})
  +e^{-a_i} (e^{m_{k+1}-b}  +e^{n_{k+1}+b}) \right)
  \notag \\
  &\qquad \qquad 
  +\sqrt{pq} \left( e^{a_{k+1}} (e^{-m_{k+1}+b} + e^{-n_{k+1}-b} ) 
+ e^{-a_{k+1}} ( e^{m_{k+1}-b}  +e^{n_{k+1}+b}) \right)
  \bigg\}
  \Big]
  \label{eq:decomp_pert_part}
\end{align}
such that Higgsing \eqref{eq:Higgs_no_defect} yields for the different parts
\begin{subequations}
\begin{align}
 -(1+pq) &\sum_{i=1}^{k}  \left( e^{a_i -a_{k+1} }+ e^{a_{k+1} -a_i } \right) 
\notag \\
&= -(1+pq) \sum_{i=1}^{k}  \left( \sqrt{pq} e^{a_i -m_{k+1}+b }+ 
\frac{1}{\sqrt{pq}} e^{m_{k+1}-b -a_i } \right) \\
%
%
\sqrt{pq}  &\sum_{l=1}^{k} \left( e^{a_{k+1}} (e^{-m_l+b}+e^{-n_l-b}) 
+ e^{-a_{k+1}} (e^{m_l-b}  +e^{n_l+b}) \right) \notag \\
&= 
\sqrt{pq}  \sum_{l=1}^{k} \left( \frac{1}{\sqrt{pq}}e^{m_{k+1}-b} 
(e^{-m_l+b}+e^{-n_l-b}) 
+ \sqrt{pq} e^{-m_{k+1}+b} (e^{m_l-b}  +e^{n_l+b}) \right)
\\
%
%
\sqrt{pq} &\sum_{i=1}^{k}  \left(e^{a_i} (e^{-m_{k+1}+b} + e^{-n_{k+1}-b})
  +e^{-a_i} (e^{m_{k+1}-b}  +e^{n_{k+1}+b}) \right)\notag \\
&=
\sqrt{pq} \sum_{i=1}^{k}  \left(e^{a_i} (e^{-m_{k+1}+b} + pq e^{-m_{k+1}+b})
  +e^{-a_i} (e^{m_{k+1}-b}  +\frac{1}{pq} e^{m_{k+1}-b}) \right)
\notag\\
&=
(1  +pq ) \sum_{i=1}^{k}  \left( \sqrt{pq}e^{a_i-m_{k+1}+b} 
  + \frac{1}{\sqrt{pq}} e^{-a_i+m_{k+1}-b}   \right)
  \\
%
%
\sqrt{pq} &\left( e^{a_{k+1}} (e^{-m_{k+1}+b} + e^{-n_{k+1}-b} ) 
+ e^{-a_{k+1}} ( e^{m_{k+1}-b}  +e^{n_{k+1}+b}) \right) \notag \\
&=
\sqrt{pq} \left(\frac{1}{\sqrt{pq}} e^{m_{k+1}-b} (e^{-m_{k+1}+b} + 
pq e^{-m_{k+1}+b} ) 
+\sqrt{pq} e^{-m_{k+1}+b} ( e^{m_{k+1}-b}  +\frac{1}{pq}e^{m_{k+1}-b}) \right) 
\notag \\
&=2 ( 1  +pq)
\end{align}
\end{subequations}
and collecting all the pieces leads to
\begin{align}
  Z_{\mathrm{pert}}^{k+1} &=   Z_{\mathrm{pert}}^{k}
  \cdot 
  \PE \Big[ \frac{1}{(1-p)(1-q)} \left( \frac{Q}{1-Q} +\frac{1}{2}\right) 
\notag 
\\
  &\qquad \bigg\{
  -(1+pq) \sum_{i=1}^{k}  \left( \sqrt{pq} e^{a_i -m_{k+1}+b }+ 
\frac{1}{\sqrt{pq}} e^{m_{k+1}-b -a_i } \right) 
  -(1+pq)
  \notag \\
  &\qquad \qquad 
  +(1 +pq) \sum_{i=1}^{k} \left( \sqrt{pq} e^{a_i-m_{k+1}+b} 
  + \frac{1}{\sqrt{pq}} e^{m_{k+1}-b-a_i}  \right)
  +2 ( 1  +pq)
  \notag \\
  &\qquad \qquad 
 +\sqrt{pq}  \sum_{l=1}^{k} \left( \frac{1}{\sqrt{pq}}e^{m_{k+1}-b} 
(e^{-m_l+b}+e^{-n_l-b}) 
+ \sqrt{pq} e^{-m_{k+1}+b} (e^{m_l-b}  +e^{n_l+b}) \right)
  \bigg\}
  \Big] \\
    &=   Z_{\mathrm{pert}}^{k}
  \cdot 
  \PE \Big[ \frac{1}{(1-p)(1-q)} \left( \frac{Q}{1-Q} +\frac{1}{2}\right)
  \bigg\{\left(1 + pq\right)\notag\\
  &\qquad \qquad
 + \sqrt{pq}  \sum_{l=1}^{k} \left( \frac{1}{\sqrt{pq}}e^{m_{k+1}-m_l} 
+ \sqrt{pq} e^{m_l-m_{k+1}}  \right) \notag \\
&\qquad \qquad
+\sqrt{pq}  \sum_{l=1}^{k} \left( \frac{1}{\sqrt{pq}}e^{m_{k+1}-n_l-2b}
+ \sqrt{pq} e^{n_l-m_{k+1}+2b}  \right)
  \bigg\}
  \Big]\,,
\end{align}
where the additional pieces are attributed to the Goldstone modes for the 
reduced global symmetry. In detail,
\begin{align}
 Z_{\mathrm{G}} &= 
 \PE \Big[ \frac{1}{(1-p)(1-q)} \left( \frac{Q}{1-Q} +\frac{1}{2}\right) 
  \bigg\{\left(1 + pq\right) 
 + \sqrt{pq}  \sum_{l=1}^{k} \left( \frac{1}{\sqrt{pq}}e^{m_{k+1}-m_l} 
+ \sqrt{pq} e^{m_l-m_{k+1}}  \right) \notag \\
&\qquad \qquad \qquad 
+\sqrt{pq}  \sum_{l=1}^{k} \left( \frac{1}{\sqrt{pq}}e^{m_{k+1}-n_l-2b}
+ \sqrt{pq} e^{n_l-m_{k+1}+2b}  \right)
  \bigg\}
  \Big]\,, \notag \\
&= 
 \PE \Big[ \frac{\sqrt{pq}}{(1-p)(1-q)} \left( \frac{Q}{1-Q} 
+\frac{1}{2}\right) 
  \bigg\{ \left(\frac{1}{\sqrt{pq}} + \sqrt{pq}\right) 
 +  \sum_{l=1}^{k} \left( \frac{1}{\sqrt{pq}}e^{m_{k+1}-m_l} 
+ \sqrt{pq} e^{m_l-m_{k+1}}  \right) \notag \\
&\qquad \qquad \qquad 
+ \sum_{l=1}^{k} \left( \frac{1}{\sqrt{pq}}e^{m_{k+1}-n_l-2b}
+ \sqrt{pq} e^{n_l-m_{k+1}+2b}  \right)
  \bigg\}
  \Big]\,.
\end{align}
\subsubsection{Higgsing: position dependent VEV}
\label{app:defect_Higgs_pert}
Inspecting the different contributions to \eqref{eq:decomp_pert_part} yields
for the position dependent Higgsing \eqref{eq:Higgs_with_defect} the following:
\begin{subequations}
\begin{align}
 -(1+pq) &\sum_{i=1}^{k}  \left( e^{a_i -a_{k+1} }+ e^{a_{k+1} -a_i } \right) 
\notag \\
&=  
-(1+pq) \sum_{i=1}^{k}  \left( \frac{1}{\sqrt{pq}} \frac{e^{a_i }}{X}  
+ \sqrt{pq} \frac{X}{e^{a_i }}   \right) 
\\
%
%
\sqrt{pq}  &\sum_{l=1}^{k} \left( e^{a_{k+1}} (e^{-m_l+b}+e^{-n_l-b}) 
+ e^{a_{k+1}} (e^{m_l-b}  +e^{n_l+b}) \right) \notag \\
&= 
\sqrt{pq}  \sum_{l=1}^{k} \left(  \sqrt{pq} X
(e^{-m_l}B+e^{-n_l}B^{-1}) 
+ \frac{1}{\sqrt{pq}X}  (e^{m_l}B^{-1}  +e^{n_l}B) \right)
\notag \\
&= 
\sqrt{pq}  \sum_{l=1}^{k} \left(  
\sqrt{pq} X B e^{-m_l}
+ \frac{1}{\sqrt{pq}X B }  e^{m_l}   
+\sqrt{pq}X e^{-n_l}
+ \frac{1}{\sqrt{pq}X}    e^{n_l}
\right)
\\
%
%
\sqrt{pq} &\sum_{i=1}^{k}  \left(e^{a_i} (e^{-m_{k+1}+b} + e^{-n_{k+1}-b})
  +e^{-a_i} (e^{m_{k+1}-b}  +e^{n_{k+1}+b}) \right)\notag \\
&=
\sqrt{pq} \sum_{i=1}^{k}  \left(e^{a_i}X^{-1}  (\frac{1}{pq} + p^rq^s )
  +e^{-a_i}X (pq+\frac{1}{p^r q^s} ) \right)
  \notag  \\
&=
\sqrt{pq} \sum_{i=1}^{k}  \left(e^{a_i}X^{-1}  (1+\frac{1}{pq} -1 + 
p^rq^s )
  +e^{-a_i}X (1+ pq-1+\frac{1}{p^r q^s} ) \right)
\notag  \\
&=
 (1+ pq) \sum_{i=1}^{k}  \left(\frac{e^{a_i}}{\sqrt{pq} X } 
  + \frac{\sqrt{pq}X}{  e^{a_i}}  \right)
  +
( 1 - p^rq^s ) \sqrt{pq} \sum_{i=1}^{k}  \left(-\frac{e^{a_i}}{X}  + 
\frac{1}{p^r q^s} \frac{X}{e^{a_i}}   \right)  \\
%
%
\sqrt{pq} &\left( e^{a_{k+1}} (e^{-m_{k+1}+b} + e^{-n_{k+1}-b} ) 
+ e^{-a_{k+1}} ( e^{m_{k+1}-b}  +e^{n_{k+1}+b}) \right) \notag \\
&=
\sqrt{pq} \left( \sqrt{pq} (\frac{1}{pq}  + p^r q^s) 
+ \frac{1}{\sqrt{pq}} ( pq  +\frac{1}{p^r q^s}) \right) 
\notag
\\
&=\frac{1}{p^r q^s} (1+p^r q^s)(1+p^{r+1} q^{s+1})
\end{align}
\end{subequations}
and collecting all the pieces leads to
\begin{align}
  Z_{\mathrm{pert}}^{k+1} &=   Z_{\mathrm{pert}}^{k}
  \cdot 
  \PE \Big[ \frac{1}{(1-p)(1-q)} \left( \frac{Q}{1-Q} +\frac{1}{2}\right) 
\notag 
\\
  &\qquad \bigg\{
-(1+pq) \sum_{i=1}^{k}  \left( \frac{1}{\sqrt{pq}} \frac{e^{a_i }}{X}  
+ \sqrt{pq} \frac{X}{e^{a_i }}   \right) 
-(1+pq)
  \notag \\
  &\qquad \qquad 
+\sqrt{pq}  \sum_{l=1}^{k} \left(  
\sqrt{pq} X B e^{-m_l}
+ \frac{1}{\sqrt{pq}X B }  e^{m_l}   
+\sqrt{pq}\frac{X}{B} e^{-n_l}
+ \frac{B}{\sqrt{pq}X}    e^{n_l}
\right) 
  \notag \\
  &\qquad \qquad 
+(1+ pq) \sum_{i=1}^{k}  \left(\frac{e^{a_i}}{\sqrt{pq} X } 
  + \frac{\sqrt{pq}X}{  e^{a_i}}  \right)
  +
( 1 - p^rq^s ) \sqrt{pq} \sum_{i=1}^{k}  \left(-\frac{e^{a_i}}{X}  + 
\frac{1}{p^r q^s} \frac{X}{e^{a_i}}   \right)
  \notag \\
  &\qquad \qquad 
+\frac{1}{p^r q^s} (1+p^r q^s)(1+p^{r+1} q^{s+1})
  \bigg\}
  \Big] \notag \\
  &=   Z_{\mathrm{pert}}^{k}
  \cdot 
  \PE \Big[ \frac{1}{(1-p)(1-q)} \left( \frac{Q}{1-Q} +\frac{1}{2}\right) 
\notag 
\\
  &\qquad \bigg\{
    -(1+pq)
  +\sqrt{pq}  \sum_{l=1}^{k} \left(  
\sqrt{pq} X B e^{-m_l}
+ \frac{1}{\sqrt{pq} X B }  e^{m_l}   
+\sqrt{pq}\frac{X}{B} e^{-n_l}
+ \frac{B}{\sqrt{pq}X}    e^{n_l}
\right)
  \notag \\
  &\qquad \qquad
+
( 1 - p^rq^s ) \sqrt{pq} \sum_{i=1}^{k}  \left(-\frac{e^{a_i}}{X}  + 
\frac{1}{p^r q^s} \frac{X }{e^{a_i}}   \right)
+\frac{1}{p^r q^s} (1+p^r q^s)(1+p^{r+1} q^{s+1})
    \bigg\}
  \Big] \,.
\end{align}
Recalling the contribution \eqref{eq:Goldstone_part} from the Goldstone 
bosons, one formally arrives at
\begin{align}
   Z_{\mathrm{pert}}^{k+1} &=   Z_{\mathrm{pert}}^{k} \cdot Z_G
   \cdot 
   \PE \Big[ \frac{1}{(1-p)(1-q)} \left( \frac{Q}{1-Q} +\frac{1}{2}\right)
   \bigg\{  
   ( 1 - p^rq^s ) \sqrt{pq} \sum_{i=1}^{k}  \left(-\frac{e^{a_i}}{X}  + 
\frac{1}{p^r q^s} \frac{X}{e^{a_i}}   \right)\notag 
\\
&\qquad\qquad 
+\frac{1}{p^r q^s} (1+p^r q^s)(1+p^{r+1} q^{s+1})
-2(1+pq)
   \bigg\} \Big]\\
   Z_{\mathrm{pert}}^{(r,s)\Deff} &=
   \PE \Big[ \frac{1}{(1-p)(1-q)} \left( \frac{Q}{1-Q} +\frac{1}{2}\right)
   \big\{  
-2(1+pq) 
+\frac{1}{p^r q^s} (1+p^r q^s)(1+p^{r+1} q^{s+1}) \notag \\
&\qquad \qquad 
+( 1 - p^rq^s ) \sqrt{pq} \sum_{i=1}^{k}  \left(-\frac{e^{a_i}}{X }  + 
\frac{1}{p^r q^s} \frac{X }{e^{a_i}}   \right)
   \big\} \Big]
   \notag   \\
   &=
   \PE \Big[ \frac{1}{(1-p)(1-q)} \left( \frac{Q}{1-Q} +\frac{1}{2}\right)
   \big\{  \frac{1}{p^r q^s} (1-p^r q^s)(1-p^{r+1} q^{s+1}) \notag \\
&\qquad \qquad 
+( 1 - p^rq^s ) \sqrt{pq} \sum_{i=1}^{k}  \left(-\frac{e^{a_i}}{X}  + 
\frac{1}{p^r q^s} \frac{X }{e^{a_i}}   \right)
   \big\} \Big]
   \notag \\
 &=
   \PE \Big[ \frac{( 1 - p^rq^s )}{(1-p)(1-q)} \left( \frac{Q}{1-Q} 
+\frac{1}{2}\right)
   \left\{  \frac{(1-p^{r+1} q^{s+1})}{p^r q^s} 
+ \sqrt{pq} \sum_{i=1}^{k}  \left(-\frac{e^{a_i}}{X}  + 
\frac{1}{p^r q^s} \frac{X }{e^{a_i}}   \right)
   \right\} \Big]
\end{align}
and \eqref{eq:defect_pert} contains the additional contributions from the codimension 
2 defect, i.e.
\begin{align}
 Z_{\mathrm{pert}}^{k+(r,s)\Deff} =  Z_{\mathrm{pert}}^{k} \cdot  
Z_{\mathrm{pert}}^{(r,s)\Deff}
\end{align}
where the Goldstone mode contribution have been removed. Note that 
$Z_{\mathrm{pert}}^{(r,s)\Deff} =1$ for $(r,s)=(0,0)$.
%
\subsection{Elliptic functions}
The non-perturbative contributions of the 6d partition function on $\T^2 \times \R^4_{\epsilon_1,\epsilon_2}$ equals the infinite sum of 2d elliptic genera. These elliptic genera are naturally composed of elliptic modular forms, whose definitions and properties are summarised in this appendix.
%
\subsubsection{Theta functions}
\label{app:theta_fct}
 There are various different definitions; here, the relevant definitions are 
recalled. Use the conventions $Q= e^{2\pi \im\, \tau}$, 
$x=e^{2\pi \im\, z}$ and the Dedekind eta function \cite[Eq.\ (A.1)]{Benini:2013xpa}
\begin{align}
 \eta(\tau) = Q^{\frac{1}{24}} \prod_{n=1}^{\infty} (1-Q^n) \,.
 \label{eq:def_Dedekind}
\end{align}
Then, the different definitions are as follows:
\begin{subequations}
\label{eq:various_theta}
\begin{alignat}{2}
&\substack{\text{\cite[Eq.\ (A.3)]{Benini:2013xpa}} \\
\text{\cite[Eq.\ (D.6)]{Nekrasov:2012xe}} \\
\text{\cite[Eq.\ (A.10)]{Kim:2014dza}}}
& 
\qquad
 \theta_1(\tau|z) &= -\im Q^{\frac{1}{8}} x^{\frac{1}{2}}
 \prod_{k=1}^\infty (1-Q^k) (1-xQ^k) (1-x^{-1} Q^{k-1})
 \,, \\
 &\scriptstyle{ 
 \text{\cite[Eq.\ (3.7)]{Haghighat:2017vch}}
 } & 
 \qquad
 \overline{\theta}_1(\tau|z) &= \im Q^{\frac{1}{8}} x^{\frac{1}{2}}
 \prod_{k=1}^\infty (1-Q^k) (1-xQ^k) (1-x^{-1} Q^{k-1})
 \,, \\
&\substack{\text{\cite[Eq.\ (D.9)]{Nekrasov:2012xe}} \\
\text{\cite[Eq.\ (3.43)]{Haghighat:2017vch}}}
&
 \qquad
 \widehat{\theta}_1(\tau|z) &= \prod_{n=0}^{\infty} (1-x Q^n) (1-Q^{n+1}) 
(1-x^{-1}Q^{n+1})
 \,, \\
 &\scriptstyle{
 \text{\cite[Eq.\ (A.4)]{Haghighat:2017vch}}
 } & 
 \qquad
 \widetilde{\theta}_1(\tau|z) &= \prod_{j=0}^{\infty} (1-x^{-1}Q^{j+1})(1-x Q^j)
 \,.
\end{alignat}
\end{subequations}
Notice that $\widehat{\theta}_1(\tau|z)$ has been called \emph{basic 
pseudo-elliptic $\theta$-function} in \cite[App.\ D]{Nekrasov:2012xe}.
For this note, the following definition is useful
\begin{align}
\boxed{
  \vartheta_1(\tau|z) \coloneqq \frac{\theta_1(\tau|z)}{Q^{\frac{1}{12}} 
\eta(\tau)} 
\quad \text{such that} \quad
\begin{cases}
&\vartheta_1(\tau|-z) =- \vartheta_1(\tau|z)\\
&\partial_z^k \vartheta_1(\tau|z)= \partial_z^k \theta_1(\tau|z) \\
 &\lim_{\tau \to \im \infty} \vartheta_1(\tau|z) = \im \left(\frac{1}{\sqrt{x}}- 
\sqrt{x} \right)
\end{cases}
\quad, \, x \equiv e^{2\pi \im z} \,.
}
\label{eq:my_theta}
 \end{align}
\paragraph{Comparison.}
The differently defined functions are related as follows:
\begin{subequations}
\label{eq:theta_defs}
\begin{align}
    \overline{\theta}_1(\tau|z) &= -\theta_1(\tau|z)  \,,\\
\widehat{\theta}_1(\tau|z)
&=\frac{x^{\frac{1}{2}}}{\im\, Q^{\frac{1}{8}} 
} \theta_1(\tau|z)
\,, \\
\widetilde{\theta}_1(\tau|z) 
&=
\frac{ x^{\frac{1}{2}} Q^{-\frac{1}{12}}  }{\im\, \eta(\tau)
} \theta_1(\tau|z) \,.
\end{align}
\end{subequations}
\paragraph{Reflection property.}
Consider the shift property following \cite[Eq.\ (A.5)]{Benini:2013xpa}:
\begin{subequations}
 \begin{align}
 \theta_1(\tau|-z)  &= -\theta_1(\tau|z) \,,\\
  \overline{\theta}_1(\tau|-z) &= 
-\overline{\theta}_1(\tau|z)  \,,\\
\widehat{\theta}_1(\tau|-z) &= 
-x^{-1}  \widehat{\theta}_1(\tau|z) \,,\\
\widetilde{\theta}_1(\tau|-z) &= -x^{-1}\widetilde{\theta}_1(\tau|z)
\,.
 \end{align}
Note that the transformation rule for $\widehat{\theta}_1$ agrees with 
\cite[Eq.\ (D.10)]{Nekrasov:2012xe}.
\end{subequations}
\paragraph{Shift properties.}
Next, compute the shift properties following \cite[Eq.\ (A.4)]{Benini:2013xpa} 
for $a,b \in \mathbb{Z}$:
\begin{subequations}
\begin{align}
 \theta_1(\tau| z+a+b\tau) &= (-1)^{a+b} x^{-b}Q^{-\frac{b^2}{2}}  
\theta_1(\tau| z)
\,,
\\
\overline{\theta}_1(\tau| z+a+b\tau) &= (-1)^{a+b} x^{-b}Q^{-\frac{b^2}{2}}  
\overline{\theta}_1(\tau| z)
\,,
\\
\widehat{\theta}_1(\tau| z+a+b\tau) &= 
(-1)^{b} x^{-b}Q^{-\frac{b(b-1)}{2}}  
\widehat{\theta}_1(\tau| z)
\,,
\\
\widetilde{\theta}_1(\tau| z+a+b\tau) &= 
\eta(\tau)
(-1)^{b} x^{-b}Q^{-\frac{b(b-1)}{2}}  
\widetilde{\theta}_1(\tau| z) 
\,.
\end{align}
\label{eq:shift_property_theta}
\end{subequations}
\paragraph{Residue.}
According to \cite[Eq.\ (B.7)]{Benini:2013nda} or \cite[Eq.\ 
(A.7)]{Benini:2013xpa}, the residue at the pole $a+b \tau$ is
\begin{subequations}
\label{eq:integral_theta}
\begin{align}
\frac{1}{2 \pi i}\oint_{u=a+b\tau} \frac{du }{\theta_1(u)} = 
\frac{(-1)^{a+b}e^{i\pi b^2 \tau}}{2\pi \eta^3}
\quad 
\Rightarrow
\quad 
\oint_{u=0} \frac{du }{\theta_1(u)} = 
\frac{i}{ \eta^3} \,,
\end{align}
which implies
\begin{align}
 \oint_{u=0} \frac{du }{\vartheta_1(u)} = 
\frac{iQ^{\frac{1}{12}} \eta }{ \eta^3} =  \frac{iQ^{\frac{1}{12}} }{ 
\eta^2}  
\end{align}
\end{subequations}
for the modified function \eqref{eq:my_theta}.
%
%
\subsubsection{Hierarchy of multiple elliptic gamma functions}
Following for instance \cite[App.\ A]{Haghighat:2017vch}, the definition of the multiple 
elliptic gamma function $G_r(z|\underline{\tau})$ includes 
\begin{align}
 G_0(z|\tau)  = \widetilde{\theta}_1(z,\tau ) 
 \qquad \text{and} \qquad 
 G_1(z|\tau,\sigma) = \Gamma(z,\tau,\sigma) \,,
\end{align}
see \eqref{eq:theta_defs} for the definition of $\widetilde{\theta}_1$.
These functions satisfy the following useful identity
\begin{align}
 G_r(z +\tau_j|\underline{\tau})= G_{r-1}(z|\underline{\tau}^-(j)) 
G_r(z|\underline{\tau}) \,,
\end{align}
such that one finds 
\begin{align}
 \log \widetilde{\theta}_1(z,\tau ) &= \log 
\Gamma(z+\epsilon_1,\tau,\epsilon_1) - \log \Gamma(z,\tau,\epsilon_1)\,.
\label{eq:shift_Gamma}
\end{align}
%
%
\subsection{Conventions for NS-limit}
The NS-limit $\epsilon_2 \to 0$ only yields a finite defect partition function 
if a suitable normalisation is chosen. In addition, the expansion coefficients 
in the $\epsilon_2$ expansion need to be defined.
\subsubsection{Normalised defect partition function}
\label{app:normalised_part_fct}
For the 6d theory with and without a defect, one has the following $q_\phi$ expansions:
\begin{align}
 Z^{\mathrm{6d}} = Z_{\mathrm{pert}}^{\mathrm{6d}} \left( 1 + \sum_{l=1}^\infty Z_{l}^{\mathrm{6d}} 
\right) 
\,, \qquad 
Z^{\mathrm{6d}+\Deff} = Z_{\mathrm{pert}+\Deff}^{\mathrm{6d}} \left( 1 + 
\sum_{l=1}^\infty Z_{l}^{\mathrm{6d}+\Deff} 
\right) \,,
\end{align}
such that the \emph{normalised defect partition function} is defined as
\begin{align}
 \widetilde{Z}^{\mathrm{6d}+\Deff} \coloneqq \frac{Z^{\mathrm{6d}+\Deff}}{ Z^{\mathrm{6d}} } 
 \equiv
  \widetilde{Z}^{\mathrm{6d}+\Deff}_{\mathrm{pert}} 
 \left( 1 + \sum_{l=1}^\infty  \widetilde{Z}^{\mathrm{6d}+\Deff}_l  q_\phi^l \right) 
 \,.
\end{align}
The $q_\phi$ expansion of the normalisation factor reads
\begin{align}
 \frac{1}{ Z^{6d} }
 =
 \frac{1}{Z_{\mathrm{pert}}^{\mathrm{6d}}} \left[ 
 1- Z_{1}^{\mathrm{6d}} \phi  
 - \left(Z_{2}^{\mathrm{6d}}- \left(Z_{1}^{\mathrm{6d}}\right)^2 \right) \phi^2 
 -\left(Z_{3}^{\mathrm{6d}} - 2 Z_{1}^{\mathrm{6d}} Z_{2}^{\mathrm{6d}} +\left(Z_{1}^{\mathrm{6d}} \right)^3 
\right) \phi^3
 +\mathcal{O}(\phi^4)
 \right]
\end{align}
and the standard expansion coefficients of the normalised defect partition function $ \widetilde{Z}^{6d+\Deff}$ are
\begin{subequations}
\begin{align}
\widetilde{Z}^{\mathrm{6d}+\Deff}_{\mathrm{pert}} &=
\frac{Z_{\mathrm{pert}}^{\mathrm{6d}+\Deff}}{ Z_{\mathrm{pert}}^{\mathrm{6d}} } 
\,,
\\
\widetilde{Z}^{\mathrm{6d}+\Deff}_1 &=
Z_{1}^{\mathrm{6d}+\Deff} -Z_{1}^{\mathrm{6d}} 
\,,
\\
\widetilde{Z}^{\mathrm{6d}+\Deff}_2 &=
Z_{2}^{\mathrm{6d}+\Deff} -Z_{2}^{\mathrm{6d}}   
-Z_{1}^{\mathrm{6d}} \left(Z_{1}^{\mathrm{6d}+\Deff} -Z_{1}^{\mathrm{6d}} \right)
\,,
\end{align}
\end{subequations}
and similarly for higher orders in $q_\phi$.
\subsubsection{Notation and expansion coefficients}
Some frequently appearing combinations of Theta functions have the following expansions: 
 \begin{subequations}
 \label{eq:expansion_coeffs}
\begin{alignat}{3}
 \frac{\vartheta_1(2\epsilon_+)}{\vartheta_1(\epsilon_1)\vartheta_1(\epsilon_2)}
 &= \frac{1}{\vartheta^\prime_1(0)} \frac{1}{\epsilon_2} + B^{(0)} + 
\mathcal{O}(\epsilon_2) 
&
\qquad &\text{with } &
B^{(0)} &= \frac{1}{\vartheta^\prime_1(0)} L(\epsilon_1)  \,,
\\
\frac{
\vartheta_1(2\epsilon_+) 
\vartheta_1(s\epsilon_2)
}{
\vartheta_1(\epsilon_1) 
\vartheta_1(\epsilon_2)
}
&= s + A^{(1)} \epsilon_2 +\mathcal{O}(\epsilon_2^2) 
&
\qquad &\text{with } &
A^{(1)} &= s L(\epsilon_1) \,.
\end{alignat}
The $\epsilon_2$ expansion of functions defined in \eqref{eq:def_path_int}, \eqref{eq:residue_calc}, \eqref{eq:defect_term}, and \eqref{eq:Ell_genus_Wilson} are given by
\begin{align}
V_{(0,s)}(u-\epsilon_+) &=1+  V_s^{(1)}(u-\epsilon_+) \cdot \epsilon_2 
+ V_s^{(2)}(u-\epsilon_+) \cdot \epsilon_2^2 
+\mathcal{O}(\epsilon_2^3) \\
\text{with } &\quad 
V_s^{(1)}(u-\epsilon_+) = s L\left( 
u-x-\tfrac{1}{2}\epsilon_1 \right)
\notag \\
&\quad 
V_s^{(2)}(u-\epsilon_+) = \frac{s}{2} 
L\left(u-x-\tfrac{1}{2}\epsilon_1\right)^2
+\frac{s}{2}(s-1) K\left(u-x-\tfrac{1}{2}\epsilon_1\right)
\,,
\notag  \\
Q^\vee(a_i-\epsilon_+) &= Q_{(0)}^\vee(a_i-\epsilon_+) 
+ Q_{(1)}^\vee(a_i-\epsilon_+) \cdot \epsilon_2 
+ \mathcal{O}(\epsilon_2^2) \\
 \text{with } &\quad 
 Q_{(0)}^\vee(a_i-\epsilon_+)  = Q^\vee(a_i-\epsilon_+) 
\big|_{\epsilon_2=0} 
 \notag \\
 &\quad 
 Q_{(1)}^\vee(a_i-\epsilon_+)  =
 Q_{(0)}^\vee(a_i-\epsilon_+)  
 \sum_{k}
 \bigg[
 L(a_j-a_k-\epsilon_1) \notag \\
 &\qquad \qquad \qquad  \qquad 
 -\tfrac{1}{2} L(a_j -\tfrac{1}{2}\epsilon_1 -m_k +b)
 -\tfrac{1}{2} L(a_j -\tfrac{1}{2}\epsilon_1 -n_k -b)
 \bigg]
 \,,
 \notag \\
W(a_i -\epsilon_+) &=1 + W_{(1)}(a_i -\epsilon_+) \cdot \epsilon_2 +W_{(2)}(a_i 
-\epsilon_+) \cdot \epsilon_2^2 + \mathcal{O}(\epsilon_2^3) \\
\text{with } &\quad W_{(1)}(a_i -\epsilon_+) =
L(u-z-\epsilon_1) - L(u-z)
\notag \\
&\quad W_{(2)}(a_i -\epsilon_+) =
\frac{1}{2} \left( K(u-z) - K(u-z-\epsilon_1) \right) \notag \\
&\qquad \qquad \qquad \qquad  +L(u-z-\epsilon_1) \left(L(u-z-\epsilon_1) - L(u-z) \right)
\,,
\notag
 \end{align}
 with $L(\cdot)$, $K(\cdot)$ as defined in \eqref{eq:def_L_and_K}. In addition, for certain relevant combinations one finds
 \begin{align}
 V_{(0,s)} (u_1-\epsilon_+) V_{(0,s)}(u_2-\epsilon_+) -1 &=
 V_{(1)}(u_1-\tfrac{1}{2}\epsilon_1,u_2-\tfrac{1}{2}\epsilon_1) \cdot 
\epsilon_2 \\
 &\qquad +V_{(2)}(u_1-\tfrac{1}{2}\epsilon_1,u_2-\tfrac{1}{2}\epsilon_1) \cdot 
\epsilon_2^2 
+\mathcal{O}(\epsilon_2^3)
\notag \\
\text{with }\quad 
V_{(1)}(u_1-\tfrac{1}{2}\epsilon_1,u_2-\tfrac{1}{2}\epsilon_1) &= 
s \left( 
L\left(u_1-x-\tfrac{1}{2} \epsilon_1\right) 
+ L\left(u_2-x-\tfrac{1}{2} \epsilon_1\right)
\right) 
\notag \\
V_{(2)}(u_1-\tfrac{1}{2}\epsilon_1,u_2-\tfrac{1}{2}\epsilon_1) &= 
\frac{s}{2} \left\{ L(u_1-x-\tfrac{1}{2}\epsilon_1)^2  +  
L(u_2-x-\tfrac{1}{2}\epsilon_1)^2  \right\}
\notag \\
&\qquad + s^2 L(u_1-x-\tfrac{1}{2}\epsilon_1) L(u_2-x-\tfrac{1}{2}\epsilon_1)
\notag \\
&\qquad 
+\frac{s}{2}(s-1) 
\left\{ 
K(u_1-x-\tfrac{1}{2}\epsilon_1) 
+  K(u_2-x-\tfrac{1}{2}\epsilon_1)
\right\}
\notag \\
 W(u_1-\epsilon_+)W(u_2-\epsilon_+) -1 &=
 W_{(1)}(u_1-\tfrac{1}{2}\epsilon_1,u_2-\tfrac{1}{2}\epsilon_1) \cdot 
\epsilon_2 \\
 &\qquad +W_{(2)}(u_1-\tfrac{1}{2}\epsilon_1,u_2-\tfrac{1}{2}\epsilon_1) \cdot 
\epsilon_2^2 
+\mathcal{O}(\epsilon_2^3)
\notag \\
\text{with }\quad 
W_{(1)}(u_1-\tfrac{1}{2}\epsilon_1,u_2-\tfrac{1}{2}\epsilon_1) &= 
 \sum_{J=1}^2
 \left[ 
L\left(u_J-z- \epsilon_1\right) 
- L\left(u_J-z\right)
\right] 
\notag \\
W_{(2)}(u_1-\tfrac{1}{2}\epsilon_1,u_2-\tfrac{1}{2}\epsilon_1) &= 
\sum_{J=1}^2
\bigg[
\frac{1}{2} 
\left(
K(u_J -z)
-K(u_J -z-\epsilon_1)
\right) 
\notag \\
& \quad \quad 
+ L(u_J -z-\epsilon_1)
\left(
L(u_J -z-\epsilon_1)
-
L(u_J -z)
\right)
\bigg]
\notag \\
&\quad \quad 
+L(u_1 -z)L(u_2 -z)
-L(u_1 -z-\epsilon_1)L(u_2 -z)
\notag \\
&\quad \quad 
-L(u_1 -z)L(u_2 -z-\epsilon_1)
+L(u_1 -z-\epsilon_1)L(u_2 -z-\epsilon_1)
\notag \\
 W(a_j -\epsilon_+)W(a_j -\epsilon_+ -\epsilon_\kappa) -1 &=
 W_{(1)}(a_j -\epsilon_+,a_j -\epsilon_+ -\epsilon_\kappa) 
\cdot 
\epsilon_2 \\
 &\qquad +W_{(2)}(a_j -\epsilon_+,a_j -\epsilon_+ -\epsilon_\kappa) \cdot 
\epsilon_2^2 
+\mathcal{O}(\epsilon_2^3)
\; ,\qquad \kappa\in\{1,2\}
\notag \\
\text{with }\quad 
W_{(1)}(a_j -\epsilon_+,a_j -\epsilon_+ -\epsilon_1) &= 
L(a_j - z - 2 \epsilon_1) - L(a_j-z)
 \notag \\
W_{(1)}(a_j -\epsilon_+,a_j -\epsilon_+ -\epsilon_2) &= 
2\left[
L(a_j - z -\epsilon_1) - L(a_j-z)
\right]
 \notag \\
W_{(2)}(a_j -\epsilon_+, a_j -\epsilon_+ -\epsilon_1) &= 
\frac{1}{2} 
\left[ 
K(a_j-z) - K(a_j -z - 2\epsilon_1)
\right]
\notag \\
&\quad \quad 
+L(a_j -z - 2\epsilon_1)
\left[ 
L(a_j -z - 2\epsilon_1)
-L(a_j -z )
\right]
\notag \\
W_{(2)}(a_j -\epsilon_+, a_j -\epsilon_+ -\epsilon_2) &= 
2
\left[ 
K(a_j-z) - K(a_j -z - \epsilon_1)
\right]
\notag \\
&\quad \quad 
+4 L(a_j -z - 2\epsilon_1)
\left[ 
L(a_j -z - \epsilon_1)
-L(a_j -z )
\right]
\notag \\
\widetilde{Q}(a_j-\tfrac{1}{2}\epsilon_1)
&= \frac{1}{\vartheta^\prime_1(0)} Q^\vee_{(0)}(a_j - \tfrac{1}{2}\epsilon_1)
\Bigg\{ 
\sum_i \bigg[
\frac{3}{2} L(a_j -\tfrac{1}{2}\epsilon_1 -m_i +b)
\notag \\
&\qquad 
+\frac{3}{2} L(a_j -\tfrac{1}{2}\epsilon_1 -n_i -b)
-2L(a_j-a_i-\epsilon_1)
\bigg]
-\sum_{i\neq j}
L(a_j -a_i)
\Bigg\}
\notag 
\end{align}
\end{subequations}
%
%
\subsection{Elliptic genera for theory without defect}
\label{app:details_pure}
For the theory without defects of Section \ref{sec:6d_no_defect}, the non-perturbative contributions can be computed via \eqref{eq:ell_genus_k-string}. In this section, the details of the 1 and 2-string calculation are presented. As detailed in \cite{Benini:2013nda,Benini:2013xpa}, the JK-residue prescription requires the choice of an auxiliary vector that determines the poles which contribute to the contour integral. While the final result is independent of the choice made, the individual residues do not have an invariant meaning. For this paper, the auxiliary vector is chosen to be $+1$ on 1-string level and $(1,1)$ on 2-string level.
\subsubsection{1-string}
\label{app:details_pure_k=1}
For the evaluation of the 1-string contribution, the residues of the following poles are relevant:
\begin{align}
\epsilon_++u-a_i=0\,.
\end{align}
Since $Q(u)=\frac{M(u)}{P_0(u)P_0(u+2\epsilon_+)}$, this choice of poles 
corresponds to the zeros of $P_0(u+2\epsilon_+)$.
Using \eqref{eq:integral_theta}, one computes 
\begin{align}
 \oint \diff u \frac{f(u)}{P_0(u+2\epsilon_+)} 
 &= \sum_{i=1}^k \frac{f(u)}{
 \prod_{j\neq i} \vartheta_1(u-a_j+\epsilon_+)}
 \bigg|_{u=a_i -\epsilon_+} 
 \oint_{u=a_i-\epsilon_+}  \frac{\diff u}{\vartheta_1(u-a_i+\epsilon_+)} 
 \notag\\
&=\sum_{i=1}^k \frac{f(u)}{
 \prod_{j\neq i} \vartheta_1(u-a_j+\epsilon_+)}
 \bigg|_{u=a_i -\epsilon_+} 
 \frac{i Q^{\frac{1}{12}} \eta}{\eta^3}
 \notag \\
&=\frac{\im Q^{\frac{1}{12}} \eta}{\eta^3} \sum_{i=1}^k \frac{f(u)}{
P_0^\vee(u+2\epsilon_+) }
 \bigg|_{u=a_i -\epsilon_+} 
=
\frac{\im Q^{\frac{1}{12}} \eta}{\eta^3} \sum_{i=1}^k \frac{f(a_i -\epsilon_+)}{
P_0^\vee(a_i+\epsilon_+) }
\end{align}
where the definitions \eqref{eq:residue_calc} have been used.
With this preparation, the elliptic genus becomes
\begin{align}
 Z_1
&= \oint \frac{\diff u}{(2\pi \im)}
 \left(\frac{2\pi 
\,\eta^3\theta_1(2\epsilon_+)}{    
\theta_1(\epsilon_1)\,\theta_1(\epsilon_2)}
\right) 
Q(u)
=
\frac{ \vartheta_1(2\epsilon_+)}{  
\vartheta_1(\epsilon_1)\,\vartheta_1(\epsilon_2)}
\sum_{i=1}^k Q^\vee(a_i-\epsilon_+)
\,,
\end{align}
using \eqref{eq:my_theta}.
\subsubsection{2-string}
\label{app:details_pure_k=2}
For $l=2$, the elliptic genus becomes
\begin{align}
 Z_2 =\frac{1}{2} \oint \frac{\diff u_1 \diff u_2}{(2\pi \im)^2} 
 \left(
 \frac{2\pi \, \eta^3 \theta_1(2\epsilon_+)
 }{\theta_1(\epsilon_1)\theta_1(\epsilon_2)}
\right)^2
D(u_1-u_2)D(u_2-u_1)
\prod_{p=1}^2
Q(u_p)
\end{align}
and the relevant poles are as follows:
\begin{compactitem}
 \item Both poles originate from $P_0(u_{p}+\epsilon_1+\epsilon_2)$ i.e.
 \begin{align}
  (u_1,u_2)=(a_i-\epsilon_+,a_j-\epsilon_+) \qquad \text{for }i\neq j \,.
 \end{align}
\item One pole from  $P_0(u_{p}+\epsilon_1+\epsilon_2)$ and one from $D(\pm 
(u_1-u_2))$, i.e.\
\begin{align}
\begin{aligned}
 (u_1,u_2)&=(a_m-\epsilon_+,a_m-\epsilon_+ -\epsilon_{1,2}) \qquad  \text{and} 
\\
 (u_1,u_2)&=(a_m-\epsilon_+ -\epsilon_{1,2},a_m-\epsilon_+) \,.
\end{aligned}
\end{align}
\end{compactitem}
In order to compute the residues, the following intermediate results are useful:
\begin{subequations}
\begin{align}
 \oint_{u=-\epsilon_2} \diff u\, f(u) D(u) 
 &=\oint_{u=-\epsilon_2} \diff u\, f(u) 
 \frac{\vartheta_1(u)\vartheta_1(u+\epsilon_1+\epsilon_2)}{
 \vartheta_1(u+\epsilon_1)\vartheta_1(u+\epsilon_2)} 
 \notag \\
&=\frac{i Q^{\frac{1}{12} } \eta}{\eta^3} f(-\epsilon_2)
 \frac{\vartheta_1(-\epsilon_2)\vartheta_1(\epsilon_1)}{
 \vartheta_1(\epsilon_1-\epsilon_2)} 
 \\
\oint_{u=-\epsilon_1} \diff u\, f(u) D(u) 
 &=\oint_{u=-\epsilon_1} \diff u\, f(u) 
 \frac{\vartheta_1(u)\vartheta_1(u+\epsilon_1+\epsilon_2)}{
 \vartheta_1(u+\epsilon_1)\vartheta_1(u+\epsilon_2)} 
 \notag \\
&=\frac{i Q^{\frac{1}{12} } \eta}{\eta^3} f(-\epsilon_1)
 \frac{\vartheta_1(-\epsilon_1)\vartheta_1(\epsilon_2)}{
 \vartheta_1(\epsilon_2-\epsilon_1)} \,,
\end{align}
as well as
\begin{align}
D(\epsilon_1)&=  
\frac{\vartheta_1(\epsilon_1)\vartheta_1(2\epsilon_1+\epsilon_2)}{
 \vartheta_1(2\epsilon_1)\vartheta_1(\epsilon_1+\epsilon_2)} 
 =
 \frac{\vartheta_1(\epsilon_1)\vartheta_1(\epsilon_1+2\epsilon_+)}{
 \vartheta_1(2\epsilon_1)\vartheta_1(2\epsilon_+)} \,,
 \\
 D(\epsilon_2)&=  
\frac{\vartheta_1(\epsilon_2)\vartheta_1(\epsilon_1+2\epsilon_2)}{
 \vartheta_1(\epsilon_1+\epsilon_2)\vartheta_1(2\epsilon_2)}
 =
 \frac{\vartheta_1(\epsilon_2)\vartheta_1(\epsilon_2+2\epsilon_+)}{
 \vartheta_1(2\epsilon_+)\vartheta_1(2\epsilon_2)} 
 \,.
\end{align}
\end{subequations}
Firstly, consider the contributions for 
$(u_1,u_2)=(a_i-\epsilon_+,a_j-\epsilon_+)$
\begin{align}
 Z_2 \supset 
 \frac{1}{2}
\left(
 \frac{  \vartheta_1(2\epsilon_+)
 }{\vartheta_1(\epsilon_1)\vartheta_1(\epsilon_2)}
\right)^2
D(a_i-a_j)D(a_j-a_i)
Q^\vee(a_i-\epsilon_+)
Q^\vee(a_j-\epsilon_+)
\,.
\end{align}
Secondly, both 
$(u_1,u_2)=(a_m-\epsilon_+,a_m-\epsilon_+ -\epsilon_{1})$
and 
$(u_1,u_2)=(a_m-\epsilon_+ -\epsilon_{1},a_m-\epsilon_+)$
yield
\begin{align}
 Z_2&\supset
\frac{1}{2} 
 \frac{  \vartheta_1(2\epsilon_+) \vartheta_1(\epsilon_1+2\epsilon_+)
 }{\vartheta_1(\epsilon_2) \vartheta_1(2\epsilon_-)  \vartheta_1(2\epsilon_1)}
 Q^\vee(a_m-\epsilon_+)
 Q(a_m-\epsilon_+-\epsilon_1)
 \,.
\end{align}
Thirdly, both $(u_1,u_2)=(a_m-\epsilon_+,a_m-\epsilon_+ -\epsilon_{2})$
and $(u_1,u_2)=(a_m-\epsilon_+ -\epsilon_{2},a_m-\epsilon_+)$ yield
\begin{align}
 Z_2&\supset
\frac{1}{2} 
 \frac{ -1\cdot \vartheta_1(2\epsilon_+) \vartheta_1(\epsilon_2+2\epsilon_+)
 }{\vartheta_1(\epsilon_1) \vartheta_1(2\epsilon_-)  \vartheta_1(2\epsilon_2)}
Q^\vee(a_m-\epsilon_+)
 Q(a_m-\epsilon_+-\epsilon_2)
 \,.
\end{align}
Summing up all the individual contributions yields
\begin{align}
 Z_2 
&=
\left(
 \frac{  \vartheta_1(2\epsilon_+)
 }{\vartheta_1(\epsilon_1)\vartheta_1(\epsilon_2)}
\right)^2
\sum_{1 \leq i< j\leq k}
D(a_i-a_j)D(a_j-a_i)
Q^\vee(a_i-\epsilon_+)
Q^\vee(a_j-\epsilon_+)
 \\
&+
\frac{  \vartheta_1(2\epsilon_+)
 }{ \vartheta_1(2\epsilon_-)  }
 \sum_{j=1}^k
 Q^\vee(a_j-\epsilon_+)
\bigg[ 
\frac{ \vartheta_1(\epsilon_1+2\epsilon_+)
 }{ \vartheta_1(\epsilon_2) \vartheta_1(2\epsilon_1)  }
 Q(a_j-\epsilon_+-\epsilon_1)
-
\frac{ \vartheta_1(\epsilon_2+2\epsilon_+)
 }{ \vartheta_1(\epsilon_1) \vartheta_1(2\epsilon_2)}
 Q(a_j-\epsilon_+-\epsilon_2)
\bigg]\notag
\end{align}
where the notation \eqref{eq:residue_calc} has been used.
%
%
\subsection{Elliptic genera for theory with codimension 2 defect}
\label{app:details_codim=2}
In Section \ref{sec:codim_2}, the theory with codiemnsion 2 defect is introduced. The non-perturbative contributions are computed via \eqref{eq:elliptic_genus_codum2}, and in this section the 1 and 2-string results are detailed. Again, choice of the auxiliary vector in the JK-residue is $+1$ on 1-string level and $(1,1)$ on 2-string level.
\subsubsection{1-string}
\label{app:details_codim=2_k=1}
The 1-string elliptic genus is given by
\begin{align}
 Z_{1}^{(0,s)\Deff}
 = \oint \frac{\diff u}{2\pi \im} \,
 Z_{\mathrm{1-loop}}^{(0,s)\Deff}(k,1)
 \equiv 
 \oint \frac{\diff u}{2\pi \im} \,
  Z_{\mathrm{1-loop}}(k,1)
\cdot 
V_{(0,s)}(u)
\end{align}
and the contour integral is evaluated by selecting the residues of the following poles:
\begin{itemize}
 \item $u=a_i -\epsilon_+ $ for $i=1,\ldots,k$
 \begin{align}
   Z_{1}^{(0,s)\Deff} 
&\supset \frac{ \vartheta_1(2\epsilon_+)}{  
\vartheta_1(\epsilon_1)\,\vartheta_1(\epsilon_2)}
\sum_{i=1}^k 
\left(
Q^\vee(a_i-\epsilon_+)
\cdot
V_{(0,s)}(a_i-\epsilon_+)
\right)
 \end{align}
 \item $u=x$
 \begin{align}
   Z_{1}^{(0,s)\Deff} &\supset
\frac{\vartheta_1(2\epsilon_+)\vartheta_1(s\epsilon_2 )}{ 
\vartheta_1(\epsilon_1)\vartheta_1(\epsilon_2)
}
Q(x)
 \end{align}
\end{itemize}
such that the elliptic genus for $l=1$ reads
\begin{align}
  Z_{1}^{(0,s)\Deff} 
&=
\frac{ \vartheta_1(2\epsilon_+)}{  
\vartheta_1(\epsilon_1)\,\vartheta_1(\epsilon_2)}
\left[
\sum_{i=1}^k \bigg(
Q^\vee(a_i-\epsilon_+)
\cdot
V_{(0,s)}(a_i-\epsilon_+)
\bigg) 
+
\vartheta_1(s\epsilon_2 )
\cdot 
Q(x)
\right]
\,. \notag 
\end{align}
Following Section \ref{app:normalised_part_fct}, the normalised 1-string contribution in the NS-limit reads
\begin{align}
 \widetilde{Z}_{1}^{(0,s)\Deff}  &= 
 Z_{1}^{(0,s)\Deff}  - Z_{1}  \notag \\
\lim_{\epsilon_2\to0} \widetilde{Z}_{1}^{(0,s)\Deff}
&=
\lim_{\epsilon_2\to0} \bigg\{
\frac{ \vartheta_1(2\epsilon_+)}{  
\vartheta_1(\epsilon_1)\,\vartheta_1(\epsilon_2)}
\sum_{i=1}^k \bigg(
Q^\vee(a_i-\epsilon_+)
\cdot
\left[
V_{(0,s)}(a_i-\epsilon_+)
-1
\right]
\bigg) 
\notag \\
&\qquad\qquad  +
\frac{\vartheta_1(2\epsilon_+)\vartheta_1(s\epsilon_2 
)}{ 
\vartheta_1(\epsilon_1)\vartheta_1(\epsilon_2)
}
Q(x)
\bigg\}
\,.
\end{align}
To further evaluate the limit, consider 
\begin{subequations}
\begin{align}
 \lim_{\epsilon_2\to0}\frac{V_{(0,s)}(a_i-\epsilon_+)
-1}{\vartheta_1(\epsilon_2)} &= 
s\frac{1}{\vartheta_1^\prime(0)}
L(a_i-x-\tfrac{1}{2}\epsilon_1)
\,,
\notag  \\
 \lim_{\epsilon_2\to0} \frac{\vartheta_1(s\epsilon_2)}{\vartheta_1(\epsilon_2)} 
&= s
 \qquad 
 \text{and}
 \qquad
 \lim_{\epsilon_2\to0} 
\frac{\vartheta_1(2\epsilon_+)}{\vartheta_1(\epsilon_1)}=1 \,,
\end{align}
\end{subequations}
such that 
\begin{align}
 \lim_{\epsilon_2\to0} \widetilde{Z}_{1}^{(0,s)\Deff}
&=
\frac{ s}{  
\vartheta_1^\prime(0)}
\sum_{i=1}^k \bigg(
Q_{(0)}^\vee(a_i-\tfrac{1}{2}\epsilon_1)
\cdot
L(a_i-x-\tfrac{1}{2}\epsilon_1)
\bigg)
+s \cdot
Q_{(0)}(x)
\\
&= s \cdot \lim_{\epsilon_2\to0} \widetilde{Z}_{1}^{(0,1)\Deff}
\,,
\notag
\end{align}
using the notation \eqref{eq:expansion_coeffs}.
Therefore, the $(0,s)$ defect part is the product of $s$ $(0,1)$ defect 
contributions.
%
%
\subsubsection{2-string}
\label{app:details_codim=2_k=2}
Consider the $l=2$ elliptic genus with defect given by
\begin{align}
 Z_2^{(0,s)\Deff} &=\frac{1}{2} \oint \frac{\diff u_1 \diff u_2}{(2\pi 
\im)^2} 
 \left(
 \frac{2\pi \, \eta^3 \theta_1(2\epsilon_+)
 }{\theta_1(\epsilon_1)\theta_1(\epsilon_2)}
\right)^2
D(u_1-u_2)D(u_2-u_1)
\prod_{p=1}^2
Q(u_p)
V_{(0,s)}(u_p)
\end{align}
and the relevant poles can be split into poles that come from the theory 
without defect such as: 
\begin{compactitem}
 \item Both poles originate from $P_0(u_{p}+\epsilon_1+\epsilon_2)$ i.e.
 \begin{align}
  (u_1,u_2)=(a_i-\epsilon_+,a_j-\epsilon_+) \qquad \text{for }i\neq j \,.
 \end{align}
\item One pole from  $P_0(u_{p}+\epsilon_1+\epsilon_2)$ and one from $D(\pm 
(u_1-u_2))$, i.e.\
\begin{align}
\begin{aligned}
 (u_1,u_2)&=(a_m-\epsilon_+,a_m-\epsilon_+ -\epsilon_{1,2}) \qquad  \text{and} 
\\
 (u_1,u_2)&=(a_m-\epsilon_+ -\epsilon_{1,2},a_m-\epsilon_+) \,.
\end{aligned}
\end{align}
\end{compactitem}
In addition, there are new poles from the defect part. These are
\begin{compactitem}
 \item One pole from  $P_0(u_{p}+\epsilon_1+\epsilon_2)$  and one from 
$V_{(0,s)}(u_p)$, i.e.\
\begin{align}
 (u_1,u_2) = (a_m-\epsilon_+,x) 
 \qquad \text{and}\qquad 
 (u_1,u_2) = (x,a_m-\epsilon_+) \,.
\end{align}
 \item One pole from  $D(\pm (u_1-u_2))$  and one from 
$V_{(0,s)}(u_p)$, i.e.\
\begin{align}
 (u_1,u_2) = (x,x-\epsilon_{1,2})
 \qquad \text{and}\qquad 
 (u_1,u_2) = (x-\epsilon_{1,2},x) \,.
\end{align}
\end{compactitem}
Now, one can work out the residues for the individual poles as before:
Firstly, consider the contributions for 
$(u_1,u_2)=(a_i-\epsilon_+,a_j-\epsilon_+)$
\begin{align}
 Z_2^{(0,s)\Deff} &\supset 
 \frac{1}{2}
\left(
 \frac{  \vartheta_1(2\epsilon_+)
 }{\vartheta_1(\epsilon_1)\vartheta_1(\epsilon_2)}
\right)^2
D(a_i-a_j)D(a_j-a_i) 
\notag \\
&\quad \cdot
Q^\vee(a_i-\epsilon_+)
Q^\vee(a_j-\epsilon_+)
V_{(0,s)}(a_i-\epsilon_+)
V_{(0,s)}(a_j-\epsilon_+) \,.
\end{align}
Secondly, both 
$(u_1,u_2)=(a_m-\epsilon_+,a_m-\epsilon_+ -\epsilon_{1})$
and 
$(u_1,u_2)=(a_m-\epsilon_+ -\epsilon_{1},a_m-\epsilon_+)$
yield
\begin{align}
 Z_2^{(0,s)\Deff}&\supset
\frac{1}{2} 
 \frac{  \vartheta_1(2\epsilon_+) \vartheta_1(\epsilon_1+2\epsilon_+)
 }{\vartheta_1(\epsilon_2) \vartheta_1(2\epsilon_-)  \vartheta_1(2\epsilon_1)}
 Q^\vee(a_m-\epsilon_+)
 Q(a_m-\epsilon_+-\epsilon_1)
\notag \\
&\qquad \cdot
V_{(0,s)}(a_m-\epsilon_+)
V_{(0,s)}(a_m-\epsilon_+ -\epsilon_{1})
\end{align}
Thirdly, both $(u_1,u_2)=(a_m-\epsilon_+,a_m-\epsilon_+ -\epsilon_{2})$
and $(u_1,u_2)=(a_m-\epsilon_+ -\epsilon_{2},a_m-\epsilon_+)$ yield
\begin{align}
 Z_2^{(0,s)\Deff}
&\supset \frac{1}{2} 
 \frac{ -1\cdot \vartheta_1(2\epsilon_+) \vartheta_1(\epsilon_2+2\epsilon_+)
 }{\vartheta_1(\epsilon_1) \vartheta_1(2\epsilon_-)  \vartheta_1(2\epsilon_2)}
Q^\vee(a_m-\epsilon_+)
Q(a_m-\epsilon_+-\epsilon_2)
\notag \\
&\qquad \cdot
V_{(0,s)}(a_m-\epsilon_+)
V_{(0,s)}(a_m-\epsilon_+ -\epsilon_{2})
\end{align}
Fourthly, both $(u_1,u_2) = (a_m-\epsilon_+,x)$ 
and  $(u_1,u_2) = (x,a_m-\epsilon_+)$ yield
\begin{align}
 Z_2^{(0,s)\Deff}
&\supset
\frac{1}{2}
\left(
\frac{\vartheta_1(2\epsilon_+)}{
\vartheta_1(\epsilon_1)
\vartheta_1(\epsilon_2)}
\right)^2
D(a_m-x-\epsilon_+)
D(x+\epsilon_+-a_m)
\notag \\
&\qquad \cdot
Q^\vee(a_m-\epsilon_+)
Q(x)
V_{(0,s)}(a_m-\epsilon+)
\vartheta_1(s \epsilon_2)
\end{align}
Fifthly, both $(u_1,u_2) = (x,x-\epsilon_{1})$
and  $(u_1,u_2) = (x-\epsilon_{1},x)$
 \begin{align}
 Z_2^{(0,s)\Deff}
&\supset
\frac{1}{2}
\frac{
\vartheta_1(2\epsilon_+)
\vartheta_1(\epsilon_1+2\epsilon_+)}{
\vartheta_1(2\epsilon_-)
\vartheta_1(\epsilon_2)
 \vartheta_1(2\epsilon_1)}
\cdot 
Q(x)
Q(x-\epsilon_1)
V_{(0,s)}(x-\epsilon_1) \vartheta_1(s\epsilon_2)
\end{align}
Lastly, both $(u_1,u_2) = (x,x-\epsilon_{2})$
and  $(u_1,u_2) = (x-\epsilon_{2},x)$ yield
\begin{align}
 Z_2^{(0,s)\Deff}
&\supset
-\frac{1}{2}
\frac{\vartheta_1(2\epsilon_+)
\vartheta_1(\epsilon_2+2\epsilon_+)
}{
\vartheta_1(\epsilon_1)
\vartheta_1(2\epsilon_-)
\vartheta_1(2\epsilon_2)
}
\cdot 
Q(x)
Q(x-\epsilon_2)
V_{(0,s)}(x-\epsilon_2) \vartheta_1(s\epsilon_2)
\end{align}
Summing up all the individual contributions leads to 
\begin{align}
 Z_2^{(0,s)\Deff}
&=
\left(
 \frac{  \vartheta_1(2\epsilon_+)
 }{\vartheta_1(\epsilon_1)\vartheta_1(\epsilon_2)}
\right)^2
\sum_{1\leq i< j \leq k}
D(a_i-a_j)D(a_j-a_i)  \\
&\qquad \qquad \qquad \qquad \qquad  \cdot
Q^\vee(a_i-\epsilon_+)
Q^\vee(a_j-\epsilon_+)
V_{(0,s)}(a_i-\epsilon_+)
V_{(0,s)}(a_j-\epsilon_+) 
\notag \\
&+
 \frac{  \vartheta_1(2\epsilon_+) 
 }{ \vartheta_1(2\epsilon_-) }
 \sum_{j=1}^k
 Q^\vee(a_j-\epsilon_+)
V_{(0,s)}(a_j-\epsilon_+)
\notag \\
&\qquad\qquad \qquad  \cdot
\bigg[
\frac{\vartheta_1(\epsilon_1+2\epsilon_+)}{
\vartheta_1(\epsilon_2)
\vartheta_1(2\epsilon_1)
}
Q(a_j-\epsilon_+-\epsilon_1)
V_{(0,s)}(a_j-\epsilon_+ -\epsilon_{1})
\notag
\\
&\qquad \qquad \qquad \qquad \qquad -
\frac{\vartheta_1(\epsilon_2+2\epsilon_+)}{
\vartheta_1(\epsilon_1)
\vartheta_1(2\epsilon_2)
}
Q(a_j-\epsilon_+-\epsilon_2)
V_{(0,s)}(a_j-\epsilon_+ -\epsilon_{2})
\bigg]
\notag \\
&+
\left(
\frac{\vartheta_1(2\epsilon_+)}{
\vartheta_1(\epsilon_1)
\vartheta_1(\epsilon_2)}
\right)^2
\vartheta_1(s \epsilon_2)
\sum_{j=1}^k
D(a_j-x-\epsilon_+)
D(x+\epsilon_+-a_j)
\notag \\
&\qquad \qquad \qquad \qquad  \cdot
Q^\vee(a_j-\epsilon_+)
Q(x)
V_{(0,s)}(a_j-\epsilon+)
\notag \\
&+
\frac{
\vartheta_1(2\epsilon_+)}{
\vartheta_1(2\epsilon_-)
}
\cdot 
Q(x)
\vartheta_1(s\epsilon_2)
\cdot 
\bigg[
 \frac{
\vartheta_1(\epsilon_1+2\epsilon_+)}{
\vartheta_1(\epsilon_2)
 \vartheta_1(2\epsilon_1)}
 Q(x-\epsilon_1)
V_{(0,s)}(x-\epsilon_1) 
\notag \\
&\qquad \qquad \qquad \qquad \qquad \qquad 
-
 \frac{
\vartheta_1(\epsilon_2+2\epsilon_+)}{
\vartheta_1(\epsilon_1)
 \vartheta_1(2\epsilon_2)}
 Q(x-\epsilon_2)
V_{(0,s)}(x-\epsilon_2)
\bigg]
\,.
\notag
\end{align}
Next, consider the normalised 2-string elliptic genus, see Appendix \ref{app:normalised_part_fct},
\begin{align}
 \widetilde{Z}_{2}^{(0,s)\Deff}  &= 
 Z_{2}^{(0,s)\Deff}  - Z_{2} 
 -  Z_{1} \left(  Z_{1}^{(0,s)\Deff}  - Z_{1} \right) 
 \qquad 
 \text{and }
 Z_{2}^{\mathrm{aux}} = Z_{2}^{(0,s)\Deff}  - Z_{2} \,.
 \end{align}
Firstly, focus on the 1-string contributions
\begin{align}
 Z_{1}^{(0,s)\Deff}  - Z_{1} 
 = \widetilde{Z}_1 \big|_{\mathrm{fin}} 
 + \widetilde{Z}_1 \big|_{\epsilon_2} \cdot \epsilon_2
 + \mathcal{O}(\epsilon_2^2)
\end{align}
with $\epsilon_2$ expansion coefficients
\begin{align}
 \widetilde{Z}_1 \big|_{\mathrm{fin}} 
 &=
 \frac{1}{\vartheta^\prime_1(0)}
 \sum_j 
 Q^\vee_{(0)} \left( a_j -\tfrac{1}{2} \epsilon_1 \right)
 V_s^{(1)}\left( a_j -\tfrac{1}{2} \epsilon_1 \right)
 +s Q_{(0)}(x)
 \,,
 \notag \\
\widetilde{Z}_1 \big|_{\epsilon_2}
&= 
\frac{1}{\vartheta^\prime_1(0)}
\sum_j Q_{(0)}^\vee \left( a_j -\tfrac{1}{2}\epsilon_1 \right)
V_s^{(2)}\left( a_j -\tfrac{1}{2}\epsilon_1 \right) 
+ \frac{1}{\vartheta^\prime_1(0)}
\sum_j Q_{(1)}^\vee \left( a_j -\tfrac{1}{2}\epsilon_1 \right)
V_s^{(1)}\left( a_j -\tfrac{1}{2}\epsilon_1 \right) 
\notag \\
&+ B^{(0)}
\sum_j Q_{(0)}^\vee \left( a_j -\tfrac{1}{2}\epsilon_1 \right)
V_s^{(1)}\left( a_j -\tfrac{1}{2}\epsilon_1 \right) 
+ A^{(1)} Q_{(0)}(x) + s Q_{(1)}(x)
\,.
\notag
\end{align}
Secondly, consider pure 2-string contributions
\begin{align}
  Z_{k=2}^{\mathrm{aux}} 
  &= I_1 + I_2 +I_3+I_4 
\end{align}
with the following four parts:
\begin{align}
I_1&=
\left(
 \frac{  \vartheta_1(2\epsilon_+)
 }{\vartheta_1(\epsilon_1)\vartheta_1(\epsilon_2)}
\right)^2
\sum_{1\leq i< j \leq k}
D(a_i-a_j)D(a_j-a_i)  
\notag \\
&\qquad \qquad \qquad \qquad \qquad \cdot
Q^\vee(a_i-\epsilon_+)
Q^\vee(a_j-\epsilon_+)
\left[V_{(0,s)}(a_i-\epsilon_+)
V_{(0,s)}(a_j-\epsilon_+) 
-1
\right]
\,,
\notag \\
I_2 &=
 \frac{  \vartheta_1(2\epsilon_+) 
 }{ \vartheta_1(2\epsilon_-) }
 \sum_{j=1}^k
 Q^\vee(a_j-\epsilon_+)
\cdot
\bigg[
\frac{\vartheta_1(\epsilon_1+2\epsilon_+)}{
\vartheta_1(\epsilon_2)
\vartheta_1(2\epsilon_1)
}
Q(a_j-\epsilon_+-\epsilon_1)
\left[
V_{(0,s)}(a_j-\epsilon_+)
V_{(0,s)}(a_j-\epsilon_+ -\epsilon_{1})
-1\right]
\notag
\\
&\qquad \qquad  \qquad \qquad  \qquad \qquad  -
\frac{\vartheta_1(\epsilon_2+2\epsilon_+)}{
\vartheta_1(\epsilon_1)
\vartheta_1(2\epsilon_2)
}
Q(a_j-\epsilon_+-\epsilon_2)
\left[
V_{(0,s)}(a_j-\epsilon_+)
V_{(0,s)}(a_j-\epsilon_+ -\epsilon_{2})
-1
\right]
\bigg]
\,,
\notag \\
I_3 &=
\left(
\frac{\vartheta_1(2\epsilon_+)}{
\vartheta_1(\epsilon_1)
\vartheta_1(\epsilon_2)}
\right)^2
\vartheta_1(s \epsilon_2)
\sum_{j=1}^k
D(a_j-x-\epsilon_+)
D(x+\epsilon_+-a_j)
Q^\vee(a_j-\epsilon_+)
Q(x)
V_{(0,s)}(a_j-\epsilon_+)
\,,
\notag \\
I_4 &=
\frac{
\vartheta_1(2\epsilon_+)}{
\vartheta_1(2\epsilon_-)
}
\vartheta_1(s\epsilon_2)
\cdot 
Q(x)
\cdot 
\bigg[
 \frac{
\vartheta_1(\epsilon_1+2\epsilon_+)}{
\vartheta_1(\epsilon_2)
 \vartheta_1(2\epsilon_1)}
Q(x-\epsilon_1) 
V_{(0,s)}(x-\epsilon_1) 
\notag \\
&\qquad \qquad \qquad \qquad \qquad \qquad -
 \frac{
\vartheta_1(\epsilon_2+2\epsilon_+)}{
\vartheta_1(\epsilon_1)
 \vartheta_1(2\epsilon_2)}
Q(x-\epsilon_2) 
V_{(0,s)}(x-\epsilon_2)
\bigg]
\,.
\notag
\end{align}
The $\epsilon_2$ expansion is defined as 
\begin{align}
 I_\mu = 
  I_\mu \big|_{(\frac{1}{\epsilon_2})^2}  \cdot \frac{1}{\epsilon_2^2}
  + I_\mu \big|_{\frac{1}{\epsilon_2}}  \cdot \frac{1}{\epsilon_2}
  + I_\mu \big|_{\mathrm{fin}}
  + I_\mu \big|_{\epsilon_2 } \cdot \epsilon_2 
  +\mathcal{O}(\epsilon_2^2) \,.
\end{align}
The inspection of the most singular terms reveals 
\begin{align}
  I_\mu \big|_{(\frac{1}{\epsilon_2})^2} =0 \;, \; \forall \mu 
  \qquad  \Rightarrow \qquad 
  \left( Z_{k=2}^{(0,s)\Deff}  - Z_{k=2} \right) 
\big|_{(\frac{1}{\epsilon_2})^2} =0 \,,
\end{align}
which is required to vanish by consistency.
The less singular expansion coefficients are given by
\begin{align}
 I_1 \big|_{\frac{1}{\epsilon_2}} &= 
 \left( \frac{1}{\vartheta^\prime_1(0)} \right)^2
 \sum_{i<j} 
 Q_{(0)}^\vee \left(a_i -\tfrac{1}{2} \epsilon_1 \right)
 Q_{(0)}^\vee \left(a_j -\tfrac{1}{2} \epsilon_1 \right)
 V^{(1)} \left( a_i -\tfrac{1}{2}\epsilon_1,a_j -\tfrac{1}{2}\epsilon_1 \right)
 \notag \\
 I_1 \big|_{\mathrm{fin}} &= 
 \left( \frac{1}{\vartheta^\prime_1(0)} \right)^2
 \sum_{i<j} 
 Q_{(0)}^\vee \left(a_i -\tfrac{1}{2} \epsilon_1 \right)
 Q_{(0)}^\vee \left(a_j -\tfrac{1}{2} \epsilon_1 \right)
 V^{(2)} \left( a_i -\tfrac{1}{2}\epsilon_1,a_j -\tfrac{1}{2}\epsilon_1 \right) 
 \notag \\
&+\left( \frac{1}{\vartheta^\prime_1(0)} \right)^2
 \sum_{i<j} 
 \left\{ 
 Q_{(0)}^\vee \left(a_i -\tfrac{1}{2} \epsilon_1 \right)
 Q_{(1)}^\vee \left(a_j -\tfrac{1}{2} \epsilon_1 \right)
 +
 Q_{(1)}^\vee \left(a_i -\tfrac{1}{2} \epsilon_1 \right)
 Q_{(0)}^\vee \left(a_j -\tfrac{1}{2} \epsilon_1 \right)
 \right\}
 V^{(1)} \left( a_i -\tfrac{1}{2}\epsilon_1,a_j -\tfrac{1}{2}\epsilon_1 \right)
 \notag \\
&+\left( \frac{1}{\vartheta^\prime_1(0)} \right)^2
 \sum_{i<j} 
 \left\{ 
 D^{(1)}(a_i -\epsilon_+) 
 + D^{(1)}(a_j -\epsilon_+) 
 \right\}
 Q_{(0)}^\vee \left(a_i - \epsilon_+ \right)
 Q_{(0)}^\vee \left(a_j - \epsilon_+ \right)
 V^{(1)} \left( a_i -\tfrac{1}{2}\epsilon_1,a_j -\tfrac{1}{2}\epsilon_1 \right) 
\notag \\
&+\frac{2 B^{(0)}}{\vartheta^\prime_1(0)}
 \sum_{i<j} 
 Q_{(0)}^\vee \left(a_i - \epsilon_+ \right)
 Q_{(0)}^\vee \left(a_j -\epsilon_+ \right)
 V^{(1)} \left( a_i -\epsilon_+,a_j -\epsilon_+ \right)
 \notag 
 \\
  I_2 \big|_{\frac{1}{\epsilon_2}} &= 
  \left( \frac{1}{\vartheta^\prime_1(0)} \right)^2
  \sum_j \left( Q_{(0)}^\vee \left(a_j -\tfrac{1}{2}\epsilon_1 \right) 
\right)^2 V_s^{(1)} (a_j -\tfrac{1}{2}\epsilon_1)
   \notag \\
 I_2 \big|_{\mathrm{fin}} &= 
 \frac{1}{\vartheta^\prime_1(0)} 
 \sum_j
 Q_{(0)}^\vee \left(a_j -\tfrac{1}{2}\epsilon_1 \right)
 Q^{(0)} \left(a_j -\tfrac{3}{2}\epsilon_1 \right)
 \left( 
 V_s^{(1)}\left(a_j -\tfrac{1}{2}\epsilon_1 \right) 
 + V_s^{(1)}\left(a_j -\tfrac{3}{2}\epsilon_1 \right)
 \right) 
\notag  \\
&- \frac{1}{\vartheta^\prime_1(0)}
\sum_j
 Q_{(0)}^\vee \left(a_j -\tfrac{1}{2}\epsilon_1 \right)
 \widetilde{Q}\left(a_j -\tfrac{1}{2}\epsilon_1 \right)
 V_s^{(1)}\left(a_j -\tfrac{1}{2}\epsilon_1 \right)
 \notag \\
&+2 \frac{B^{(0)}}{\vartheta^\prime_1(0)}
\sum_j
\left( Q_{(0)}^\vee \left(a_j -\tfrac{1}{2}\epsilon_1 \right)
\right)^2
 V_s^{(1)}\left(a_j -\tfrac{1}{2}\epsilon_1 \right)
 \notag \\
&+ \left( \frac{1}{\vartheta^\prime_1(0)} \right)^2
\sum_j
\left( 
2 \frac{\vartheta^\prime_1(\epsilon_1)}{\vartheta_1(\epsilon_1)}
Q_{(0)}^\vee \left(a_j -\tfrac{1}{2}\epsilon_1 \right)
+
Q_{(1)}^\vee \left(a_j -\tfrac{1}{2}\epsilon_1 \right)
\right)
Q_{(0)}^\vee \left(a_j -\tfrac{1}{2}\epsilon_1 \right)
V_s^{(1)} \left(a_j-\tfrac{1}{2}\epsilon_1 \right)
\notag \\
&+ \frac{1}{2 (\vartheta^\prime_1(0))^2} 
\sum_j 
\left( Q_{(0)}^\vee \left(a_j -\tfrac{1}{2}\epsilon_1 \right) \right)^2
V^{(2)}\left( a_j -\epsilon_+,a_j -\epsilon_+ -\epsilon_2 \right)
\notag 
\\
  I_3 \big|_{\frac{1}{\epsilon_2}} &= 
  \frac{1}{\vartheta^\prime_1(0)} 
  s Q_{(0)}(x)
  \sum_j 
  Q_{(0)}^\vee \left(a_j -\tfrac{1}{2}\epsilon_1 \right)
   \notag \\
 I_3 \big|_{\mathrm{fin}} &= 
 \frac{1}{\vartheta^\prime_1(0)} 
 s Q_{(0)}(x)
 \sum_j
 \left[ 
 D^{(1)} (a_j -x - \tfrac{1}{2}\epsilon_1)
 +  D^{(1)} (x + \tfrac{1}{2}\epsilon_1 - a_j)
 \right]
 Q^\vee_{(0)}\left( a_j - \tfrac{1}{2} \epsilon_1 \right)
 \notag \\
&+ \frac{1}{\vartheta^\prime_1(0)}
s
\sum_j
\left[ 
Q^\vee_{(1)}\left( a_j - \tfrac{1}{2} \epsilon_1 \right) 
Q_{(0)}(x)
+
Q^\vee_{(0)}\left( a_j - \tfrac{1}{2} \epsilon_1 \right) 
Q_{(1)}(x)
+
Q^\vee_{(0)}\left( a_j - \tfrac{1}{2} \epsilon_1 \right) 
Q_{(0)}(x)
V_{s}^{(1)}(a_j-\tfrac{1}{2}\epsilon_1)
\right]
\notag \\
&+ 2 B^{(0)} s
\sum_j
Q^\vee_{(0)}\left( a_j - \tfrac{1}{2} \epsilon_1 \right) 
Q_{(0)}(x)
\notag
\\
  I_4 \big|_{\frac{1}{\epsilon_2}} &= 0
   \notag \\
 I_4 \big|_{\mathrm{fin}} &= 
 s Q_{(0)}(x) 
 \left[ 
 Q_{(0)}(x-\epsilon_1)
 + \frac{s-1}{2} Q_{(0)}(x) 
 \right] 
 \,.
 \notag
\end{align}
Another consistency check is given by the vanishing of the $\tfrac{1}{\epsilon_2}$ terms if one considers the pure 2-string terms together with the product of the 1-string contributions. Explicitly, one finds
\begin{align}
 \left( Z_{2}^{(0,s)\Deff} - Z_2 \right)\big|_{\frac{1}{\epsilon_2}}
- Z_1 \big|_{\frac{1}{\epsilon_2}} \cdot  \left( Z_{1}^{(0,s)\Deff} - 
Z_1 \right) \big|_{\mathrm{fin}} =0 \,,
\end{align}
as expected. Recalling the notation \eqref{eq:expansion_coeffs}, the full normalised 
2-string elliptic genus for the codimension 2 defect in the 
NS-limit is given by
\begin{align}
  \widetilde{Z}_{l=2}^{(0,s)\Deff}  &= 
  \frac{s(s+1)}{2}   
  \sum_{j=1}^k 
  \left( 
  \frac{Q_{(0)}^\vee(a_j - \tfrac{\epsilon_1}{2})}{
  \vartheta_1^\prime(0)}
  L(a_j -x - \tfrac{\epsilon_1}{2}) 
  \right)^2
  \notag \\
&\qquad 
-\frac{s}{2} 
\sum_{j=1}^k
\left( 
\frac{Q_{(0)}^\vee(a_j - \tfrac{\epsilon_1}{2}) }{
\vartheta_1^\prime(0)}
\right)^2
 K(a_j -x - \tfrac{\epsilon_1}{2}) 
 \notag \\
&\qquad 
+\frac{s^2}{2}
\sum_{\substack{i,j=1\\ i\neq j}}^k
\frac{  Q_{(0)}^\vee(a_i - \tfrac{\epsilon_1}{2})}{
\vartheta_1^\prime(0)}
\frac{ Q_{(0)}^\vee(a_j - \tfrac{\epsilon_1}{2})}{
\vartheta_1^\prime(0)}
  L(a_i -x - \tfrac{\epsilon_1}{2})
  L(a_j -x - \tfrac{\epsilon_1}{2}) 
  \notag \\
&\qquad 
+2 s \cdot L(\epsilon_1) 
\sum_{j=1}^k \left(
\frac{Q_{(0)}^\vee(a_j - \tfrac{\epsilon_1}{2}) }{
\vartheta_1^\prime(0)}
\right)^2
L(a_j -x- \tfrac{\epsilon_1}{2})
\notag \\
&\qquad
+ s  
\sum_{\substack{i,j=1 \\ i\neq j}}^{k} 
\frac{Q_{(0)}^\vee(a_i - \tfrac{\epsilon_1}{2}) }{
\vartheta_1^\prime(0)}
\frac{ Q_{(0)}^\vee(a_j - \tfrac{\epsilon_1}{2})}{
\vartheta_1^\prime(0)}
  L(a_i -x - \tfrac{\epsilon_1}{2})
  \bigg[ 
  L(a_i - a_j +\epsilon_1)
  -L(a_i - a_j )
 \notag \\
  &\qquad \qquad \qquad \qquad \qquad \qquad \qquad \qquad 
  +L(a_j - a_i +\epsilon_1)
  -L(a_j - a_i )
  \bigg]
  \notag \\
&\qquad
+s
\sum_{j=1}^k  
\frac{Q_{(0)}^\vee(a_j - \tfrac{\epsilon_1}{2})}{
\vartheta_1^\prime(0)}
Q_{(0)}(a_j - \tfrac{3\epsilon_1}{2})
\left[
L(a_j -x - \tfrac{\epsilon_1}{2})
+
L(a_j -x - \tfrac{3\epsilon_1}{2})
\right]
\notag \\
&\qquad 
+s 
\sum_{j=1}^k 
\left( 
\frac{Q_{(0)}^\vee(a_j - \tfrac{\epsilon_1}{2}) }{
\vartheta_1^\prime(0)}
\right)^2
L(a_j -x - \tfrac{\epsilon_1}{2})
\bigg[
\sum_{i=1}^k L(a_j -a_i -\epsilon_1)
+ \sum_{\substack{i=1\\ i\neq j}}^k L(a_j -a_i) 
\notag \\
  &\qquad \qquad \qquad \qquad \qquad \qquad \qquad \qquad 
-\sum_{i=1}^k \left( 
L(a_j -\tfrac{\epsilon_1}{2} -m_i +b)
+L(a_j -\tfrac{\epsilon_1}{2} -n_i -b) 
\right)
\bigg]
\notag \\
&\qquad 
+s\cdot Q_{(0)}(x)
\sum_{j=1}^l 
\frac{Q_{(0)}^\vee(a_j - \tfrac{\epsilon_1}{2})}{
\vartheta_1^\prime(0)}
\bigg[
L(a_j -x +\tfrac{\epsilon_1}{2} )
- L(a_j -x -\tfrac{\epsilon_1}{2} )
\notag \\
  &\qquad \qquad \qquad \qquad \qquad \qquad \qquad \qquad 
+L(x- a_j +\tfrac{3\epsilon_1}{2} )
- L(x-a_j  +\tfrac{\epsilon_1}{2} )
+ s L(a_j -x -\tfrac{\epsilon_1}{2} )
\bigg] 
\notag \\
&\qquad 
+s Q_{(0)}(x) 
\left( 
Q_{(0)}(x-\epsilon_1)
-\frac{1-s}{2}
Q_{(0)}(x)
\right)
\end{align}
the computation has been check against the NS-limit performed with 
\texttt{Mathematica} for $k=2,3$.
%
%
\subsection{Shift operator acting on defect partition function}
The shift operator $Y$ defined in \eqref{eq:def_shift_op} acts on the codimension 2 defect fugacity $x$. In the appendix, the action on the perturbative and non-perturbative part of the partition function is derived.
\subsubsection{Perturbative contribution}
\label{app:shift_perturbative}
The normalised perturbative part \eqref{eq:pert_part_NS_limit} for an 
$(0,s)$ defect can be written as 
\begin{align}
\widetilde{Z}_{\mathrm{pert}}^{(0,s)\Deff} &= 
    \PE \bigg[ 
\frac{s}{2(1-p)} \left( \frac{1+Q}{1-Q} \right)
   \left\{ (1-p)
   +\sqrt{p} \sum_{i=1}^{k}  
   \left( 
   \frac{X}{A_i }  - \frac{A_i}{X } 
\right) 
   \right\} \bigg]
   \quad \text{with} \; A_i = e^{a}
   \notag \\
&= 
    \PE \bigg[ 
\frac{1}{(1-p)} \left( \frac{1+Q}{1-Q} \right)
   \left\{ (1-p)
   +\sqrt{p} \sum_{i=1}^{k}  
   \left( 
   \frac{X }{A_i }  - \frac{A_i}{X } 
\right) 
   \right\} \bigg]^{\frac{s}{2}} \notag \\
&= 
   \prod_{j,h=0}^{\infty} \PE \bigg[ 
 \left( Q^j+Q^{j+1} \right)
   \left\{ 1
   + p^{h+\frac{1}{2}} \sum_{i=1}^{k}  
   \left( 
   L_i^{-1}  - L_i 
\right) 
   \right\} \bigg]^{\frac{s}{2}}
   \quad \text{with} \; L_i = \frac{A_i}{X} \,.
\end{align}
Focusing only on the $X$-dependent part, one proceeds further 
\begin{align}
 f(X) &= 
 \prod_{i=1}^k \prod_{j,h=0}^{\infty} \PE \Big[ 
 \left( Q^j+Q^{j+1} \right)
    p^{h+\frac{1}{2}}   
   \left( 
   L_i^{-1}  - L_i 
\right) \Big]^{\frac{1}{2}} 
\notag \\
&= 
 \prod_{i=1}^k \prod_{j,h=0}^{\infty} \PE \Big[ 
    Q^j p^{h+\frac{1}{2}} L_i^{-1}  - Q^j p^{h+\frac{1}{2}} L_i 
   Q^{j+1} p^{h+\frac{1}{2}} L_i^{-1}  - Q^{j+1} p^{h+\frac{1}{2}} L_i
 \Big]^{\frac{1}{2}}
 \notag \\
&= 
 \prod_{i=1}^k \prod_{j,h=0}^{\infty} 
 \sqrt{
 \frac{(1-  Q^j p^{h+\frac{1}{2}} L_i) (1- Q^{j+1} p^{h+\frac{1}{2}} 
L_i)}{
 (1- Q^j p^{h+\frac{1}{2}} L_i^{-1}) (1-  Q^{j+1} p^{h+\frac{1}{2}} L_i^{-1})}
 }
 \notag \,,
\end{align}
which can be expressed in different forms:
\begin{compactitem}
 \item As elliptic Gamma functions
 \begin{align}
   f(X)
   &= 
 \prod_{i=1}^k 
  \sqrt{
  \left(
 \prod_{j,h=0}^{\infty}
 \frac{
  (1- Q^{j+1} p^{h+1} \frac{L_i}{\sqrt{p}}) 
 }{
 (1- Q^j p^{h}  \frac{\sqrt{p}}{L_i} ) 
 }
 \right)^2
 \cdot
 \prod_{h=0}^{\infty}
\left(
1-p^h \frac{\sqrt{p}}{L_i} 
\right)
\left(
1-p^{h+1} \frac{L_i}{\sqrt{p}} 
\right)
 } \notag\\
&= 
 \prod_{i=1}^k
  \sqrt{
  \left(
 \Gamma( x+\tfrac{1}{2}\epsilon_1 -a_i,\tau,\epsilon_1)
 \right)^2
 \cdot
 \widetilde{\theta}_1( x+\tfrac{1}{2}\epsilon_1 -a_i,\epsilon_1)
 } \,,
 \label{eq:pert_part_Gamma_fct}
 \end{align}
 and the silly looking notation turns out to be useful to resolve a potential sign issues.
\item As inverse of Gamma functions
\begin{align}
 f(X) &= \prod_{i=1}^k 
 \sqrt{
 \left(
 \prod_{j,h=0}^\infty
 \frac{
 (1-p^h Q^j (\sqrt{p}L_i))
 }{
 (1-p^{h+1} Q^{j+1} \frac{1}{\sqrt{p}L_i})
 }
 \right)^2
 \cdot
 \prod_{h=0}^\infty
 \frac{1}{
 (1-p^h (\sqrt{p}L_i))
 (1-p^{h+1} \frac{1}{\sqrt{p}L_i})}
 }
 \notag\\
 &= \prod_{i=1}^k \sqrt{ \frac{1}{\widetilde{\theta}_1 
(a_i-x+\tfrac{1}{2}\epsilon_1,\epsilon_1)} \cdot 
\left(\frac{1}{
\Gamma(a_i-x+\tfrac{1}{2}\epsilon_1,\tau,\epsilon_1)
}
\right)^2
}\,,
\label{eq:pert_part_inv_Gamma_fct}
\end{align}
and the clumpsy looking notation is kept on purpose.
\end{compactitem}
The perturbative part becomes
\begin{align}
 \widetilde{Z}_{\mathrm{pert}}^{(0,s)\Deff} &= 
 \left( \PE\left[\frac{1+Q}{1-Q}\right]
 \cdot f(X) \right)^{\tfrac{s}{2}} \,.
\end{align}
Using the shift property in \eqref{eq:shift_property_theta} and the expression 
in terms of elliptic Gamma functions \eqref{eq:pert_part_Gamma_fct} and 
\eqref{eq:pert_part_inv_Gamma_fct}, one can straightforwardly show that
\begin{align}
Y \sqrt{
 \widetilde{\theta}_1( y_i,\epsilon_1)
  \cdot 
  \left(\Gamma( y_i,\tau,\epsilon_1) \right)^2
  }
 &= 
 \sqrt{
 \widetilde{\theta}_1( y_i-\epsilon_1,\epsilon_1)
  \cdot 
 \left(
 \Gamma( y_i-\epsilon_1,\tau,\epsilon_1)
 \right)^2
 }
 \quad \text{with} 
 \quad 
 y_i = x+\tfrac{1}{2}\epsilon_1 -a_i
 \notag \\
&= 
\sqrt{
 - e^{(y_i-\epsilon_1)}
 \widetilde{\theta}_1( y_i,\epsilon_1)
  \cdot
 \left( 
 \frac{ \Gamma( y_i,\tau,\epsilon_1)
 }{
 \widetilde{\theta}_1(y-\epsilon_1,\tau)}
 \right)^2
 }
 \qquad \text{using \eqref{eq:shift_Gamma}}
\notag \\
&= 
\sqrt{
 - e^{(y_i-\epsilon_1)}
 \widetilde{\theta}_1( y_i,\epsilon_1)
  \cdot
 \left( 
 \frac{i Q^{\frac{1}{12}}\eta(\tau)
 }{ e^{\frac{1}{2}(y_i-\epsilon_1)}
 \theta_1(y-\epsilon_1,\tau)}
 \right)^2
 \left( 
  \Gamma( y_i,\tau,\epsilon_1)
 \right)^2
 }
\notag \\
&= 
\sqrt{
 \left( 
 \frac{1 }{  \vartheta_1(y-\epsilon_1,\tau)}
 \right)^2
 }
 \cdot
 \sqrt{
 \widetilde{\theta}_1( y_i,\epsilon_1)
 \left( 
  \Gamma( y_i,\tau,\epsilon_1)
 \right)^2
 }
 \label{eq:shift_perturbative_check1}
\end{align}
and similarly
\begin{align}
 Y \sqrt{\frac{1}{\widetilde{\theta}_1 
(z_i,\epsilon_1)} \cdot 
\frac{1}{\left( 
\Gamma(z_i,\tau,\epsilon_1)
\right)^2}
}
&=
\sqrt{
\frac{1}{\widetilde{\theta}_1 
(z_i+\epsilon_1,\epsilon_1)} \cdot 
\frac{1}{
\left(
\Gamma(z_i+\epsilon_1,\tau,\epsilon_1)
\right)^2
}
}
\quad \text{with} 
 \quad 
 z_i = a_i-x+\tfrac{1}{2}\epsilon_1
 \notag \\
&=
\sqrt{
\frac{1}{- e^{-z_i}\widetilde{\theta}_1 
(z_i,\epsilon_1)} \cdot 
\frac{1}{
\left(
\widetilde{\theta}_{1}(z_i,\tau)
\Gamma(z_i,\tau,\epsilon_1)
\right)^2}
}
\qquad \text{using \eqref{eq:shift_Gamma}}
\notag \\
&=
\sqrt{
\frac{1}{- e^{-z_i}} \cdot 
\left(\frac{i Q^{\frac{1}{12} }\eta(\tau) }{ 
e^{\frac{z_i}{2}}\theta_{1}(z_i,\tau)}\right)^2
\frac{1}{
\left(
\widetilde{\theta}_1 
(z_i,\epsilon_1)
\Gamma(z_i,\tau,\epsilon_1)
\right)^2}
}
\notag \\
&=
\sqrt{ 
\left(\frac{ 1 }{ 
\vartheta_{1}(z_i,\tau)}\right)^2
}
\cdot
\sqrt{
\frac{1}{
\widetilde{\theta}_1
(z_i,\epsilon_1)
\left(
\Gamma(z_i,\tau,\epsilon_1)
\right)^2}
}
 \label{eq:shift_perturbative_check2}
\end{align}
such that both calculations \eqref{eq:shift_perturbative_check1} and 
\eqref{eq:shift_perturbative_check2} lead to \eqref{eq:shift_op_pert_part}.
\subsubsection{Elliptic genus}
\label{app:shift_elliptic_genus}
The defect part \eqref{eq:def_defect_terms} can be written as 
\begin{align}
\mathcal{V}_1^{(0,s)} &= 
s \cdot \partial_u \log \vartheta_1(u-x) 
= s \cdot\partial_u \log \left[ \frac{\im\, \eta 
Q^{\tfrac{1}{12}}}{e^{\tfrac{u-x}{2}}} \widetilde{\theta}_1(u-x)\right]
\qquad 
\text{using \eqref{eq:theta_defs}} \notag\\
&= s \cdot\partial_u\left( \log\widetilde{\theta}_1(u-x) +    \log 
\left[ \im\, 
\eta 
Q^{\tfrac{1}{12}}\right]
- \frac{1}{2}(u-x)
\right)
\notag \\
&= s \cdot \partial_u\left( \log 
\Gamma(u-x+\epsilon_1,\tau,\epsilon_1) - \log \Gamma(u-x,\tau,\epsilon_1)
+    \log \left[ \im\, \eta Q^{\tfrac{1}{12}}\right]
- \frac{1}{2}(u-x)
\right)
\notag \\
&= s \cdot \partial_u\left( \log 
\Gamma(u-x+\epsilon_1,\tau,\epsilon_1) - \log \Gamma(u-x,\tau,\epsilon_1)
- \frac{1}{2}(u-x)
\right)
\end{align}
where the $\log[\im\, \eta Q^{\frac{1}{12}}]$ term vanishes due to the derivative.
Then, the shift has the following effect:
\begin{align}
 \mathcal{V}_1^{(0,s)}(x\to x-\epsilon_1)
 &= s \cdot \partial_u\left( \log 
\Gamma(u-x+2\epsilon_1,\tau,\epsilon_1) - \log 
\Gamma(u-x+\epsilon_1,\tau,\epsilon_1)
- \frac{1}{2}(u-x+\epsilon_1)
\right)
\notag \\
 &=  s \cdot \partial_u\bigg( \log 
\Gamma(u-x+\epsilon_1,\tau,\epsilon_1)
+\log \widetilde{\theta}_1(u-x+\epsilon_1,\tau) \notag \\
&\qquad - \log \Gamma(u-x,\tau,\epsilon_1)
-\log \widetilde{\theta}_1(u-x,\tau)
- \frac{1}{2}(u-x+\epsilon_1)
\bigg)
\quad \text{using \eqref{eq:shift_Gamma}}
\notag \\
 &= \mathcal{V}_1^{(0,s)}(x) 
 + s \cdot 
 \partial_u\left( 
\log \widetilde{\theta}_1(u-x+\epsilon_1,\tau) -\log 
\widetilde{\theta}_1(u-x,\tau)
- \frac{1}{2}\epsilon_1
\right)
\notag \\
&= \mathcal{V}_1^{(0,s)}(x) 
+ s \cdot 
\partial_u\left( 
\log \left[ 
\frac{\widetilde{\theta}_1(u-x+\epsilon_1,\tau)}{\widetilde{\theta}_1(u-x,
\tau)} \cdot e^{- \frac{1}{2}\epsilon_1} \right] 
\right)
\notag \\
&= \mathcal{V}_1^{(0,s)}(x) 
+ s \cdot 
\partial_u\left( 
\log \left[ 
\frac{\vartheta_1(u-x+\epsilon_1,\tau)}{\vartheta_1(u-x,
\tau)} \right] 
\right)
\qquad 
\text{using \eqref{eq:theta_defs}, \eqref{eq:my_theta}}
\notag \\ 
&=
\mathcal{V}_1^{(0,s)}(x) 
- s \cdot 
\partial_x\left( 
\log \left[ 
\frac{\vartheta_1(u-x+\epsilon_1,\tau)}{\vartheta_1(u-x,
\tau)} \right] 
\right)
\,.
\end{align}
Therefore, it follows that
\begin{align}
 \int \diff u \ \rho_\ast (u) \mathcal{V}_1^{(0,s)}(x\to x-\epsilon_1) 
 &=
 \int \diff u \ \rho_\ast (u) \mathcal{V}_1^{(0,s)}(x)
 - s \cdot 
 \int \diff u \ \rho_\ast (u) 
 \partial_x\left( 
\log \left[ 
\frac{\theta_1(u-x+\epsilon_1,\tau)}{\theta_1(u-x-b,
\tau)} \right] 
\right) \notag \\
 &=
 \int \diff u \ \rho_\ast (u) \mathcal{V}_1^{(0,s)}(x)
 +
\left(
 \log \left[ \frac{\mathcal{Y}(x-\epsilon_1)}{\mathcal{Y}(x)} 
\right]
\right)^s
\end{align}
 and one arrives at \eqref{eq:shift_op_string_part}.
%
%
\subsection{Elliptic genera for theory with codimension 4 defect}
\label{app:details_codim=4}
In Section \ref{sec:Wilson_surface}, the theory in the presence of a codimension 4 defect has been considered. The elliptic genus can be computed via \eqref{eq:ell_genus_k-string_w/_Wilson_surface}, and the 1 and 2-string computations are detailed here. The chosen auxiliary vector in the JK-residue is $+1$ on 1-string level and $(1,1)$ on 2-string level.
\subsubsection{1-string}
\label{app:details_codim=4_k=1}
For 1-string contribution, one needs to evaluate the contour integral of 
\eqref{eq:ell_genus_k-string_w/_Wilson_surface} for $l=1$, i.e.
\begin{align}
Z^{\rm Wilson}_1 
&=\oint \frac{\diff u}{2\pi \im}\left(\frac{2\pi 
\,\eta^3\theta_1(2\epsilon_+)}{\theta_1(\epsilon_1)\,\theta_1(\epsilon_2)}
\right)\cdot
Q(u)
\cdot W(u)
\,.
\end{align}
Similar to the codimension 2 defect computation \eqref{eq:elliptic_genus_k=1}, 
there are two types of poles:
\begin{compactitem}
 \item $u=a_i -\epsilon_+ $ for $i=1,\ldots,k$
 \begin{align}
   Z_{1}^{\mathrm{Wilson}} \supset
\frac{\vartheta_1(2\epsilon_+)}{\vartheta_1(\epsilon_1)\,\vartheta_1(\epsilon_2)
}
\sum_{i=1}^k \left(
Q^\vee(a_i -\epsilon_+)
\cdot W(a_i-\epsilon_+) 
\right)  \,.
\label{eq:ell_genus_1-string_w/_Wilson_surface_pt1}
 \end{align}
 \item $u=z+\epsilon_+$
 \begin{align}
   Z_{1}^{\mathrm{Wilson}} &\supset
   Q(z+\epsilon_+) \,.
\label{eq:ell_genus_1-string_w/_Wilson_surface_pt2}
 \end{align}
\end{compactitem}
In total, the $l=1$ genus reads
\begin{align}
  Z_{1}^{\mathrm{Wilson}} 
&=
\frac{\vartheta_1(2\epsilon_+)}{\vartheta_1(\epsilon_1)\,\vartheta_1(\epsilon_2)
}
\sum_{i=1}^k 
Q^\vee(a_i -\epsilon_+)
\cdot W(a_i-\epsilon_+)
+
Q(z+\epsilon_+) \,,
\end{align}
where the notation \eqref{eq:residue_calc} has been used.
The normalised 1-string contribution in the NS-limit is derived as follows:
\begin{align}
 \widetilde{Z}_{1}^{\mathrm{Wilson}}  &=  Z_{1}^{\mathrm{Wilson}}  -  
Z_{1} \notag \\
 \lim_{\epsilon_2 \to 0} \widetilde{Z}_{1}^{\mathrm{Wilson}} 
  &=
   \lim_{\epsilon_2 \to 0}
   \left(
\frac{\vartheta_1(2\epsilon_+)}{\vartheta_1(\epsilon_1)\,\vartheta_1(\epsilon_2)
}
\sum_{i=1}^k 
Q^\vee(a_i -\epsilon_+)
\cdot \left[ W(a_i-\epsilon_+)-1 \right]
+
 Q(z+\epsilon_+)
 \right)
\notag \\
&= \frac{1}{\vartheta_1^\prime(0)}
\sum_{i=1}^k 
Q^\vee_{(0)}(a_i -\tfrac{1}{2}\epsilon_1)
\cdot \left[ L(a_i-z-\epsilon_1) - L(a_i-z)
\right]
+ Q_{(0)}(z+\tfrac{1}{2}\epsilon_1)
\,,
\end{align}
using \eqref{eq:def_L_and_K}, \eqref{eq:expansion_coeffs}, and
\begin{align}
 \lim_{\epsilon_2 \to 0}
 \frac{W(a_i-\epsilon_+)-1}{\vartheta_1(\epsilon_2)}
 = \frac{1}{\vartheta_1^\prime(0)}
 \left[
 L(a_i-z-\epsilon_1) - L(a_i-z)
 \right] \,.
\end{align}
\subsubsection{2-string}
\label{app:details_codim=4_k=2}
Consider the following $l=2$ elliptic genus 
\begin{align}
 Z_2^{\mathrm{Wilson}} &=\frac{1}{2} \oint \frac{\diff u_1 \diff u_2}{(2\pi 
\im)^2} 
 \left(
 \frac{2\pi \, \eta^3 \theta_1(2\epsilon_+)
 }{\theta_1(\epsilon_1)\theta_1(\epsilon_2)}
\right)^2
D(u_1-u_2)D(u_2-u_1)
\prod_{p=1}^2
Q(u_p)
W(u_p)
\end{align}
and the relevant poles can be split into poles that come from the theory 
without defect such as: 
\begin{compactitem}
 \item Both poles originate from $P_0(u_{p}+\epsilon_1+\epsilon_2)$ i.e.
 \begin{align}
  (u_1,u_2)=(a_i-\epsilon_+,a_j-\epsilon_+) \qquad \text{for }i\neq j \,.
 \end{align}
\item One pole from  $P_0(u_{p}+\epsilon_1+\epsilon_2)$ and one from $D(\pm 
(u_1-u_2))$, i.e.\
\begin{align}
\begin{aligned}
 (u_1,u_2)&=(a_m-\epsilon_+,a_m-\epsilon_+ -\epsilon_{1,2}) \qquad  \text{and} 
\\
 (u_1,u_2)&=(a_m-\epsilon_+ -\epsilon_{1,2},a_m-\epsilon_+) \,.
\end{aligned}
\end{align}
\end{compactitem}
In addition, there are new poles from the codimension 4 defect part. These are
\begin{compactitem}
 \item One pole from  $P_0(u_{p}+\epsilon_1+\epsilon_2)$  and one from 
$W(u_p)$, i.e.\
\begin{align}
 (u_1,u_2) = (a_m-\epsilon_+,z+\epsilon_+) 
 \qquad \text{and}\qquad 
 (u_1,u_2) = (z+\epsilon_+,a_m-\epsilon_+) \,.
\end{align}
 \item One pole from  $D(\pm (u_1-u_2))$  and one from 
$V_{(0,s)}(u_p)$, i.e.\
\begin{align}
 (u_1,u_2) = (z+\epsilon_+,z+\epsilon_+-\epsilon_{1,2})
 \qquad \text{and}\qquad 
 (u_1,u_2) = (z+\epsilon_+-\epsilon_{1,2},z+\epsilon_+) \,.
\end{align}
\end{compactitem}
Now, one can work out the residues for the individual poles as before:
Firstly, consider the contributions for 
$(u_1,u_2)=(a_i-\epsilon_+,a_j-\epsilon_+)$
\begin{align}
 Z_2^{\mathrm{Wilson}} &\supset 
 \frac{1}{2}
\left(
 \frac{  \vartheta_1(2\epsilon_+)
 }{\vartheta_1(\epsilon_1)\vartheta_1(\epsilon_2)}
\right)^2
D(a_i-a_j)D(a_j-a_i) \notag \\
&\quad \cdot
Q^\vee(a_i-\epsilon_+)
Q^\vee(a_j-\epsilon_+)
W(a_i-\epsilon_+)
W(a_j-\epsilon_+)
\,.
\end{align}
Secondly, both 
$(u_1,u_2)=(a_m-\epsilon_+,a_m-\epsilon_+ -\epsilon_{1})$
and 
$(u_1,u_2)=(a_m-\epsilon_+ -\epsilon_{1},a_m-\epsilon_+)$
yield
\begin{align}
 Z_2^{\mathrm{Wilson}}&\supset
\frac{1}{2} 
 \frac{  \vartheta_1(2\epsilon_+) \vartheta_1(\epsilon_1+2\epsilon_+)
 }{\vartheta_1(\epsilon_2) \vartheta_1(2\epsilon_-)  \vartheta_1(2\epsilon_1)}
Q^\vee(a_m-\epsilon_+)
Q(a_m-\epsilon_+-\epsilon_1)
\notag \\
&\qquad 
\cdot
W(a_m-\epsilon_+)
W(a_m-\epsilon_+ -\epsilon_{1})
\,.
\end{align}
Thirdly, both $(u_1,u_2)=(a_m-\epsilon_+,a_m-\epsilon_+ -\epsilon_{2})$
and $(u_1,u_2)=(a_m-\epsilon_+ -\epsilon_{2},a_m-\epsilon_+)$ yield
\begin{align}
 Z_2^{\mathrm{Wilson}}
&\supset -\frac{1}{2} 
 \frac{  \vartheta_1(2\epsilon_+) \vartheta_1(\epsilon_2+2\epsilon_+)
 }{\vartheta_1(\epsilon_1) \vartheta_1(2\epsilon_-)  \vartheta_1(2\epsilon_2)}
Q^\vee(a_m-\epsilon_+)
Q(a_m-\epsilon_+-\epsilon_2)
\notag \\
&\qquad \cdot
W(a_m-\epsilon_+)
W(a_m-\epsilon_+ -\epsilon_{2})
\,.
\end{align}
Fourthly, both $(u_1,u_2) = (a_m-\epsilon_+,z+\epsilon_+)$ 
and  $(u_1,u_2) = (z+\epsilon_+,a_m-\epsilon_+)$ yield
\begin{align}
 Z_2^{\mathrm{Wilson}}
&\supset
\frac{1}{2}
\frac{\vartheta_1(2\epsilon_+)}{
\vartheta_1(\epsilon_1)
\vartheta_1(\epsilon_2)}
D(a_m-z-2\epsilon_+)
D(z+2\epsilon_+-a_m)
\notag \\
&\qquad 
\cdot
Q^\vee(a_m-\epsilon_+)
Q(z+\epsilon_+)
W(a_m-\epsilon_+)
\,.
\end{align}
Fifthly, both $(u_1,u_2) = (z+\epsilon_+,z+\epsilon_+-\epsilon_{1})$
and  $(u_1,u_2) = (z+\epsilon_+-\epsilon_{1},z+\epsilon_+)$
 \begin{align}
 Z_2^{\mathrm{Wilson}}
&\supset
\frac{1}{2}
\frac{
\vartheta_1(\epsilon_1)
\vartheta_1(\epsilon_1+2\epsilon_+)}{
\vartheta_1(2\epsilon_-)
 \vartheta_1(2\epsilon_1)}
\cdot 
Q(z+\epsilon_+-\epsilon_1)
Q(z+\epsilon_+)
W(z+\epsilon_+-\epsilon_1)
=0 \,,
\end{align}
because  $W(z+\epsilon_+-\epsilon_1)=0$.
Lastly, both $(u_1,u_2) = (z+\epsilon_+,z+\epsilon_+-\epsilon_{2})$
and  $(u_1,u_2) = (z+\epsilon_+-\epsilon_{2},z+\epsilon_+)$ yield
\begin{align}
 Z_2^{\mathrm{Wilson}}
&\supset
-\frac{1}{2}
\frac{\vartheta_1(\epsilon_2)
\vartheta_1(\epsilon_2+2\epsilon_+)
}{
\vartheta_1(2\epsilon_-)
\vartheta_1(2\epsilon_2)
}
\cdot 
Q(z+\epsilon_+-\epsilon_2)
Q(z+\epsilon_+)
W(z+\epsilon_+-\epsilon_2)
=0 \,,
\end{align}
because $W(z+\epsilon_+-\epsilon_2)=0$. 
Summing up all the individual contributions leads to 
\begin{align}
 Z_2^{\mathrm{Wilson}}
&=
\left(
 \frac{  \vartheta_1(2\epsilon_+)
 }{\vartheta_1(\epsilon_1)\vartheta_1(\epsilon_2)}
\right)^2
\sum_{1\leq i< j \leq k}
D(a_i-a_j)D(a_j-a_i)  \\
&\qquad \qquad \qquad \qquad \qquad  \cdot
Q^\vee(a_i-\epsilon_+)
Q^\vee(a_j-\epsilon_+)
W(a_i-\epsilon_+)
W(a_j-\epsilon_+) 
\notag \\
&+
 \frac{  \vartheta_1(2\epsilon_+) 
 }{ \vartheta_1(2\epsilon_-) }
 \sum_{j=1}^k
 Q^\vee(a_j-\epsilon_+)
W(a_j-\epsilon_+)
\notag \\
&\qquad\qquad \qquad  \cdot
\bigg[
\frac{\vartheta_1(\epsilon_1+2\epsilon_+)}{
\vartheta_1(\epsilon_2)
\vartheta_1(2\epsilon_1)
}
Q(a_j-\epsilon_+-\epsilon_1)
W(a_j-\epsilon_+ -\epsilon_{1})
\notag
\\
&\qquad \qquad \qquad \qquad \qquad -
\frac{\vartheta_1(\epsilon_2+2\epsilon_+)}{
\vartheta_1(\epsilon_1)
\vartheta_1(2\epsilon_2)
}
Q(a_j-\epsilon_+-\epsilon_2)
W(a_j-\epsilon_+ -\epsilon_{2})
\bigg]
\notag \\
&+
\frac{\vartheta_1(2\epsilon_+)}{
\vartheta_1(\epsilon_1)
\vartheta_1(\epsilon_2)}
\sum_{j=1}^k
D(a_j-z-2\epsilon_+)
D(z+2\epsilon_+-a_j)
\notag \\
&\qquad \qquad \qquad \qquad  \cdot
Q^\vee(a_j-\epsilon_+)
Q(z+\epsilon_+)
W(a_j-\epsilon_+)
\notag 
\end{align}
For the evaluation of the normalised partition function in the NS-limit, the computation is split into several steps as above:
\begin{align}
 Z_2^{\mathrm{Wilson}} - Z_2 = J_1 + J_2 +J_3
\end{align}
with the following parts:
\begin{align}
 J_1 &=
 \left(
 \frac{  \vartheta_1(2\epsilon_+)
 }{\vartheta_1(\epsilon_1)\vartheta_1(\epsilon_2)}
\right)^2
\sum_{1\leq i< j \leq k}
D(a_i-a_j)D(a_j-a_i)  \\
&\qquad \qquad \qquad \qquad \qquad  \cdot
Q^\vee(a_i-\epsilon_+)
Q^\vee(a_j-\epsilon_+)
\left[ W(a_i-\epsilon_+)
W(a_j-\epsilon_+) 
-1 \right]
\,,
 \notag \\
  J_2 &=
  \frac{  \vartheta_1(2\epsilon_+) 
 }{ \vartheta_1(2\epsilon_-) }
 \sum_{j=1}^k
 Q^\vee(a_j-\epsilon_+)
 \\
&\qquad\qquad \qquad  \cdot
\bigg[
\frac{\vartheta_1(\epsilon_1+2\epsilon_+)}{
\vartheta_1(\epsilon_2)
\vartheta_1(2\epsilon_1)
}
Q(a_j-\epsilon_+-\epsilon_1)
\left[ 
W(a_j-\epsilon_+)
W(a_j-\epsilon_+ -\epsilon_{1})
-1
\right]
\notag
\\
&\qquad \qquad \qquad \qquad \qquad -
\frac{\vartheta_1(\epsilon_2+2\epsilon_+)}{
\vartheta_1(\epsilon_1)
\vartheta_1(2\epsilon_2)
}
Q(a_j-\epsilon_+-\epsilon_2)
\left[
W(a_j-\epsilon_+)
W(a_j-\epsilon_+ -\epsilon_{2})
-1
\right]
\bigg]
\,,
\notag \\
 J_3 &=
 \frac{\vartheta_1(2\epsilon_+)}{
\vartheta_1(\epsilon_1)
\vartheta_1(\epsilon_2)}
\sum_{j=1}^k
D(a_j-z-2\epsilon_+)
D(z+2\epsilon_+-a_j)
\notag \\
&\qquad \qquad \qquad \qquad  \cdot
Q^\vee(a_j-\epsilon_+)
Q(z+\epsilon_+)
W(a_j-\epsilon_+)
\,,
\end{align}
and the $\epsilon_2$ expansion yields
\begin{align}
 J_1 \big|_{\frac{1}{\epsilon_2}} &= 
 \left(\frac{1}{\vartheta^\prime_1(0)}\right)^2
 \sum_{i<j} 
 Q_{(0)}^\vee (a_i -\tfrac{1}{2}\epsilon_1)
 Q_{(0)}^\vee (a_j -\tfrac{1}{2}\epsilon_1)
 W_{(1)}(a_i -\tfrac{1}{2}\epsilon_1,a_j -\tfrac{1}{2}\epsilon_1)
 \notag \\
 J_1 \big|_{\mathrm{fin}} &= 
 \left(\frac{1}{\vartheta^\prime_1(0)}\right)^2
 \sum_{i<j} 
 Q_{(0)}^\vee (a_i -\tfrac{1}{2}\epsilon_1)
 Q_{(0)}^\vee (a_j -\tfrac{1}{2}\epsilon_1)
 W_{(2)}(a_i -\tfrac{1}{2}\epsilon_1,a_j -\tfrac{1}{2}\epsilon_1)
 \notag
 \\
&+ \left(\frac{1}{\vartheta^\prime_1(0)}\right)^2
\sum_{i<j} 
\Bigg\{
Q_{(0)}^\vee (a_i -\tfrac{1}{2}\epsilon_1)
 Q_{(1)}^\vee (a_j -\tfrac{1}{2}\epsilon_1)
 +
 Q_{(1)}^\vee (a_i -\tfrac{1}{2}\epsilon_1)
 Q_{(0)}^\vee (a_j -\tfrac{1}{2}\epsilon_1)
\Bigg\}
W_{(1)}(a_i -\tfrac{1}{2}\epsilon_1,a_j -\tfrac{1}{2}\epsilon_1)
\notag \\
&+ \left(\frac{1}{\vartheta^\prime_1(0)}\right)^2
\sum_{i<j}
\left( 
D^{(1)}(a_i-a_j)
+
D^{(1)}(a_j-a_i)
\right)
 Q_{(0)}^\vee (a_i -\tfrac{1}{2}\epsilon_1)
 Q_{(0)}^\vee (a_j -\tfrac{1}{2}\epsilon_1)
 W_{(1)}(a_i -\tfrac{1}{2}\epsilon_1,a_j -\tfrac{1}{2}\epsilon_1)
\notag \\
&+ 2\frac{B^{(0)}}{\vartheta^\prime_1(0)} 
\sum_{i<j}
 Q_{(0)}^\vee (a_i -\tfrac{1}{2}\epsilon_1)
 Q_{(0)}^\vee (a_j -\tfrac{1}{2}\epsilon_1)
 W_{(1)}(a_i -\tfrac{1}{2}\epsilon_1,a_j -\tfrac{1}{2}\epsilon_1)
 \,,
 \notag \\
 J_2 \big|_{\frac{1}{\epsilon_2}} &= 
 \frac{1}{2} \left( \frac{1}{\vartheta^\prime_1(0)} \right)^2
 \sum_{j} 
 Q^\vee_{(0)} (a_j -\epsilon_+ -\epsilon_2)
 Q^\vee_{(0)} (a_j -\epsilon_+)
 W_{(1)}(a_j-\epsilon_+,a_j-\epsilon_+-\epsilon_2)
 \notag \\
 J_2 \big|_{\mathrm{fin}} &= 
 \frac{1}{\vartheta^\prime_1(0)} 
 \sum_j 
 Q^\vee_{(0)}(a_j-\tfrac{1}{2}\epsilon_1)
 Q^{(0)}(a_j-\tfrac{3}{2} \epsilon_1)
 W_{(1)}(a_j-\epsilon_+,a_j-\epsilon_+-\epsilon_1)
 \notag \\
&-\frac{1}{2\vartheta^\prime_1(0)}
 \sum_j
 Q^\vee_{(0)}(a_j-\tfrac{1}{2}\epsilon_1)
 \widetilde{Q}(a_j-\tfrac{1}{2}\epsilon_1)
 W_{(1)}(a_j-\epsilon_+,a_j-\epsilon_+-\epsilon_2)
 \notag \\
&+\frac{B^{(0)}}{\vartheta^\prime_1(0)}
\sum_j 
\left(
Q^\vee_{(0)}(a_j-\tfrac{1}{2}\epsilon_1)
\right)^2
W_{(1)}(a_j-\epsilon_+,a_j-\epsilon_+-\epsilon_2) 
\notag \\
&+\frac{1}{2} \left( \frac{1}{\vartheta^\prime_1(0)} \right)^2
\sum_j 
\bigg[
2 L(\epsilon_1) 
Q^\vee_{(0)} (a_j -\tfrac{1}{2}\epsilon_1)
+ Q^\vee_{(1)} (a_j -\tfrac{1}{2}\epsilon_1)
\bigg]
Q^\vee_{(0)} (a_j -\tfrac{1}{2}\epsilon_1)
W_{(1)}(a_j-\epsilon_+,a_j-\epsilon_+-\epsilon_2) 
\notag \\
&+\frac{1}{2( \vartheta^\prime_1(0))^2}
\sum_j 
\left( 
Q^\vee_{(0)} (a_j -\tfrac{1}{2}\epsilon_1)
\right)^2
W_{(2)}(a_j-\epsilon_+,a_j-\epsilon_+-\epsilon_2) 
\,,
\notag \\
J_3 \big|_{\frac{1}{\epsilon_2}} &= 
\frac{1}{\vartheta^\prime_1(0)} Q_{(0)}(z+\epsilon_+)
\sum_j Q^\vee_{(0)}(a_j-\epsilon_+)
 \notag \\
 J_3 \big|_{\mathrm{fin}} &= 
 \frac{1}{\vartheta^\prime_1(0)} 
 \sum_j 
 \left( 
 D^{(1)}(a_j-z-2\epsilon_+)
 +
 D^{(1)}(z+2\epsilon_+ - a_j)
 \right)
 Q^\vee_{(0)}(a_j-\tfrac{1}{2}\epsilon_1)
 Q_{(0)}(z+\epsilon_+)
 \notag \\
&+ \frac{1}{\vartheta^\prime_1(0)} \sum_j
\bigg[
Q^\vee_{(1)}(a_j-\tfrac{1}{2}\epsilon_1)
Q_{(0)}(z+\epsilon_+)
+
Q^\vee_{(0)}(a_j-\tfrac{1}{2}\epsilon_1)
Q_{(1)}(z+\epsilon_+)
\notag \\
&\qquad +
Q^\vee_{(0)}(a_j-\tfrac{1}{2}\epsilon_1)
Q_{(0)}(z+\epsilon_+)
W_{(1)}(a_j-\epsilon_+)
\bigg]
\notag \\
&+B^{(0)} Q_{(0)}(z+\epsilon_+) 
\sum_j 
Q^\vee_{(0)}(a_j-\epsilon_+)
\,.
\notag
\end{align}
With the conventions \eqref{eq:def_L_and_K} and \eqref{eq:expansion_coeffs}, the normalised 2-string 
elliptic genus in presence of a codimension 4 defect reads
\begin{align}
 \widetilde{Z}_2^{\mathrm{Wilson}} &=
 \sum_{\substack{i,j=1\\ i\neq j}}^k  
 \frac{Q_{(0)}^\vee(a_i - \tfrac{\epsilon_1}{2}) }{\vartheta_1^\prime(0)}
 \frac{Q_{(0)}^\vee(a_j - \tfrac{\epsilon_1}{2}) }{\vartheta_1^\prime(0)}
 \bigg[
 \frac{1}{2} L(a_i-z)L(a_j-z) \notag \\
 &\qquad \qquad \qquad \qquad -L(a_i-z)L(a_j-z-\epsilon_1)
 +\frac{1}{2} L(a_i-z-\epsilon_1)L(a_j-z-\epsilon_1)
 \bigg] \notag \\
&+\frac{1}{2} \sum_{j=1}^k 
\left( \frac{Q_{(0)}^\vee(a_j - \tfrac{\epsilon_1}{2}) }{\vartheta_1^\prime(0)} 
\right)^2 
\left[ 
K(a_j-z)-K(a_j-z-\epsilon_1)
\right] \notag  \\
&+\frac{1}{2} \sum_{j=1}^k 
\left( \frac{Q_{(0)}^\vee(a_j - \tfrac{\epsilon_1}{2}) }{\vartheta_1^\prime(0)} 
\right)^2 
L(a_j-z-\epsilon_1)
\left[ 
L(a_j-z-\epsilon_1)-L(a_j-z)
\right] \notag  \\
&+2 L(\epsilon_1) \sum_{j=1}^k 
\left( \frac{Q_{(0)}^\vee(a_j - \tfrac{\epsilon_1}{2}) }{\vartheta_1^\prime(0)} 
\right)^2
\left[ 
L(a_j-z-\epsilon_1)-L(a_j-z)
\right] \notag  \\
&+ \sum_{j=1}^k 
\left( \frac{Q_{(0)}^\vee(a_j - \tfrac{\epsilon_1}{2}) }{\vartheta_1^\prime(0)} 
\right)^2
\left[ 
L(a_j-z-\epsilon_1)-L(a_j-z)
\right] 
\bigg[
\sum_{i=1}^k L(a_j -a_i -\epsilon_1) 
+\sum_{\substack{i=1\\ i\neq j}}^k L(a_j -a_i -\epsilon_1)  
\notag  \\
&\qquad \qquad \qquad \qquad
-\sum_{i=1}^k \left( L(a_j-\tfrac{\epsilon_1}{2} -m_i +b)
+ L(a_j-\tfrac{\epsilon_1}{2} -n_i -b)\right)
\bigg]
\notag  \\
&+ \sum_{\substack{i,j=1\\ i \neq j}}^k 
\frac{Q_{(0)}^\vee(a_i - \tfrac{\epsilon_1}{2}) }{\vartheta_1^\prime(0)} 
\frac{Q_{(0)}^\vee(a_j - \tfrac{\epsilon_1}{2}) }{\vartheta_1^\prime(0)} 
\left[ 
L(a_i-z-\epsilon_1)-L(a_i-z)
\right] 
\notag  \\
&\qquad \qquad \qquad \qquad
\cdot \bigg[
L(a_i -a_j +\epsilon_1) 
- L(a_i -a_j )
+L(a_j -a_i +\epsilon_1) 
- L(a_j -a_i )  
\bigg]
\notag  \\
&+\sum_{j=1}^k 
\frac{Q_{(0)}^\vee(a_j - \tfrac{\epsilon_1}{2}) }{\vartheta_1^\prime(0)} 
Q_{(0)}(a_j - \tfrac{3\epsilon_1}{2})
\left[ 
L(a_j-z-2\epsilon_1)-L(a_j-z)
\right] 
\notag \\
&+
Q_{(0)}(z + \tfrac{\epsilon_1}{2})
\sum_{j=1}^k 
\frac{Q_{(0)}^\vee(a_j - \tfrac{\epsilon_1}{2}) }{\vartheta_1^\prime(0)}
\left[ 
L(z-a_j+2\epsilon_1)-L(z-a_j+\epsilon_1)
\right] 
\,,
\end{align}
which has been checked against the explicit NS-limit for $k=2,3$ via 
\texttt{Mathematica}.
%
%
\subsection{Computation of \texorpdfstring{$P(x+\epsilon_1)$}{P(x)} 
coefficients}
In Section \ref{sec:difference_eq}, the function $P(x)$ appeared in the derivation of the difference equation \eqref{eq:diff_equation}. The main focus of Section \ref{sec:comparison} is to argue that $P$ is related to the expectation value of a Wilson surface. Here, the details of the 1 and 2-string comparison are presented.
\subsubsection{1-string}
\label{app:details_P1}
Consider the prediction \eqref{eq:prediction_P1}, then start by computing
\begin{align}
(Y^{-1}-1) \widetilde{Z}_1^{(0,1)\Deff}&=
 \frac{ 1}{  \vartheta_1^\prime(0)}
\sum_{i=1}^k 
Q_{(0)}^\vee(a_i-\tfrac{1}{2}\epsilon_1)
\cdot
\left[
L(a_i-x-\tfrac{3}{2}\epsilon_1)
-
L(a_i-x-\tfrac{1}{2}\epsilon_1)
\right] \\
&\qquad \qquad \qquad +
Q_{(0)}(x+\epsilon_1)
-
Q_{(0)}(x) \notag
\end{align}
such that the addition of $Q_{(0)}(x)$ results in \eqref{eq:prediction_P1_calc}.
\subsubsection{2-string}
\label{app:details_P2}
Work out the 2-string prediction \eqref{eq:prediction_P2} with the results from 
above. To begin with, set $s=1$ then detail $(Y^{-1}-1)Z_2$ with 
$Y^{-1}f(x)=f(x+\epsilon_1)$
\begin{align}
  (Y^{-1}-1)\widetilde{Z}_{k=2}^{(0,1)\Deff}  &=
  \sum_{j=1}^k 
  \left( 
  \frac{Q_{(0)}^\vee(a_j - \tfrac{\epsilon_1}{2})}{
  \vartheta_1^\prime(0)} \right)^2
  \left(
  L(a_j -x - \tfrac{3\epsilon_1}{2}) 
  \right)^2
  -\left(
  L(a_j -x - \tfrac{\epsilon_1}{2}) 
  \right)^2
  \notag \\
&\qquad 
-\frac{1}{2} 
\sum_{j=1}^k
\left( 
\frac{Q_{(0)}^\vee(a_j - \tfrac{\epsilon_1}{2}) }{
\vartheta_1^\prime(0)}
\right)^2
\left[
 K(a_j -x - \tfrac{3\epsilon_1}{2}) 
 - K(a_j -x - \tfrac{\epsilon_1}{2}) 
 \right]
 \notag \\
&\qquad 
+\frac{1}{2}
\sum_{\substack{i,j=1\\ i\neq j}}^k
\frac{  Q_{(0)}^\vee(a_i - \tfrac{\epsilon_1}{2})}{
\vartheta_1^\prime(0)}
\frac{ Q_{(0)}^\vee(a_j - \tfrac{\epsilon_1}{2})}{
\vartheta_1^\prime(0)} \notag \\
&\qquad \qquad \qquad \qquad 
\cdot\big[
  L(a_i -x - \tfrac{3\epsilon_1}{2})
  L(a_j -x - \tfrac{3\epsilon_1}{2}) 
  -
   L(a_i -x - \tfrac{\epsilon_1}{2})
  L(a_j -x - \tfrac{\epsilon_1}{2}) 
  \big]
  \notag \\
&\qquad 
+2  \cdot L(\epsilon_1) 
\sum_{j=1}^k \left(
\frac{Q_{(0)}^\vee(a_j - \tfrac{\epsilon_1}{2}) }{
\vartheta_1^\prime(0)}
\right)^2
\left[
L(a_j -x- \tfrac{3\epsilon_1}{2})
-
L(a_j -x- \tfrac{\epsilon_1}{2})
\right]
\notag \\
&\qquad
+  
\sum_{\substack{i,j=1 \\ i\neq j}}^{k} 
\frac{Q_{(0)}^\vee(a_i - \tfrac{\epsilon_1}{2}) }{
\vartheta_1^\prime(0)}
\frac{ Q_{(0)}^\vee(a_j - \tfrac{\epsilon_1}{2})}{
\vartheta_1^\prime(0)}
  L(a_i -x - \tfrac{\epsilon_1}{2})
  \bigg[ 
  L(a_i - a_j +\epsilon_1)
  -L(a_i - a_j )
 \notag \\
  &\qquad \qquad \qquad \qquad \qquad \qquad \qquad \qquad 
  +L(a_j - a_i +\epsilon_1)
  -L(a_j - a_i )
  \bigg]
  \notag \\
&\qquad
+
\sum_{j=1}^k  
\frac{Q_{(0)}^\vee(a_j - \tfrac{\epsilon_1}{2})}{
\vartheta_1^\prime(0)}
Q_{(0)}(a_j - \tfrac{3\epsilon_1}{2})
\left[
L(a_j -x - \tfrac{3\epsilon_1}{2})
+
L(a_j -x - \tfrac{\epsilon_1}{2})
\right]
\notag \\
&\qquad 
+
\sum_{j=1}^k 
\left( 
\frac{Q_{(0)}^\vee(a_j - \tfrac{\epsilon_1}{2}) }{
\vartheta_1^\prime(0)}
\right)^2
\left[
L(a_j -x - \tfrac{3\epsilon_1}{2})
-
L(a_j -x - \tfrac{\epsilon_1}{2})
\right]
\notag \\
  &\qquad \qquad 
\bigg[
\sum_{i=1}^k L(a_j -a_i -\epsilon_1)
+ \sum_{\substack{i=1\\ i\neq j}}^N L(a_j -a_i)
-\sum_{i=1}^k \left( 
L(a_j -\tfrac{\epsilon_1}{2} -m_i +b)
+L(a_j -\tfrac{\epsilon_1}{2} -n_i -b) 
\right)
\bigg]
\notag \\
&\qquad 
+ \sum_{j=1}^k 
\frac{Q_{(0)}^\vee(a_j - \tfrac{\epsilon_1}{2})}{
\vartheta_1^\prime(0)}
\Bigg\{
Q_{(0)}(x+\epsilon_1)
\bigg[
L(a_j -x -\tfrac{\epsilon_1}{2} )
+L(x- a_j +\tfrac{5\epsilon_1}{2} )
- L(x-a_j  +\tfrac{3\epsilon_1}{2} )
\bigg]
\notag \\
  &\qquad \qquad \qquad \qquad \qquad \qquad 
-Q_{(0)}(x)
\bigg[
 + L(a_j -x +\tfrac{\epsilon_1}{2} )
+L(x- a_j +\tfrac{3\epsilon_1}{2} )
- L(x-a_j  +\tfrac{\epsilon_1}{2} )
\bigg] 
\Bigg\}
\notag \\
&\qquad 
+Q_{(0)}(x)
\left[
Q_{(0)}(x+\epsilon_1)
-
Q_{(0)}(x-\epsilon_1)
\right] \,.
\end{align}
Next, one needs to work out the following contribution:
\begin{align}
 (Y-1)\widetilde{Z}_{1}^{(0,1)\Deff} 
 &=
\sum_{i=1}^k
 \frac{ Q_{(0)}^\vee(a_i-\tfrac{1}{2}\epsilon_1) }{  
\vartheta_1^\prime(0)}
\cdot
\left[
L(a_i-x+\tfrac{1}{2}\epsilon_1)
-
L(a_i-x-\tfrac{1}{2}\epsilon_1)
\right] \notag \\
&\qquad \qquad \qquad 
+
Q_{(0)}(x-\epsilon_1)
-
Q_{(0)}(x)  \,,
\end{align}
such that 
\begin{align}
 Q_{(0)}(x)(Y-1)\widetilde{Z}_{1}^{(0,1)\Deff} 
 &=
 Q_{(0)}(x) 
\sum_{i=1}^k
\frac{ Q_{(0)}^\vee(a_i-\tfrac{1}{2}\epsilon_1) }{  
\vartheta_1^\prime(0)}
\cdot
\left[
L(a_i-x+\tfrac{1}{2}\epsilon_1)
-
L(a_i-x-\tfrac{1}{2}\epsilon_1)
\right] \notag \\
&\qquad \qquad \qquad 
+
Q_{(0)}(x)
\left(
Q_{(0)}(x-\epsilon_1)
-
Q_{(0)}(x) 
\right) \,.
\end{align}
In addition, one needs the following contribution:
\begin{align}
 \widetilde{Z}_{1}^{(0,1)\Deff} 
 \cdot (Y^{-1}-1)\widetilde{Z}_{1}^{(0,1)\Deff} 
 &=
 \left(
 \sum_{j=1}^k
\frac{Q_{(0)}^\vee(a_j-\tfrac{1}{2}\epsilon_1) }{  
\vartheta_1^\prime(0)}
\cdot
L(a_j-x-\tfrac{1}{2}\epsilon_1)
+ Q_{(0)}(x)
\right) \notag \\
 &\cdot \Bigg(
\sum_{i=1}^k 
\frac{ Q_{(0)}^\vee(a_i-\tfrac{1}{2}\epsilon_1)
}{  \vartheta_1^\prime(0)}
\cdot
\left[
L(a_i-x-\tfrac{3}{2}\epsilon_1)
-
L(a_i-x-\tfrac{1}{2}\epsilon_1)
\right] \notag \\
&\qquad \qquad \qquad +
Q_{(0)}(x+\epsilon_1)
-
Q_{(0)}(x) \Bigg)\notag\\
&= 
\sum_{i,j=1}^k
\frac{Q_{(0)}^\vee(a_j-\tfrac{1}{2}\epsilon_1) }{  
\vartheta_1^\prime(0)}
\frac{ Q_{(0)}^\vee(a_i-\tfrac{1}{2}\epsilon_1)
}{  \vartheta_1^\prime(0)}
L(a_j-x-\tfrac{1}{2}\epsilon_1)
 \notag \\
&\qquad \qquad 
\cdot \left[
L(a_i-x-\tfrac{3}{2}\epsilon_1)
-
L(a_i-x-\tfrac{1}{2}\epsilon_1)
\right] \notag \\
&\qquad + Q_{(0)}(x) 
\cdot
\sum_{i=1}^k 
\frac{ Q_{(0)}^\vee(a_i-\tfrac{1}{2}\epsilon_1)
}{  \vartheta_1^\prime(0)}
\cdot
\left[
L(a_i-x-\tfrac{3}{2}\epsilon_1)
- 2
L(a_i-x-\tfrac{1}{2}\epsilon_1)
\right] 
\notag \\
&\qquad +Q_{(0)}(x+\epsilon_1)
\cdot
\sum_{j=1}^k
\frac{Q_{(0)}^\vee(a_j-\tfrac{1}{2}\epsilon_1) }{  
\vartheta_1^\prime(0)}
\cdot
L(a_j-x-\tfrac{1}{2}\epsilon_1)
\notag  \\
&\qquad + Q_{(0)}(x) 
\left(
Q_{(0)}(x+\epsilon_1)
-
Q_{(0)}(x)
\right) \,.
\end{align}
Combining the individual terms, one finds \eqref{eq:prediction_P2_calc}
%
%
 \bibliographystyle{JHEP}     
 {\footnotesize{
 \bibliography{references}
 }}
\end{document}